\documentclass[
lettersize,journal]{IEEEtran}
\pdfoutput=1
\usepackage{amsmath,amsfonts}
\usepackage{algorithmic}
\usepackage{algorithm}
\usepackage{array}
\usepackage[caption=false,font=normalsize,labelfont=sf,textfont=sf]{subfig}
\usepackage{textcomp}
\usepackage{stfloats}
\usepackage{url}
\usepackage{verbatim}
\usepackage{pdfpages}

\usepackage{graphicx}
\usepackage[noadjust]{cite}
\usepackage{tikz}
 \usepackage[normalem]{ulem}
\usepackage{hyperref}
\usetikzlibrary{spy}
\usepackage{amsmath}
\usepackage{amssymb}
\usepackage{tikz} % To generate the plot from csv
\usepackage{pgfplots}
\usepackage{tabularx}
\usepackage{xcolor}
\usepackage{color}
\pgfplotsset{compat=newest}
\usepackage{mathtools}
\definecolor{orcidlogocol}{HTML}{A6CE39}
\tikzset{
  orcidlogo/.pic={
    \fill[orcidlogocol] svg{M256,128c0,70.7-57.3,128-128,128C57.3,256,0,198.7,0,128C0,57.3,57.3,0,128,0C198.7,0,256,57.3,256,128z};
    \fill[white] svg{M86.3,186.2H70.9V79.1h15.4v48.4V186.2z}
                 svg{M108.9,79.1h41.6c39.6,0,57,28.3,57,53.6c0,27.5-21.5,53.6-56.8,53.6h-41.8V79.1z M124.3,172.4h24.5c34.9,0,42.9-26.5,42.9-39.7c0-21.5-13.7-39.7-43.7-39.7h-23.7V172.4z}
                 svg{M88.7,56.8c0,5.5-4.5,10.1-10.1,10.1c-5.6,0-10.1-4.6-10.1-10.1c0-5.6,4.5-10.1,10.1-10.1C84.2,46.7,88.7,51.3,88.7,56.8z};
  }
}
\newcommand\orcid[1]{\href{https://orcid.org/#1}{\mbox{\scalerel*{
\begin{tikzpicture}[yscale=-1,transform shape]
\pic{orcidlogo};
\end{tikzpicture}
}{|}}}}
\newcommand\myeq{\stackrel{\mathclap{\normalfont\mbox{(a)}}}{=}}
\newcommand\myeqb{\stackrel{\mathclap{\normalfont\mbox{(b)}}}{=}}

\newcommand\myineqlb{\stackrel{\mathclap{\normalfont\mbox{(b)}}}{\leq}}
\newcommand\myineqla{\stackrel{\mathclap{\normalfont\mbox{(a)}}}{\leq}}

\newcommand\myineqb{\stackrel{\mathclap{\normalfont\mbox{(b)}}}{\geq}}
\newcommand\myineqc{\stackrel{\mathclap{\normalfont\mbox{(c)}}}{\geq}}
\definecolor{clr1}{rgb}{0.0, 0.0, 1.0}
\definecolor{clr2}{rgb}{0.96, 0.73, 1.0}
\definecolor{clr3}{rgb}{1.0, 0.01, 0.24}
\definecolor{clr4}{rgb}{0.0, 0.5, 0.0}
\definecolor{clr5}{rgb}{1.0, 0.49, 0.0}
\definecolor{clr6}{rgb}{0.1, 0.1, 0.44}

\begin{document}

% \title{Impact of limited memory in the access points 
% on the optimal cell-free massive MIMO fronthaul architecture
% }
\title{Cell-free Massive MIMO with Sequential Fronthaul Architecture and Limited Memory Access Points
 }
\author{\IEEEauthorblockN{Vida Ranjbar, Robbert Beerten, Marc Moonen, Sofie Pollin}%
% \IEEEauthorblockA{\textit{Department of Electrical Engineering, KU Leuven, Belgium} \\
% Email: \{vida.ranjbar, robbert.beerten, marc.moonen, sofie.pollin\}@kuleuven.be}%
\thanks{
Some parts of this work have been accepted for potential publication at IEEE Global Communications Conference
(Globecom) 2023 \cite{VR_GCW2023}.

The authors are with the Department of electrical engineering (ESAT), KU Leuven, Belgium (\{vida.ranjbar, robbert.beerten, marc.moonen, sofie.pollin\}@kuleuven.be).
}

}

% \markboth{IEEE TRANSACTION ON WIRELESS COMMUNICATION}%
% {Shell \MakeLowercase{\textit{et al.}}: A Sample Article Using IEEEtran.cls for IEEE Journals}
%\vspace{5em

\maketitle% 
\begin{abstract}
Cell-free massive multiple-input multiple-output (CFmMIMO) is a paradigm that can improve users' spectral efficiency (SE) far beyond traditional cellular networks. Increased spatial diversity in CFmMIMO is achieved by spreading the antennas into small access points (APs), which cooperate to serve the users. Sequential fronthaul topologies in CFmMIMO, such as the daisy chain and multi-branch tree topology, have gained considerable attention recently. In such a processing architecture, each AP must store its received signal vector in the memory until it receives the relevant information from the previous AP in the sequence to refine the estimate of the users' signal vector in the uplink.
In this paper, we adopt vector-wise and element-wise compression on the raw or pre-processed received signal vectors to store them in the memory. 
We investigate the impact of the limited memory capacity in the APs on the optimal number of APs. 
We show that with no memory constraint, having single-antenna APs is optimal, \textcolor{black}{especially as the number of users grows}.
However, a limited memory at the APs restricts the depth of the sequential processing pipeline. Furthermore, we investigate the relation between the memory capacity at the APs and the rate of the fronthaul link.
\end{abstract}
%TC:ignore
%%%%%%%%%%%%%%%%%% 1 Introduction %%%%%%%%%%%
\section{Introduction}
Massive multiple-input multiple-output (mMIMO) technology is one of the critical enablers for ambitious future wireless communication networks. It enhances users' spectral efficiency (SE), system energy efficiency (EE), and reliability with low-cost hardware at both receiver and
transmitter \cite{ErikMassive_MIMO_NG}. 
Massive MIMO makes it possible to separate the users spatially rather than using time/frequency orthogonalization, thanks to the spatial diversity it brings. Reusing the time/frequency resources increases the average throughput of users in the network. 
Cell-free massive multiple-input multiple-output (CFmMIMO) is a new paradigm that can further improve the SE of the users. In CFmMIMO, the antennas are spread into many small access points (APs) rather than being collocated in large base stations, as in traditional cellular networks. The APs cooperate to serve the nearby users, which mitigates the adverse effect of large-scale fading on the users' signal-to-interference-plus-noise-ratio (SINR) and provides uniform service to all the users \cite{nayebi_asimolar49, Ngocellfree_vs_smallcells}. 
In other words, in a distributed network with the antennas distributed in small APs where multiple APs are serving each user, the probability of having a poor channel gain is lower than in a traditional cellular and small-cell system where only one AP serves each user \cite{nayebi_asimolar49, Ngocellfree_vs_smallcells}, a consequence of the so-called macro diversity phenomenon. 

Initially, a CFmMIMO network was conceived as having multiple distributed APs, each with one/multiple antennas, randomly distributed in an area and cooperatively serving all the users \cite{nayebi_asimolar49, Ngocellfree_vs_smallcells,bjornsonmaking}. Two main operational methods were introduced in the uplink. First, the APs can locally process their received signal vector and estimate the users' signal vector. Then, the local estimates are sent to a central processing unit (CPU) to estimate the users' signal vector globally. Second, the APs can act as relays, forwarding their received signal vector and local channel estimates to the CPU. The CPU then solves a network-wide optimization problem to estimate the users' signal vector.

\begin{figure}[!ht]
    \centering
    \includegraphics[scale=0.5]{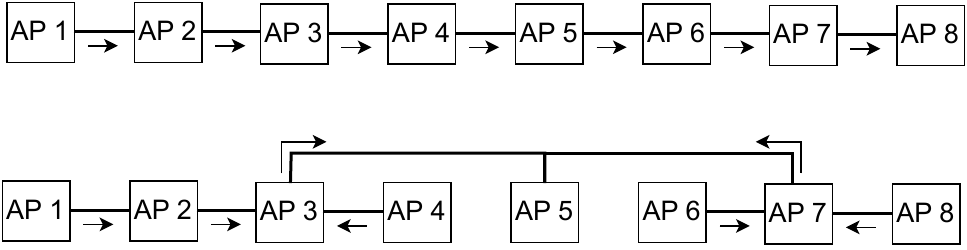}
    \caption{Data flow in Daisy chain (Top) vs multi-branch tree (Bottom) fronthaul topology.}
    \label{fig_FT}
\end{figure}

Recently, sequential fronthaul has been introduced in CFmMIMO networks in which the estimate of the users' signal vector is refined through the sequential fronthaul \cite{Interdonato2019, shaik2020, Shaik2021, ke_Dmimo_kalman, ranjbar2022}, possibly without an intermediate CPU. In such a sequential fronthaul, the APs along the sequence need to store their raw/pre-processed received signal vector until they receive the estimate of users' signal vector from the previous APs in the sequence to be able to refine them. Thus, the delay of the final estimation of the users' signal vector and the memory available at the last AP becomes a bottleneck. On-chip cache memory can be a good candidate for the memory requirement at the APs. On-chip caches are fast, energy-efficient, and usually of low capacity, making efficient usage of their capacity important. Storing the local received signal vector in a limited capacity memory may impose a compression noise on the stored received signal vector, especially at the APs toward the end of the sequence due to the large number of received signal vectors to be stored. As a result, the users' signal estimation quality will be adversely affected. This paper studies the impact of low-capacity memory on the average per-user SE in the uplink of a CFmMIMO, with the assumption of perfect channel state information (CSI), using either vector-wise or element-wise compression to store the received signal vectors in the memory. The fronthaul is a daisy chain, and each AP along the fronthaul refines users' signal estimates sequentially, starting from \textcolor{black}{first} AP to the last AP.
%(explained in algorithm \ref{algSRLS})
%\robbert{I would omit reference to the algorithm as this is only one example of how to use sequential fronthaul}
 The limited memory problem in a daisy chain fronthaul is discussed in subsection \ref{2B}. This paper considers sequential fronthaul similar to Fig. \ref{fig_FT}, which illustrates the daisy chain vs. multi-branch tree fronthaul topology for eight APs. 
% The limited capacity memory at the APs connected in a sequential fronthaul distorts the stored received signal vectors at the APs. To the authors' knowledge, this is the first paper to consider limited memory capacity at the APs in the CFmMIMO with sequential fronthaul topologies. However, in what follows, we will elaborate on other non-idealities affecting users' signal vector estimation in CFmMIMO.
\textcolor{black}{
Suppose that the processing level corresponds to the level of the tree in which a reference AP resides. The processing level in a given tree structure determines the number of the stored received signal vectors.  
In a daisy-chain topology, each AP has a distinct processing level. However,
in the multi-branch tree in Fig. \ref{fig_FT}, multiple APs can be on the same processing level. For example, in Fig. \ref{fig_FT},  AP $1$ is at the first processing level, AP $2$, AP $4$, AP $6$ and AP $8$ are at the second processing level, AP $3$ and AP $7$ are the third processing level, and finally AP $5$ is at the forth (last) processing level.}
%Furthermore, the required rate on the fronthaul links, given the memory capacity at the APs, is studied.
\subsection{Related works}
Aside from limited capacity memory, non-idealities such as limited bandwidth fronthaul links, hardware impairments, and low-resolution analog to digital converters (ADC) in (CF) mMIMO networks affect the users' signal vector estimation quality. The impact of such non-idealities on the users' SE and EE is discussed in \cite{dis_comp, HardwareBjörnson, HuLRADC,optimalADC_verenzuela, Xiongletter, YouzhiBitallocation,bashar_maxmin_2019,basharicc2019, Bashar_2021_uniform_q,bashar_EE_uq_2019, masoumiperformance,maryopi_2019}, among others.
The authors in \cite{HardwareBjörnson} show in which scenarios the correlation between the distortion vector element in a massive MIMO network with hardware impairment has a negligible impact on the users' SE. The ADC bit allocation among antennas in a massive MIMO network is discussed in \cite{optimalADC_verenzuela, HuLRADC, Xiongletter, YouzhiBitallocation}. In \cite{optimalADC_verenzuela}, the SE maximization is considered to find the optimal number of ADC bits of different antennas, subject to a constraint on the total number of bits or power consumption. The authors in \cite{YouzhiBitallocation} consider adaptive intra-AP and inter-AP bit allocation for ADCs at the APs. First, the authors consider channel estimation in the uplink and minimize the sum-weighted 
normalized mean square error
(SWNMSE) to find optimal bit allocation among the antennas of one AP (intra-AP) and subsequently among the antennas of different APs (inter-AP). Furthermore, the ADC bits are allocated to the APs during data transmission based on the solution of a sum-SE maximization problem. In general, the solutions to the optimization problems show that in the case of intra-AP bit allocation, equal bit allocation to the antennas is optimal as all the antennas of one AP experience the same large-scale fading. 
However, when it comes to inter-AP bit allocation, different APs are assigned different numbers of bits based on their overall channel gain to the users. Intuitively, the APs with larger overall channel gains for users are allocated more bits. 
Furthermore, increasing the number of users reduces the variance of the allocated bits to different APs, as the variance of the received signal between antennas of different APs reduces with increasing users. 
\subsection{Contribution}
To the best of our knowledge, this paper together with its short conference version \cite{VR_GCW2023} are the first to identify the problem of limited memory availability in sequential processing and fronthaul topologies. Our main contributions are then summarized as follows:
\begin{itemize}
    \item We consider a CFmMIMO network with sequential fronthaul where the APs have limited memory along the sequence. The sequential fronthaul can be a daisy chain or organized in a multi-branch tree.
    \item We formulate a maximization problem on the upper bound of users' sum-SE to find the optimal per-AP compression noise covariance matrix for both element-wise/vector-wise compression 
    under a limited memory constraint. This covariance matrix could be used to design optimal compression methods that are not considered here. 
    \item To use the  
    limited memory capacity more efficiently, vector-wise compression is favored over element-wise compression. However, element-wise compression is used in practice due to its lower complexity. To avoid vector-wise compression without compromising performance, we propose pre-processing the local received signal vectors with the principal component analysis (PCA) method to decorrelate the dimension of the received signal vector and then compressing the elements of the pre-processed vector element-wisely.  
   % Specifically, we show that the average per-user SE under local vector-wise compression of the received signal vector and under local element-wise compression of the PCA pre-processed received signal vector are equivalent.
   \item \textcolor{black}{We consider two general memory models: 1) fixed per-AP and 2) fixed total memory constraints. We relate the memory capacity at the APs to the maximum rate of the fronthaul links.}
    \item We provide simulation results on the impact of the limited memory at the APs on the optimal number of APs in the network with a daisy chain fronthaul topology. Having a fixed total memory budget, we compare equal or linearly increasing memory allocation schemes to the APs and compare the optimal  
    processing length of the sequential fronthaul in both cases. In addition, we consider a multi-branch tree topology as an alternative fronthaul topology and 
    study the impact of limited memory capacity in scenarios where processing along the branches can be parallelized.
    \item In contradiction to most existing state-of-the-art, we find that spatial diversity does not come free in a sequential processing topology. We must take into account that APs toward the end of this sequence need to store an exceedingly large number of received signal vectors and thus negate the effect of increased macro diversity by introducing significant compression noise on the stored received signal vectors.
    
% PCA maps the received vector into other spaces where the dimensions of the vector are uncorrelated. In other words, the correlation matrix becomes the identity matrix. As the elements of a multivariate Gaussian random vector with uncorrelated elements are also independent, the element-wise compression of the pre-processed received vector is equivalent to vector compression of the received vector. 
    
    % \item We show that, unlike in an unlimited memory capacity case where the scenario with a single antenna per AP gives the highest sum-SE, with the limited memory capacity, there is an optimal sequence length/number of antennas per AP for a given environment and with random users' location. Additionally, the effect of the number of users has been shown on the optimal length of the sequence.  \sofie{I would put this as a second list: you show that when the memory constraint is considered, unlimited distribution of single-antenna AP is not optimal. And then, for the case where you have multiple antennas per AP, you need to do vector compression of the correlated per antenna signals. Then, you show how to avoid vector compression by doing PCA first. I find this a more logical order.}
    % \sofie{grammar check not perfectly done}
    % \sofie{You should put the main conclusions, you tend to list a lot of details.}
    % \sofie{Is a 4th contribution maybe that you compare different memory allocation methods? }

\end{itemize}
% \robbert{Try to clearly define what your contributions are and then base the story 
% of the journal around that:\vida{I agree, did that}}
\subsection{Outline}
The remainder of this paper is organized as follows. In Section \ref{sec2}, the overall system model and the problem is defined. In Section \ref{section3}, the processing and compression of the received signal vector in the memory in each AP is elaborated on. Section \ref{sec4} introduces the memory capacity and allocation models in the considered fronthaul topologies. Furthermore, it elaborates on the maximum rate of the fronthaul links connecting two APs in a daisy chain fronthaul with different memory models. Section \ref{sec5} analyses the simulation results, and Section \ref{sec6} concludes the paper.
\subsection{Notation}
Vectors and matrices are denoted with boldface lower-case and upper-case letters, respectively. Transpose and conjugate transpose operations are denoted by superscripts $^{\text{T}}$ and $^{\text{H}}$, respectively. For two matrices $\mathbf{A}$ and $\mathbf{B}$, $\mathbf{A}\succeq\mathbf{B}$ means that $\mathbf{A}-\mathbf{B}$ is positive semi-definite. A zero-mean multi-variate circularly symmetric complex Gaussian distribution with covariance matrix $\mathbf{X}$ is represented as $\mathcal{C}\mathcal{N}(\mathbf{0}, \mathbf{X})$. The mean of $\mathbf{x}$ is denoted by $\mathbb{E}\{\mathbf{x}\}$, $\mathcal{H}(\mathbf{x})$ is the differential entropy of $\mathbf{x}$, and $I(\mathbf{x};\hat{\mathbf{x}})$ is the mutual information between $\mathbf{x}$ and $\hat{\mathbf{x}}$. Euclidean norm of $\mathbf{x}$ is shown as $\lVert \mathbf{x} \rVert$. Furthermore, $\mathbf{X}=\text{blkdiag}(\mathbf{X}_1,\hdots,\mathbf{X}_L)$ is a block-diagonal matrix with matrices $\mathbf{X}_i, \forall i$ as diagonal blocks. $\text{diag}(\mathbf{x})$ denotes a diagonal matrix with the elements of $\mathbf{x}$ as its diagonal elements and $\text{diag}(\mathbf{X})$ denotes a diagonal matrix with the
same diagonal elements as $\mathbf{X}$. 
$\det(\mathbf{A})$ returns the determinant of the square matrix $\mathbf{A}$. $N\times N$ identity matrix is shown as $\mathbf{I}_N$. $\mathbf{X}^{1/2}$ refers to the square root of matrix $\mathbf{X}$ and superscript $^{\text{H}/2}$ indicates the conjugate transpose of square root. %Symbol $\lfloor x \rfloor$ indicates the floor of $x$.
% Subscripts $_l$, $_v$, $_e$, and $_q$ denote the AP $l$ index, vector-wise compression, element-wise compression, and compression noise, respectively.
% In this paper, we denote vectors and matrices with boldface lower-case letters and boldface upper-case letters, respectively.  Transpose and conjugate transpose operations are denoted by superscripts $^{\text{T}}$ and $^{\text{H}}$, respectively. Symbol $\mathcal{H}$ represents the entropy function and $I$ mutual information.

\begin{table*}
%\begin{center}
\parbox{1\textwidth}{
\centering
\caption{Parameters}
\begin{tabular}{|c | c||c | c|} 
 \hline
 Parameter Description & Symbols&Parameter Description & Symbol \\ [0.5ex] 
 \hline\hline
 Number of users & $K$&Coherence time&$T_c$  \\ 
 \hline
 Number of APs & $L$&Coherence bandwidth& $B_c$  \\
 \hline
 Number of antennas per AP & $N$&Number of samples in one Coherence Block& $\tau_c$ \\
 \hline
 Total memory capacity&$C_{T}$&Number of subcarriers of an OFDM symbol& $N_{sc}$  \\
 \hline
  Memory capacity per AP&$C_{AP}$&OFDM Symbol time& $T_s$\\
 \hline
 Memory capacity per subcarrier in each AP & $C_{sc}$&Total bandwidth&$B$\\
 \hline
 User's transmit power & $p$&Receiver noise power at the APs &$\sigma^2$\\
 \hline

\end{tabular}
}
\end{table*}
%%%%%%%%%%%%%%%%%% 2 System model %%%%%%%%%%%

\section{System model and Problem statement}\label{sec2}
 %The multi-hop path %inspires the localization of the 
% enables serializing local uplink processing in each AP. Then, the local estimate of users' signal vector is refined in every AP along the sequential fronthaul. 
 % Just like distributed MMSE \cite{Shaik2021}, RLS reaches the MMSE solution in the final AP. 
 Distributed processing is a necessity in CFmMIMO, as it avoids overloading a single AP/CPU with massive computations, and it
enables truly scalable implementations and hence large-scale deployments \cite{bjornsonmaking, robbe, bjornsonscalable}. 
Distributed processing has become even more attractive with the growing interest in the sequential fronthaul topologies \cite{Shaik2021, shaik2020, ke_Dmimo_kalman, ranjbar2022}. 
% Unlike in the canonical star topology, in sequential topologies, the path between any AP and the CPU could consist of multiple hops rather than  
% a single one \cite{bjornsonscalable,cellfreevssmallcell}.
The multi-hop path between any AP and the CPU in sequential fronthaul %inspires the localization of the 
 enables serializing local processing in each AP. 
\subsection{Recursive Least-squares (RLS) for uplink signal estimation}\label{secII-A}
We consider distributed uplink signal estimation using the RLS method in a CFmMIMO network with daisy chain fronthaul topology, inspired by \cite{sanchez2018, Sánchez_daisy_2020, Shaik2021}. There are $L$ APs, each having $N$ antennas and a limited memory to store the received signal vectors in the uplink. The APs are connected in a daisy chain topology, 
 all jointly serving each of the $K$ users in the uplink. The received signal vector at AP $l$ is defined as follows:
\begin{equation}
\mathbf{y}_l=\mathbf{H}_l\mathbf{s}+\mathbf{n}_l,
    \label{eqRecSigAPl}
\end{equation}
where $\mathbf{s}\sim \mathcal{C}\mathcal{N}(\mathbf{0},p \mathbf{I}_K)$ is the users' signal vector and $\mathbf{n}_l\sim \mathcal{C}\mathcal{N}(\mathbf{0},\sigma^2\mathbf{I}_N)$ is the noise vector at AP $l$. Matrix $\mathbf{H}_l\in \mathbb{C}^{N\times K}$ is the local 
channel matrix between AP $l$ and the users. Each column of $\mathbf{H}_l$  is a multi-variate circularly symmetric complex Gaussian random vector, distributed as follows:
\begin{equation}
    \mathbf{H}_{l[:,k]}\sim\mathcal{C}\mathcal{N}(\mathbf{0},\mathbf{R}_{kl}),
    \label{chnAPl}
\end{equation}
where subscript $[:,k]$ represent the $k^{th}$ column of $\mathbf{H}_{l}$ and $\mathbf{R}_{kl}$ is the spatial correlation matrix \cite{massivemimobook} and $\beta_{kl}=\frac{\text{trace}(\mathbf{R}_{kl})}{N}$. %\robbert{This vector is received $\tau_u$ times per coherence block.}
% The covariance matrix of vector $\mathbf{y}_l$ is defined as follows:
% \begin{equation}
% \begin{aligned}
%     \mathbf{R}^y_{l}&=\mathbb{E}\{(\mathbf{y}_l-\mathbb{E}\{\mathbf{y}_l\})(\mathbf{y}_l-\mathbb{E}\{\mathbf{y}_l\})^{\text{H}}|\mathbf{H}_l\}\myeq\\&\mathbb{E}\{\mathbf{y}_l\mathbf{y}_l^{\text{H}}|\mathbf{H}_l\}=\mathbf{U}_l\underbrace{(p\mathbf{\Sigma}_l\mathbf{\Sigma}_l^{\text{H}}+\sigma^2\mathbf{I}_N)}_{\tilde{\mathbf{\Sigma}}_l}\mathbf{U}_l^{\text{H}},
%     \label{eq33}
%     \end{aligned}
% \end{equation}
The network-wide channel matrix, received vector, and noise vector are given as $\mathbf{H}=\begin{bmatrix}
    \mathbf{H}_1^T,&\hdots,& \mathbf{H}_L^T
\end{bmatrix}^T$, $\mathbf{y}=\begin{bmatrix}
    \mathbf{y}_1^T,&\hdots,& \mathbf{y}_L^T
\end{bmatrix}^T$ and $\mathbf{n}=\begin{bmatrix}
    \mathbf{n}_1^T,&\hdots,& \mathbf{n}_L^T
\end{bmatrix}^T$, respectively.
\textcolor{black}{We assume a block fading model in which the channel matrix $\mathbf{H}$ remains constant in a coherence interval of $\tau_c=B_cT_c$ samples, with $T_c$ and $B_c$ the coherence time and coherence bandwidth of the channel, respectively \cite{marzetta_larsson_yang_ngo_2016}. Out of $\tau_c$ samples in one coherence block, $\tau_u$ samples are used for uplink transmission.}
 
%The AP compresses the received signal vector with limited bits to store them in the memory. 
AP $l$ stores a compressed version of the received signal vector in the memory.
As a result of compression, a noise vector is added to the received signal vector. Section \ref{section3} elaborates on the compression model using rate-distortion theory.
% Suppose that after some operation such as compression for storage, the received vector $\mathbf{y}_l$ is mapped to the vector $\hat{\mathbf{y}}_l$ defined as below:
For now, assume that the compressed version of $\mathbf{y}_l$ is represented as $\hat{\mathbf{y}}_l$ and is formulated as follows:
\begin{equation}
\begin{aligned}
\hat{\mathbf{y}}_l=\mathbf{y}_l+\mathbf{q}_l=\mathbf{H}_l\mathbf{s}+\mathbf{n}_l+\mathbf{q}_l,
\end{aligned}
\label{recsigAPl_quntz}
\end{equation}
where $\mathbf{z}_l=\mathbf{n}_l+\mathbf{q}_l$ is a spatially correlated noise vector with zero mean and covariance matrix $\mathbf{Z}_l=\mathbb{E}\{\mathbf{z}_l\mathbf{z}_l^\text{H}\}$. \textcolor{black}{Vector $\mathbf{q}_l$ is the compression noise vector.}
 % Note that the noise at different antennas of one AP can be correlated, and consequently, $\mathbf{z}_l$ at the AP $l$ can be a colored noise.
 The network-wide compressed received signal and noise vector can be expressed as $\hat{\mathbf{y}}=\begin{bmatrix}
    \hat{\mathbf{y}}_1^{\text{T}}, & \hdots,&\hat{\mathbf{y}}_L^{\text{T}}
\end{bmatrix}^{\text{T}}$ and $\mathbf{z}=\begin{bmatrix}
    \mathbf{z}_1^{\text{T}},& \hdots, &\mathbf{z}_L^{\text{T}}
\end{bmatrix}^{\text{T}}$, respectively. The noise vectors among the APs are assumed to be independent, i.e., $\mathbf{Z}=\mathbb{E}\{\mathbf{z}\mathbf{z}^\text{H}\}=\text{blkdiag}(
    \mathbf{Z}_1,\hdots,\mathbf{Z}_L)
$. 

The APs sequentially refine the estimate of the users' signal vector based on the local 
CSI using the RLS algorithm given in Algorithm (\ref{algSRLS}). As stated in Algorithm  
\ref{algSRLS}, once per coherence block, AP $l$ exchanges $\mathbf{\Gamma}_l$ with AP $l+1$ and once per each uplink sample, i.e., $\tau_u$ times per coherence block, 
AP $l$ 
exchanges 
its local soft estimate of users' signal, i.e.,
$\hat{\mathbf{s}}_{l}$, with AP $l+1$ so AP $l+1$ can refine $\hat{\mathbf{s}}_{l}$ and 
update it to 
$\hat{\mathbf{s}}_{l+1}$.  
By updating the estimate of the users' signal vector using the RLS method, the estimate of the users' signal vector in the last AP is given as follows:
 \begin{equation}
    \hat{\mathbf{s}}=\hat{\mathbf{s}}_L=(\mathbf{H}^{\text{H}}\mathbf{Z}^{-1}\mathbf{H}+\frac{1}{p}\mathbf{I}_K)^{-1}\mathbf{H}^{\text{H}}\mathbf{Z}^{-1}\hat{\mathbf{y}}.
 \label{UEsigEst}
  \end{equation}
  %Therefore, we use the formulation in (\ref{UEsigEst}) to calculate the sum-SE function in Section \ref{section3}.
 % where it gets updated in each AP and exchanged among subsequent APs until the last AP,i.e., AP $L$.
\begin{algorithm}[h!]
\caption{ RLS algorithm for users' signal vector estimation}
 \label{algSRLS}
\begin{algorithmic}[1]
 \STATE \textbf{Initialize:}
 \STATE \hspace*{\algorithmicindent}\parbox[t]{.8\linewidth}{\raggedright  $\mathbf{\Gamma}_0=p\mathbf{I}_K$}
 \STATE \hspace*{\algorithmicindent}\parbox[t]{.8\linewidth}{\raggedright   $\hat{\mathbf{s}}^n_{0}=\mathbf{0}_{K\times 1}, \forall n\in[1:\tau_u]$}
\FOR{$l = 1 \dots L$}
\STATE{$\mathbf{\Gamma}_l=\mathbf{\Gamma}_{l-1}-\mathbf{\Gamma}_{l-1}\mathbf{H}_l^{\text{H}}\mathbf{Z}_l^{-\text{H}/2}(\mathbf{I}_N+\mathbf{Z}_l^{-1/2}\mathbf{H}_l\mathbf{\Gamma}_{l-1}\mathbf{H}_l^{\text{H}}\mathbf{Z}_l^{-\text{H}/2})^{-1}\mathbf{Z}_l^{-1/2}\mathbf{H}_l\mathbf{\Gamma}_{l-1}^{\text{H}}$}
\FOR{ $n = 1 \dots \tau_u$} 
        \STATE{$
    \hat{\mathbf{s}}^n_l=\hat{\mathbf{s}}^n_{l-1}+\mathbf{\Gamma}_l\mathbf{H}_l^{\text{H}}\mathbf{Z}_l^{-\text{H}/2}(\hat{\mathbf{y}}^n_l-\mathbf{Z}_l^{-1/2}\mathbf{H}_l\hat{\mathbf{s}}^n_{l-1}).
    $}
               
\ENDFOR
\ENDFOR
\end{algorithmic}
\end{algorithm}\\

Superscript $n$ in Algorithm \ref{algSRLS} is used to differentiate amongst the different
uplink samples. However, for the sake of readability, we remove superscript $n$ from the users' signal vector anywhere else in the paper.
Finally, in the last AP, users' signal estimation error, i.e., ${\mathbf{e}}=\mathbf{s}-\hat{\mathbf{s}}$, covariance matrix is as follows:
  \begin{equation}
      \mathbf{\Gamma}_L=\mathbb{E}\{{\mathbf{e}}{{\mathbf{e}}}^{\text{H}}\}=(\mathbf{H}^{\text{H}}\mathbf{Z}^{-1}\mathbf{H}+\frac{1}{p}\mathbf{I}_K)^{-1}.
      \label{sig_est_error}
  \end{equation}
  Note that $\hat{\mathbf{s}}_l$ is the LS estimate of users' signal having the received signal vector and CSI from AP $1$ to AP $l$. 
  %Similarly, ${\mathbf{e}}=\mathbf{s}-\hat{\mathbf{s}}$
% In RLS algorithm for uplink signal estimation, AP $l$:
% \begin{itemize}
%     \item calculates a $K\times K$ matrix once in each coherence block, as follows:
%     \begin{equation}
%         \mathbf{\Gamma}_l=\mathbf{\Gamma}_{l-1}-\mathbf{\Gamma}_{l-1}\mathbf{H}_l^{\text{H}}\mathbf{Z}_l^{-\text{H}/2}(\mathbf{I}_N+\mathbf{Z}_l^{-1/2}\mathbf{H}_l\mathbf{\Gamma}_{l-1}\mathbf{H}_l^{\text{H}}\mathbf{Z}_l^{-\text{H}/2})^{-1}\mathbf{H}_l\mathbf{Z}_l^{-1/2}\mathbf{\Gamma}_{l-1}^{\text{H}}.
%         \label{recsigAPl_quntz}
%     \end{equation}
%     \item For each sample in one coherence block, AP $l$ refines the estimates of the users' signal as below:
%     \begin{equation}
%         \hat{\mathbf{s}}_l=\hat{\mathbf{s}}_{l-1}+\mathbf{\Gamma}_l\mathbf{H}_l^{\text{H}}\mathbf{Z}_l^{-\text{H}/2}(\hat{\mathbf{y}}_l-\mathbf{Z}_l^{-1/2}\mathbf{H}_l\hat{\mathbf{s}}_{l-1}).
%         \label{UEsigEst}
%     \end{equation}
% \end{itemize}

% In a sequential fronthaul, aside from CSI, the sequential processing requires each AP in the sequence to store their raw/pre-processed received signal vector until it receives 
% the necessary information
% from the previous AP in the sequence. Thus, the number of received signal vectors to be stored grows linearly with the number of APs in the network. We elaborate on limited memory capacity in Section \ref{2c}. 
we assume perfect CSI in this paper, which is realistic with a large enough transmit power per user during pilot transmission and in an indoor or semi-static environment where the coherence block is large enough to accommodate a unique pilot per user \cite{marzetta_larsson_yang_ngo_2016}. \textcolor{black}{However, in the simulation section, we provide a subsection on the imperfect CSI.}
%\robbert{Since we assume a limited number of APs and thus limited lengths of our sequential fronthaul, we also assume a limited number of users, hence a unique pilot per user is reasonable.}

 The authors in \cite{ranjbar2022,Shaik2021,ke_Dmimo_kalman} assume an unlimited memory capacity in each AP. However, under the realistic assumption of limited memory
capacity in each AP, two questions arise: how to optimally compress and store 
received signal vector in memory and what is the effect of this 
compression on the average per-user SE? 

\subsection{Limited memory capacity at the APs}\label{2B}
\begin{figure}[t]
    \centering
    \includegraphics[scale=0.7]{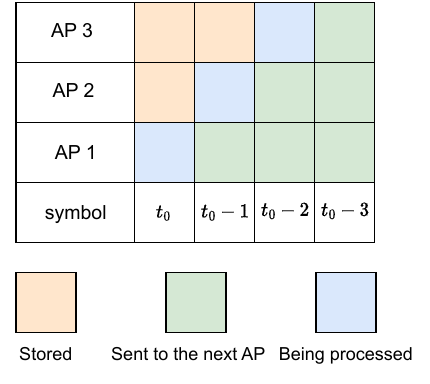}
    \caption{Sequential processing and storage in a CFmMIMO network with daisy chain fronthaul topology}
    \label{fig2_seq}
\end{figure}

In a sequential daisy chain fronthaul, each AP stores its raw/pre-processed received signal vector until it receives the estimate of the users' signal vector from the previous AP in the sequence. Then, it refines the estimate of the users' signal vector by co-processing it with their own local received signal vector, as shown in the line $7$ of Algorithm \ref{algSRLS}. 
%Hence, the local received signal vectors are stored at the APs until the estimate from the previous AP is received. 
To make the problem more tangible, consider a CFmMIMO network with a daisy chain fronthaul topology. There are $N_{sc}$ subcarriers for each OFDM (orthogonal frequency-division multiplexing) symbol. Suppose that $\mathbf{Y}_l^{t_0}\in\mathbb{C}^{N\times N_{sc}}$ is a symbol matrix with each column corresponding to the received signal vector at AP $l$ for a particular subcarrier of symbol $t_0$.
%\robbert{For clarity I would consider putting $s^{t_0}_l$ here.  }\vida{Well, one OFDM symbol has $N_{sc}$ samples so we have $N_{sc}\times K$ signal to detect. Besides I used to show sample number with $\mathbf{s}^{n}$}
%Then consider a reference subcarrier of symbol $t_0$. 
% \sofie{I think should be: Then consider a reference subcarrier of a symbol $n_0$. (As the symbols are counted in time, and the subcarrier is counted in frequency. In the next sentence, when you are referring to $n_0-1$ I think you are taking the same subcarrier from the previous symbol.)}.
When AP $1$ is processing $\mathbf{Y}_1^{t_0}$, the rest of the APs have their corresponding matrices, i.e., $\mathbf{Y}_l^{t_0}, \forall l\in\{2,\hdots, L\}$ in their memory. Similarly, when AP $2$ is processing $\mathbf{Y}_2^{t_0-1}$,
the corresponding matrix at AP $l$, 
 i.e., $\mathbf{Y}_l^{t_0-1}, \forall l\in\{3,\hdots,L\}$, is stored at the memory. Accordingly, the number of the symbol matrix stored at AP $l$ is $l-1$, meaning that the number of received signal vectors stored at AP $l$ is $(l-1)N_{sc}$, which increases linearly from one AP to the next by $N_{sc}$. \textcolor{black}{Fig. \ref{fig2_seq} demonstrates the sequential processing and storage at the APs.} It is worth mentioning that processors are designed to process at least one symbol during a symbol duration to have a stable system. In other words, the rate of the locally received signal vectors entering the memory should be lower or, in the worst case, the same as the rate at which the processor is processing them. Therefore, the number of vectors stored in the AP's memory increases by $N_{sc}$ from one AP to the next.

The memory used to store the received signal vectors is usually a fast on-chip cache memory \cite{jesusthesis}, close to the processing unit\footnote{Note that local CSI is also stored in the memory, next to the received signal vectors. However, the size of the local CSI that needs to be stored in the memory is the same in all APs, and hence, we only focus on storing the received signal vector in this paper.}.
For an example of an AP processor implementation, the reader is referred to \cite{arm}.

%%%%%%%%%%%%% 3 %%%%%%%%%%%%%%%%
\section{Local received signal vector compression at the APs}\label{section3}
In this section, we consider three options for storing the local received signal vectors:%\footnote{Note that we consider a frequency domain representation of the IQ signals, which means after FFT and CP-removal.}:
\begin{enumerate}
\item Vector-wise compression of the received signal vector, i.e., joint compression of the received signal vector elements.
    \item Element-wise compression of the received signal vector.
    \item Element-wise compression of PCA pre-processed received signal vector. 
\end{enumerate}
% \robbert{We can do this either element-wise or vector-wise.}

\begin{figure}[h]
     \centering
     \pgfplotsset{width=8.4cm,compat=1.18}
\pgfplotsset{every x tick label/.append style={font=\small, yshift=0.5ex},every y tick label/.append style={font=\small, xshift=0.5ex},
every axis legend/.append style={
at={(0.12,1)},
anchor=north west,font=\large
}}

%\begin{tikzpicture}
\begin{tikzpicture}[scale=0.8]
\begin{axis}[
domain=0:4,
grid=major,
ylabel=$I(r;\hat{r})$,
xlabel= $Q$,
xmin=0,
xmax=1,
xtick style={color=black},
xticklabel style={yshift=-2.5pt},
ymin=0,
ymax=6,
ytick={0.5,1,2,3,4,5,6,7,8},
mark size=4.0pt,
%yticks={1,2,3,5,6},
%legend columns=3,
%legend style={nodes={scale=0.9, transform shape},at={(1.83,1)},anchor=north},
]

       \addplot [black]
        table[x=C,y=B,col sep=comma] {Data_for_figures/MI.csv};
        
        \addlegendentry{$I(r,\hat{r})$ in (11)}
       
        \addplot [cyan]
        table[x=C,y=A,col sep=comma] {Data_for_figures/MI.csv};
        \addlegendentry{$I(r,\hat{r})$ in (10)}

        \draw [dashed] (0.33,-1) -- (0.33,4);
        \draw[->](0.33,3)--(0.2,3);

        % \addplot [orange]
        % table[x=D,y=C,col sep=comma] {Data_for_figures/MI.csv};
        % \addlegendentry{$I(r,\hat{r})$ using uniform quantization in (10)}
        %\addplot [orange]
       % table[x=D,y=1,col sep=comma] {Data_for_figures/MI.csv};
        %\addlegendentry{$I(r,\hat{r})$ using uniform quantization in (10)}

\end{axis}

\end{tikzpicture}
     \caption{Mutual information of $r$ and $\hat{r}$ using two test channels in (\ref{tstch}) and (\ref{eq_opt_tstch}) %\robbert{The numbers on these axis are very small now}).
     % \robbert{I think the references to the equations are no longer correct in the legend}
     }
     \label{figtstch}
 \end{figure}
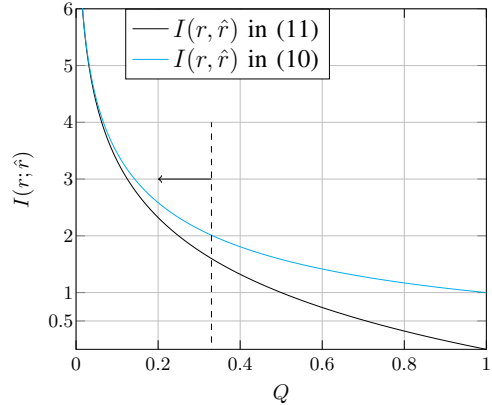

% \robbert{Before we discuss the different options for compression in more detail, we outline our 
% information theoretic framework which is based upon the rate distortion function  \cite{CovTho}.
% The rate distortion function is given in (8), it relies on an input vector $\mathbf{x}$ which is 
% distorted by $d$ sources of noise leading to the distorted vector $\hat{\mathbf{x}}$.
% The receiver has some side information $\mathbf{z}$.
% }
To model the compression of the received signal vector at each AP, we use rate-distortion theory \cite{CovTho}. Based on rate-distortion theory, the rate-distortion function of a random variable gives the minimum number of bits needed to compress realizations of the random variable given a distortion constraint.
The rate-distortion theory can also be extended to a random vector and covariance distortion constraint \cite{zahedi2014}. If there are $d$ sources, each generating a unique element of the circularly symmetric complex Gaussian random vector $\mathbf{x}$ with dimension $d$, the extension to the rate-distortion function of vector $\mathbf{x}$ is given as below: 
\begin{equation}
    \begin{aligned}    R(\mathbf{Q})=\min_{f(\hat{\mathbf{x}}|\mathbf{x}),\textcolor{black}{\mathbb{E}}\{(\mathbf{x}-\hat{\mathbf{x}})(\mathbf{x}-\hat{\mathbf{x}})^{\text{H}}\}\preceq\mathbf{Q}} \textit{I}(\mathbf{x};\hat{\mathbf{x}}),
    \label{Rat_dis_fun}
\end{aligned}
\end{equation} 
where $\hat{\mathbf{x}}$ is the compressed version of $\mathbf{x}$, $f(\hat{\mathbf{x}}|{\mathbf{x}})$ is the conditional probability distribution function (PDF) of vector $\hat{\mathbf{x}}$, $\mathbf{Q}$ is the compression noise covariance matrix (target distortion) and $R(\mathbf{Q})$ is the rate-distortion function.
The mutual information between the vector and its compressed version, i.e., $\textit{I}(\mathbf{x};\hat{\mathbf{x}})$ can be lower bounded as follows:
\begin{equation}
\begin{aligned}
  \textit{I}(\mathbf{x};\hat{\mathbf{x}}) &= \mathcal{H}(\hat{\mathbf{x}})-\mathcal{H}(\hat{\mathbf{x}}|\mathbf{x}) \\ 
  & \myeq \mathcal{H}(\hat{\mathbf{x}})-\mathcal{H}(\hat{\mathbf{x}}-\mathbf{x}|\mathbf{x})  \\
  &\myineqb \mathcal{H}(\hat{\mathbf{x}})-\mathcal{H}(\hat{\mathbf{x}}-\mathbf{x}) \\
  & \myineqc \mathcal{H}(\hat{\mathbf{x}})-\mathcal{H}(\mathcal{C}\mathcal{N}(\mathbf{0},\mathbb{E}\{(\hat{\mathbf{x}}-\mathbf{x})(\hat{\mathbf{x}}-\mathbf{x})^{\text{H}}\}) \\ 
  &= \mathcal{H}(\hat{\mathbf{x}})-\log_2 (\pi e)^d \det(\mathbf{Q}).
  \label{mutualinfo}
\end{aligned}
\end{equation}
Note that in this paper, for the vector-wise compression, we have $d=N$, and for the element-wise compression, we have $d=1$. In (\ref{mutualinfo}), equality $\myeq$ holds because $-\mathbf{x}$ is a pure translation in $\mathcal{H}(\hat{\mathbf{x}}-\mathbf{x}|\mathbf{x})$ and entropy is translation invariant \cite{elgamal_kim_2011}.
Inequality $\myineqb$ comes from the fact that conditioning reduces entropy. 
Inequality $\myineqc$ results from the maximum differential entropy, which states that given a variance, the Gaussian distribution will maximize the entropy, which can be generalized to a multivariate Gaussian distribution \cite{CovTho}. To represent the relation between vector $\mathbf{x}$ and its compressed version, i.e., vector $\hat{\mathbf{x}}$, the concept of test channel is used.
The test channel that achieves the lower bound $\myineqc$ in (\ref{mutualinfo}) is as follows \cite{jointKang,masoumiperformance}:
 \begin{equation}
     \hat{\mathbf{x}}=\mathbf{x}+\mathbf{q},
     \label{tstch}
 \end{equation}
where compression noise $\mathbf{q}\sim\mathcal{C}\mathcal{N}(\mathbf{0},\mathbf{Q})$ 
is independent of $\mathbf{x}$. For the rest of the paper, we use the aforementioned test channel in (\ref{tstch}) due to mathematical simplicity. 
However, it is worth mentioning that the optimal test channel to lower bound the mutual information in (\ref{mutualinfo}) is as follows:
\begin{equation}
\mathbf{x}=\hat{\mathbf{x}}+\mathbf{q},
\label{eq_opt_tstch}
\end{equation}
where $\mathbf{q}$ and $\hat{\mathbf{x}}$ are independent and $\mathbf{q}\sim \mathcal{C}\mathcal{N}(\mathbf{0},\mathbf{Q})$. The test channel in (\ref{eq_opt_tstch}) results in a smaller value for $\textit{I}(\mathbf{x};\hat{\mathbf{x}})$ which means that for a given number of bits, it results in a smaller compression noise power compared to the test channel in (\ref{tstch}).
However, using the optimal test channel in (\ref{eq_opt_tstch}), the solutions to the maximization problems in the following sections are not mathematically straightforward. %simple. 
Hence, the test channel in (\ref{tstch}) is used in this paper. In what follows, we show that, besides simplicity, the test channel in (\ref{tstch}) is close in performance to the optimal test channel in (\ref{eq_opt_tstch}) for the number of compression bits that we consider in this paper.
%Besides mathematical %simplicity
%elegance, the test channel in (\ref{tstch}) is more efficient than the compression methods used in reality, such as the uniform compression of real and complex parts of the elements of the received signal vector individually. 
For clarity and illustrative purposes, consider the following 
example. Suppose that we have a complex circularly symmetric Gaussian random variable $r\sim\mathcal{C}\mathcal{N}(0, p_r)$ \footnote{To make the comparison between the two test channels in (\ref{tstch}) and (\ref{eq_opt_tstch}) simple, we considered compressing a complex circularly symmetric Gaussian \textbf{scalar} random variable.}. In this example and without loss of generality, we assume $p_r=1$. By compressing the random variable using test channel in (\ref{tstch}), $I(r;\hat{r})$ is given as:
 \begin{equation}
     I(r;\hat{r})=\log_2(\frac{1}{Q}+1).
     \label{mutinf_tstch9}
 \end{equation}
 With the optimal test channel in (\ref{eq_opt_tstch}), $I(r;\hat{r})$ is given as:
 \begin{equation}
     I(r;\hat{r})=\log_2(\frac{1}{Q}),
     \label{mutinf_opttstch}
 \end{equation}
 where $Q$ is the compression noise variance. 
 In Fig. \ref{figtstch}, we plot and compare the mutual information functions in (\ref{mutinf_tstch9}) and (\ref{mutinf_opttstch}) (corresponding to the number of compression bits) versus the compression noise power for the two test channels in (\ref{tstch}) and (\ref{eq_opt_tstch}). It is clear that the lower the number of bits, the higher the compression noise power. The optimal test channel in (\ref{eq_opt_tstch}) has a compression noise power that is bounded by the signal power, here $p_r$, and in case 0 bits are allocated to compress $r$, the compression noise is, in fact, the signal itself. The test channel in (\ref{tstch}) performs worse for a very low number of bits ($<2$ bit per signal), where $Q>0.33$. In our simulations, however, we usually allocate more than 2 bits per signal. Hence, we are usually working in the part of the curve where both test channels have very similar performance, i.e., to the left of the dashed line in Fig. \ref{figtstch}. 
 
 Therefore, using the test channel in (\ref{tstch}) not only gives tractable solutions to the problems maximizing the upper bound on users' sum-SE, but also the simulation results can be generalized to the practical implementations.
 % \sofie{For me, this link with uniform compression is confusing. I would only plot the two test channels. Say that the mutual information corresponds to the number of bits too, and you assume >1. (see suggestion above) }
  % Furthermore, conditioned on having perfect CSI, the elements of compressed received signal vectors at the APs are circularly symmetric Gaussian random vectors.

% Assume that the local received vector at AP $l$ is as follows:
% \begin{equation}
% \mathbf{y}_l=\mathbf{H}_l\mathbf{s}+\mathbf{n}_{l},
%   \label{eq8}
% \end{equation}
% where $\mathbf{n}_l\in\mathcal{N}_c(\mathbf{0},\sigma^2\mathbf{I}_N)$ is white receiver noise at AP $l$. In what follows, we consider compression of the local received vectors in each AP.

% In Sections \ref{sec3_A}, \ref{sec3_B} and \ref{sec3_C}, given one of the compression option introduced in the beginning of this section, the memory bit allocation to the local received signal vector in each AP is defined from a sum-SE maximization problem constrained by the limited memory capacity of the AP. However, due to the high complexity of vector-wise compression, we use PCA as an orthogonalization technique on the received signal vector to decorrelate the dimension and to be able to compress the pre-processed received signal vector element-wisely without compromising performance.

% Then, the impact of limited memory capacity on the number of APs under different scenarios is investigated and discussed.  
\subsection{Option 1: Vector-wise compression of the received signal vector}\label{sec3_A}
In this option, each AP compresses its received signal vector vector-wisely.
Following the rate-distortion argument at the beginning of this section, for AP $l$, we replace $\mathbf{x}$ with $\mathbf{y}_l$, and then the compressed vector is represented as $\hat{\mathbf{y}}_{vl}$. Using the test channel in (\ref{tstch}), the relation between $\mathbf{y}_l$ and its compressed version $\hat{\mathbf{y}}_{vl}$ is given as follows:
\begin{equation}
\hat{\mathbf{y}}_{vl}=\mathbf{y}_l+\mathbf{q}_{vl}=\mathbf{H}_l\mathbf{s}+\mathbf{n}_l+\mathbf{q}_{vl},
  \label{rec_sig_apl_Vectquantz}
\end{equation}
where $\mathbf{y}_l$ is defined in (\ref{eqRecSigAPl}), $\mathbf{z}_{vl}=\mathbf{n}_l+\mathbf{q}_{vl}$ and $\mathbf{q}_{vl} \sim \mathcal{C}\mathcal{N}(\mathbf{0}, \mathbf{Q}_{vl})$ represents the compression noise introduced by the joint compression of the vector elements.
Loosely speaking, the differential entropy of a random variable such as $\hat{\mathbf{y}}_{vl}$ measures the uncertainty of the random variable and the number of bits required to represent it. The mutual information between $\mathbf{y}_l$ and $\hat{\mathbf{y}}_{vl}$ shows how much  
of the uncertainty of one of them is reduced due to knowledge of the other one. Hence, the mutual information can also measure how many bits are required to represent $\hat{\mathbf{y}}_{vl}$ when $\mathbf{y}_l$ is known \cite{CovTho}.
%\robbert{when $\mathbf{y}_l$ is known}
\textcolor{black}{The relation between the number of compression bits per received signal vector, i.e., $C_{sc}$, and the compression noise covariance matrix $\mathbf{Q}_{vl}$ at AP $l$, conditioned on the local CSI is as follows \cite{CovTho}:}
%\sofie{ok, I would indeed mention this explicitly in the text too. Explain things a bit for the reader!}
% \begin{equation}
% \begin{split}
%     C_{sc}=I(\mathbf{y}_l;\hat{\mathbf{y}}_l|\mathbf{H}_l)=\mathcal{H}_l(\hat{\mathbf{y}}_l|\mathbf{H}_l)-\mathcal{H}_l(\hat{\mathbf{y}}_l|\mathbf{y}_l,\mathbf{H}_l)\myeq\log_2 2\pi e \det({p\mathbf{H}_l\mathbf{H}_l^{\text{H}}+\sigma^2\mathbf{I}_N+\mathbf{Q}_{vl}})-\log_2 2\pi e \det(\mathbf{Q}_{vl})=\\ \log_2 \det({\mathbf{Q}_{vl}^{-1}(p\mathbf{H}_l\mathbf{H}_l^{\text{H}}+\sigma^2\mathbf{I}_N)+\mathbf{I}_N})
%     \label{mutualinfo}
% \end{split}
% \end{equation}
\begin{equation}
\begin{aligned}
    C_{sc}& =I(\mathbf{y}_l;\hat{\mathbf{y}}_{vl}|\mathbf{H}_l) \\
    & =\mathcal{H}(\hat{\mathbf{y}}_{vl}|\mathbf{H}_l)-\mathcal{H}(\hat{\mathbf{y}}_{vl}|\mathbf{y}_l,\mathbf{H}_l) \\
    & = \log_2 \det({\mathbf{Q}_{vl}^{-1}(p\mathbf{H}_l\mathbf{H}_l^{\text{H}}+\sigma^2\mathbf{I}_N)+\mathbf{I}_N}),
    \label{bitsREl_to_mutlinfo}
\end{aligned}
\end{equation}
where the vectors $\{\hat{\mathbf{y}}_{vl}|\mathbf{H}_l\}$ and $\{\hat{\mathbf{y}}_{vl}|\mathbf{y}_l,\mathbf{H}_l\}$ are multi-variate circularly symmetric Gaussian random vectors, as $\mathbf{s}$, $\mathbf{n}_l$ and $\mathbf{q}_{vl}$ are independent multi-variate circularly symmetric Gaussian random vectors. 
The network-wide compressed received signal vector is as follows:
\begin{equation}
    \begin{aligned}    
    \hat{\mathbf{y}}_v=\mathbf{y}+\mathbf{q}_v=\mathbf{H}\mathbf{s}+\underbrace{\mathbf{n}+\mathbf{q}_v}_{\mathbf{z}_v},
    \end{aligned}
\label{eq14}
\end{equation}
% can be sequn it can be proved that the estimate of users' signal in the last AP can be formulated as:
where $\hat{\mathbf{y}}_v=\begin{bmatrix}
        \hat{\mathbf{y}}^{\text{T}}_{v1},& \hdots,& \hat{\mathbf{y}}^{\text{T}}_{vL}
    \end{bmatrix}^{\text{T}}$, $\mathbf{q}_v=\begin{bmatrix}
        \mathbf{q}^{\text{T}}_{v1},& \hdots, & \mathbf{q}^{\text{T}}_{vL}
\end{bmatrix}^{\text{T}}$,
and $\mathbf{z}_v=\begin{bmatrix}\mathbf{z}_{v1}^{\text{T}},&\hdots,& \mathbf{z}_{vL}^{\text{T}}\end{bmatrix}^{\text{T}}$ is the receiver plus compression noise vector with covariance matrix defined as $\mathbf{Z}_v=\mathbb{E}\{\mathbf{z}_{v}\mathbf{z}_{v}^{\text{H}}\}=\text{blkdiag}(\mathbf{Z}_{v1},\hdots,\mathbf{Z}_{vL})=\text{blkdiag}(\mathbf{Q}_{v1}+\sigma^2\mathbf{I}_N,\hdots,\mathbf{Q}_{vL}+\sigma^2\mathbf{I}_N)$. Following the discussion in Section \ref{secII-A}, the LS estimate of the users' signal vector in the last AP can be formulated as follows:
\begin{equation}
 \hat{\mathbf{s}}_v= \mathbf{C}_v\hat{\mathbf{y}}_v, 
 \label{UE_sig_est_using_quantz}
\end{equation}
where the combining matrix $\mathbf{C}_v\in \mathbb{C}^{K\times N}$ given $\mathbf{H}$ is formulated as follows:
\begin{equation}
\mathbf{C}_v=(\mathbf{H}^{\text{H}}\mathbf{Z}_v^{-1}\mathbf{H}+\frac{1}{p}\mathbf{I}_K)^{-1}\mathbf{H}^{\text{H}}\mathbf{Z}_v^{-1}.
\label{combvec_VC}
\end{equation}
% The block-diagonality of matrix $\mathbf{Z}_v$ is due to the fact that the received vector of any two APs is independent so that they can be optimally compressed individually--> not true because of s.
Having an estimate of the users' signal vector as in (\ref{UE_sig_est_using_quantz}),  the sum-SE %\footnote{Note that the sum-SE in \ref{sumRate_VC} is an upper bound on sum-SE when each user's signal is detected individually. In other words $\mathcal{I}(\mathbf{C}_{v[k,:]}\hat{\mathbf{y}}_v;\mathbf{s}_k)\leq\mathcal{I}(\sum_{k}\mathbf{C}_v\hat{\mathbf{y}}_v;\mathbf{s})$.} 
of users is formulated as follows:
\begin{equation}
\begin{aligned}
    R_{v}&=\frac{\tau_u}{\tau_c}\mathcal{I}(\hat{\mathbf{s}}_v;\mathbf{s})\\&=\frac{\tau_u}{\tau_c}\bigl(\mathcal{H}(\mathbf{s})-\mathcal{H}(\mathbf{s}|\hat{\mathbf{s}}_v)\bigr) \\
   % &\text{\robbert{This equality below is hard to see for me}}\\&
   % \text{\vida{I had the detailed version, marc told me to remove the details as it is standard result}} \\
    &\myeq\frac{\tau_u}{\tau_c}\log_2\det(p\mathbf{H}\mathbf{H}^{\text{H}}\mathbf{Z}_v^{-1}+\mathbf{I}_{NL})\\
     %&\text{\robbert{from here on should be $\mathbf{I}_N$, right?}\vida{yes} \\
    &\myineqlb\frac{\tau_u}{\tau_c}\log_2 \prod_{l=1}^{L}\det(p\mathbf{H}_l\mathbf{H}_l^{\text{H}}\mathbf{Z}_{vl}^{-1}+\mathbf{I}_{N})\\
    &= \frac{\tau_u}{\tau_c}\sum_{l=1}^{L} \log_2\det(p\mathbf{H}_l\mathbf{H}_l^{\text{H}}\mathbf{Z}_{vl}^{-1}+\mathbf{I}_{N})\\&= \frac{\tau_u}{\tau_c}
    \sum_{l=1}^{L} \log_2\det(p\mathbf{H}_l\mathbf{H}_l^{\text{H}}(\mathbf{Q}_{vl}+\sigma^2\mathbf{I}_N)^{-1}+\mathbf{I}_{N}),
   \end{aligned}
    \label{sumRate_VC}  
\end{equation}
% The proof of $\myeq$ and $\myeqb$ in (\ref{UE_sig_est_using_quantz}) has been elaborated more in section B of \nameref{apdb}.
%comes from the worst case uncorrelated noise theorem in \cite{Hassibi_training_2003}.
% The equality $\myeqb$ can be proved using same approach as at APpendix I.
where $R_v$ can be achieved 
%\robbert{i.e. equality of the bound in $(b)$, }
%\textcolor{green}{actually I meant $R_v$, the equality in (a)}
if sequential interference cancellation (SIC) is used to estimate/detect users' signal in the last AP \cite{Tse_Viswanath_2005}.
\footnote{Sum-SE of the users when the users are detected jointly is more than the case in which the users are detected individually (i.e., $\sum_{k}\mathcal{I}(\hat{\mathbf{s}}_k;\mathbf{s}_k)\leq\mathcal{I}(\hat{\mathbf{s}};\mathbf{s})$ \cite{CovTho})}. Furthermore, $\myeq$ and $\myineqlb$ 
are proved in \nameref{apda}.
%being a block-diagonal matrix plus the fact that the determinant of the positive semi-definite matrix $\mathbf{H}_l\mathbf{H}_l^{\text{H}}$ is always equal or smaller than the determinant of a diagonal matrix with the same diagonal elements \cite{horn_johnson_2012}.
The upper bound in equation (\ref{sumRate_VC}) is for the instantaneous sum-SE in a particular coherence block. 
% By taking the average of the upper-bound, we can formalize the ergodic SE as follow:
% \begin{equation}
%     R^{E}_r=\mathbb{}_{\mathbf{H}}\{R\}.
%     \label{rec_sig_PCA_pp_EC}
% \end{equation}
It is the summation of $L$ functions, each dependent only on the compression noise covariance matrix \textcolor{black}{and local channel matrix} of a single AP. Additionally, each AP compresses its local received signal vector in isolation from other APs. Therefore, the maximization of the upper bound function can be decomposed into $L$ smaller optimization problems to be solved in $L$ APs. Therefore, the maximization sub-problem at AP $l$ is defined as follows:
%\robbert{target variable should be $\mathbf{Q}_{vl}$ instead of $\mathbf{Q}_{l}$\vida{yes}}
%%checked until here.
\begin{equation}
     \begin{aligned}
\arg \max_{\mathbf{Q}_{vl}^{-1}\succeq\mathbf{0}} \quad  &\log_2
    \det(p\mathbf{H}_l\mathbf{H}_l^{\text{H}}(\mathbf{Q}_{vl}+\sigma^2\mathbf{I}_N)^{-1}+\mathbf{I}_N)\\    =&\log_2\det(p\mathbf{H}_l\mathbf{H}_l^{\text{H}}+\mathbf{Q}_{vl}+\sigma^2\mathbf{I}_N)-\\&\log_2\det(\mathbf{Q}_{vl}+\sigma^2\mathbf{I}_N)\\=& \log_2 \det({\mathbf{Q}_{vl}^{-1}(p\mathbf{H}_l\mathbf{H}_l^{\text{H}}+\sigma^2\mathbf{I}_N)+\mathbf{I}_N})-\\&\log_2 \det(\sigma^2\mathbf{Q}_{vl}^{-1}+\mathbf{I}_N)\\
\textrm{s.t.} \quad & C_{sc}=\log_2 \det({\mathbf{Q}_{vl}^{-1}(p\mathbf{H}_l\mathbf{H}_l^{\text{H}}+\sigma^2\mathbf{I}_N)+\mathbf{I}_N}).
\end{aligned} 
\label{sumRate_max_prob_VC}
    \end{equation}
The problem defined in (\ref{sumRate_max_prob_VC}) is an NLP (non-linear programming) problem. The constraint does not define a convex set,
%\robbert{affine set\vida{Convex set is right. our equality constraint should be affine but again, in a convex problem, they define a convex set over which we should search for solution}},
 so the problem is not convex.
 However, the closed-form globally optimal matrix 
 %\robbert{which leads to the globally optimal solution} 
 is derived in (\ref{opt_Qmat_VC}).
%Therefore, the Karush–Kuhn–Tucker (KKT) conditions are only necessary for optimality \cite{boyd_conv}.
%Note that by concave problem, we mean a maximization problem with a concave objective function and a feasible convex set.
%Thus, KKT conditions only guarantee necessary condition for optimality. For sufficient condition for optimality, we make sure that 
% Defining the singular value decomposition of $\mathbf{H}_l$ as:
% \begin{equation}
% \mathbf{H}_l=\mathbf{U}_l\mathbf{\Sigma}_l\mathbf{C}_l^{\text{H}},
% \label{Tot_bit_rel_to_mutlinfo_PCA_EC}
% \end{equation}
% where $\mathbf{\Sigma}_l$ is an $N\times K$ matrix with the singular values on the diagonal. $\mathbf{U}_l$ has the left singular vectors
% in its columns and $\mathbf{C}_l$ has the right singular vectors in its columns. 
%Singular value decomposition of $\mathbf{H}_l$ defined in (\ref{PCA_on_rec_sig}) and

To present the optimal solution, we need the singular value decomposition (SVD) of matrix $\mathbf{H}_l$, which is formulated as follows:
\begin{equation}
\mathbf{H}_l=\mathbf{U}_l\mathbf{\Sigma}_l\mathbf{V}_l^{\text{H}}. 
\label{svdchan}
\end{equation}
Accordingly, the eigenvalue decomposition of matrix $\mathbf{H}_l\mathbf{H}_l^{\text{H}}$ is as follows:
\begin{equation}
\mathbf{H}_l\mathbf{H}_l^{\text{H}}=\mathbf{U}_l\mathbf{\Sigma}_l\mathbf{\Sigma}_l^{\text{H}}\mathbf{U}_l^{\text{H}}.
\label{eigHH'}
\end{equation}
The columns of matrices $\mathbf{U}_l\in \mathbb{C}^{N\times N}$ and $\mathbf{V}_l\in \mathbb{C}^{K\times K}$ are the left and right singular vectors of $\mathbf{H}_l$, respectively, and  
 so the columns of $\mathbf{U}_l$ are also the eigenvectors of matrix $\mathbf{H}_l\mathbf{H}_l^{\text{H}}$. Furthermore, $\mathbf{\Sigma}_l\in \mathbb{C}^{N\times K}$ is a rectangular matrix with the sorted singular values of the $\mathbf{H}_l$ on its diagonal elements.
% $\mathbf{Q}_{vl}$ is positive semi-definite, with eigenvalue decomposition as follows:
%     \begin{equation}
% \mathbf{Q}_{vl}=\mathbf{U}_{vlq}\mathbf{\Sigma}_{vlq}\mathbf{U}_{vlq}^{\text{H}}.
% \label{opt_Qmat_VC}
% \end{equation}
The optimal matrix $\mathbf{Q}_{vl}^{-1}$ is given as:
\begin{equation}
    {(\mathbf{Q}_{vl}^{-1})}^o=\mathbf{U}_l{(\mathbf{\Sigma}_{vlq}^{-1})}^o\mathbf{U}_l^{\text{H}},
    \label{opt_Qmat_VC}
\end{equation}
where the $i^{th}$ diagonal element of ${(\mathbf{\Sigma}_{vlq}^{-1})}^o$ is calculated as:
\begin{equation}
\begin{aligned}
       \lambda^o_{vlqi}=\max(0,\frac{1}{\mu_{vl}^o}(\frac{1}{\sigma^2}-\frac{1}{p\lambda_{li}^2+\sigma^2})-\frac{1}{\sigma^2}), \forall i.
        \label{eig_opt_Qmat_VC}
        \end{aligned}
    \end{equation}
    where $\lambda_{li}$ is the $i^{th}$ singular value of $\mathbf{H}_l$. \textcolor{black}{The solution in (\ref{eig_opt_Qmat_VC}) is reverse water filling on the eigenvalues of received signal vector covariance matrix \cite{CovTho}.}
The proof is elaborated on in \nameref{apdb}.
Note that the Lagrange multiplier $\mu_{vl}^o$ is selected to meet the equality constraint in (\ref{sumRate_max_prob_VC}). 
% \textcolor{black}{Based on eq. \ref{eig_opt_Qmat_VC}, the higher the eigenvalue $\lambda_{il}^2$, the higher the corresponding $\lambda^o_{vlqi}$, which means that more bits are allocated for quantizing the corresponding eigen direction in $\mathbf{y}_l$. For very low $\lambda_{il}$, most probably $\lambda^o_{vlqi}=0$, meaning no bits are allocated for quantizing the corresponding direction.}
\subsection{Option 2: Element-wise compression of the received signal vector}\label{sec3_B}
Each element of the received signal vector is compressed individually in this option. 
The number of bits allocated to compression of the $i^{th}$ element is denoted by $b_i$ and $C_{sc}=\sum_{i=1}^{N}b_i$. The compressed $i^{th}$ element of the received signal vector at AP $l$ is as follows:
\begin{equation}
   \hat{y}_{eli}=y_{li}+q_{eli}=\mathbf{H}_{l[i,:]}\mathbf{s}+n_{li}+q_{eli},
   \label{rec_sig_apl_EC}
\end{equation}
where $n_{li}$ is the $i^{th}$ element of noise vector $\mathbf{n}_l$, $z_{eli}=n_{li}+q_{eli}$ with $q_{eli}\sim\mathcal{C}\mathcal{N}(0,\sigma_{eli}^2), \forall i\in\{1,\hdots,N\}$ and $\mathbf{q}_{el}=\begin{bmatrix}
    {q}_{el1},&\cdots, & {q}_{elN}
\end{bmatrix}^{\text{T}}$
is the compression noise vector with covariance matrix $\mathbf{Q}_{el}=\mathbb{E}\{\mathbf{q}_{el}\mathbf{q}_{el}^\text{H}\}$. Vector $\hat{\mathbf{y}}_{el}=\begin{bmatrix}
    \hat{y}_{el1},& \hdots,& \hat{y}_{elN}
    \end{bmatrix}^{\text{T}}$ is the compressed received signal vector at AP $l$. The receiver plus compression noise vector is  $\mathbf{z}_{el}=\begin{bmatrix}
    {z}_{el1},&\cdots, & {z}_{elN}
\end{bmatrix}^{\text{T}}$. 
%\robbert{The rate over the compression channel is $b_i$}

The relation between $b_i$ and the compression noise of the $i^{th}$ element is as follows \cite{CovTho}:
% \robbert{I think this could use some extra
% explanation because $b_i$ \vida{b_i is the number of bit for compressing each dimesion of received vector as mentioend.}is not the 
% compression noise but the rate over the compression channel, right? Hence: We assume compression depth of $b_i$ on the transmission over the  compression
% channel when we then link this to the mutual
% information over the compression channel, we can determine the achievable
% rate given a specific compression depth.} : 
\begin{equation}
\begin{aligned}
    b_i=\mathcal{I}(y_{li};\hat{y}_{eli}|\mathbf{H}_{l[i,:]})=&\mathcal{H}(\hat{y}_{eli}|\mathbf{H}_{l[i,:]})-\mathcal{H}(\hat{y}_{eli}|y_{li},\mathbf{H}_{l[i,:]})\\=&\log_2(1+\frac{p\lVert \mathbf{H}_{l[i,:]}\rVert^2+\sigma^2}{\sigma_{eli}^{2}}),
    \label{bit_rel_to_mutlinfo_EC}
    \end{aligned}
\end{equation}
where subscript $[i,:]$ represent the $i^{th}$ row of $\mathbf{H}_{l}$.
The diagonal elements of the covariance matrix $\mathbf{Q}_{el}$ are known and equal to $\sigma_{eli}^{2}, \forall i\in\{1,\hdots,N\}$. 
%\robbert{Perhaps you could use complex conjugate to clarify that you 
%are working with scalars here:
% $\sigma_{eli}^{2}=\mathbb{E}\{q_{eli}^*q_{eli}\}, \forall i\in\{1,\hdots,N\}$ }
The off-diagonal elements of matrix $\mathbf{Q}_{el}$ are unknown.

We define diagonal matrix $\mathbf{P}_l$ with the variance of the elements of the received signal vector $\{\mathbf{y}_l|\mathbf{H}_l\}$ as its diagonal elements,
\begin{equation}
    \mathbf{P}_l = \text{diag}\left(p\lVert \mathbf{H}_{l[1,:]}\rVert^2+\sigma^2, \cdots, p\lVert \mathbf{H}_{l[N,:]}\rVert^2+\sigma^2\right),
    \label{eq25}
\end{equation}
and the diagonal matrix $\mathbf{Q}_{el}^d$ with variance of the elements of the compression noise vector $\mathbf{q}_{el}$ as its diagonal elements, as follows:
\begin{equation}
    \mathbf{Q}_{el}^d = \text{diag}\left(\sigma_{el1}^{2}, \dots, \sigma_{elN}^{2}\right).
    \label{com_nois_cov}
\end{equation}
We can relate $C_{sc}$ to the variance of the compression noise vector elements as follows:
\begin{equation}
\begin{aligned}
    C_{sc} =\sum_{i=1}^{N}b_i=&\log_2\prod_{i=1}^N(1+\frac{p\lVert \mathbf{H}_{l[i,:]}\rVert^2+\sigma^2}{\textcolor{black}{\sigma_{eli}^{2}}}) 
     \\=&\log_2\det(\mathbf{I}_N+  \mathbf{P}_l  {\mathbf{Q}^{d}_{el}}^{-1}). 
    \label{TOTbit_rel_to_mutlinfo_EC}
\end{aligned}
\end{equation}

The values for $b_i,\forall i$ (and consequently $\sigma^2_{eli}$) can be selected heuristically. One simple approach is compressing the vector elements with the same number of bits. A better approach is to select the number of bits based on the variance of each element of the received signal vector. 

The APs refine the users' signal vector based on the element-wise compressed received signal vector. The network-wide compressed received signal vector, in this case, is as follows:
 \begin{equation}
 \begin{aligned}
    \hat{\mathbf{y}}_{e}&=\mathbf{y}+\mathbf{q}_e=\mathbf{H}\mathbf{s}+\underbrace{\mathbf{n}+\mathbf{q}_e}_{\mathbf{z}_e},
\label{rec_sig_EC}
\end{aligned}
\end{equation}
where $\hat{\mathbf{y}}_{e}=\begin{bmatrix}
        \hat{\mathbf{y}}^{\text{T}}_{e1},& \hdots,& \hat{\mathbf{y}}_{eL}^{\text{T}}
    \end{bmatrix}^{\text{T}}$, $\mathbf{q}_e=\begin{bmatrix}
        \mathbf{q}_{e1}^{\text{T}},& \hdots,& \mathbf{q}_{eL}^{\text{T}}
\end{bmatrix}^{\text{T}}$ and $\mathbf{z}_e=\begin{bmatrix}\mathbf{z}_{e1}^{\text{T}},&\hdots,& \mathbf{z}_{eL}^{\text{T}}\end{bmatrix}^{\text{T}}$ with covariance matrix $\mathbf{Z}_e=\mathbb{E}\{\mathbf{z}_e\mathbf{z}_e^\text{H}\}=\text{blkdiag}(\mathbf{Z}_{e1},\hdots,\mathbf{Z}_{eL})=\text{blkdiag}(\mathbf{Q}_{e1}+\sigma^2\mathbf{I}_N,\hdots,\mathbf{Q}_{eL}+\sigma^2\mathbf{I}_N)$.
In element-wise compression, the correlation between compression noise elements in one AP, i.e., the off-diagonal elements of $\mathbf{Q}_{el}, \forall l$, are unknown.
 Therefore, while computing the combining matrix to estimate users' signal vector, $\mathbf{Q}_{el}, \forall l$ is assumed to be a diagonal matrix, which adversely affects the estimation quality.
Following the discussion in Section \ref{secII-A}, the estimation of the users' signal vector results in equations (\ref{ue_sig_est_EC}) and (\ref{comb_vec_EC})%, where $\mathbf{Q}_{el}^d$ be the diagonal matrix with the same diagonal elements as $\mathbf{Q}_{el}$.
%The estimate of users' signal vector 
, as follows:
\begin{equation}
 \hat{\mathbf{s}}_e= \mathbf{C}_e\hat{\mathbf{y}}_e,
 \label{ue_sig_est_EC}
\end{equation}
where the combining matrix $\mathbf{C}_e$ is formulated as follows:
\begin{equation}
\mathbf{C}_e=(\mathbf{H}^{\text{H}}{(\mathbf{Z}_e^d)}^{-1}\mathbf{H}+\frac{1}{p}\mathbf{I}_K)^{-1}\mathbf{H}^{\text{H}}{(\mathbf{Z}_e^d)}^{-1},
\label{comb_vec_EC}
\end{equation}
where $\mathbf{Z}_e^d=\text{blkdiag}(\mathbf{Z}_{e1}^d,\hdots,\mathbf{Z}_{eL}^d)=\text{blkdiag}(\mathbf{Q}_{e1}^d+\sigma^2\mathbf{I}_N,\hdots,\mathbf{Q}_{eL}^d+\sigma^2\mathbf{I}_N)$ and $\mathbf{Q}_{el}^d$ is the diagonal matrix with the same diagonal elements as $\mathbf{Q}_{el}$.
 The solution to the optimization problem in option $2$ follows the same steps in section \ref{sec3_A} and \nameref{apdb}. However, deriving an upper bound on users' sum-SE similar to (\ref{sumRate_VC}) is not straightforward due to the lack of knowledge of the off-diagonal elements of $\mathbf{Z}_e$ and thus
    %In option $2$, the user's sum-SE 
    can be simplified as follows:
    %\robbert{I think it should be $\mathbf{I}_N$ in the last line}
%     \begin{equation}
% \begin{aligned}
%     R_{EC}&=\mathcal{I}(\mathbf{C}_e\hat{\mathbf{y}}_e;\mathbf{s}|\mathbf{C}_e,\mathbf{H})\\&=\mathcal{H}(\mathbf{C}_e{\hat{\mathbf{y}}_e}|\mathbf{C}_e,\mathbf{H})-\mathcal{H}(\mathbf{C}_e{\hat{\mathbf{y}}_e}|\mathbf{s},\mathbf{C}_e,\mathbf{H})\\&=\log_2\det(p\mathbf{H}\mathbf{H}^{\text{H}}\mathbf{Z}_e^{-1}+\mathbf{I}_{NL})\\&\myineqla\sum_{l=1}^L\log_2\det(p\mathbf{H}_l\mathbf{H}_l^{\text{H}}\mathbf{Z}_{el}^{-1}+\mathbf{I}_{N}).
%    \end{aligned}
%     \label{comb_vec_EC}  
% \end{equation}

\begin{equation}
\begin{aligned}
    R_{e}&=\frac{\tau_u}{\tau_c}\log_2\det(p\mathbf{H}\mathbf{H}^{\text{H}}{(\mathbf{Z}_e^d)}^{-1}+\mathbf{I}_{NL})\\&\myineqla\frac{\tau_u}{\tau_c}\sum_{l=1}^L\log_2\det(p\mathbf{H}_l\mathbf{H}_l^{\text{H}}{(\mathbf{Z}_{el}^d)}^{-1}+\mathbf{I}_{N})\\&\myineqlb\frac{\tau_u}{\tau_c}\sum_{l=1}^L\log_2\det(p \text{diag}(\mathbf{H}_l\mathbf{H}_l^{\text{H}}){(\mathbf{Z}_{el}^{d}})^{-1}+\mathbf{I}_{N})\\&=\frac{\tau_u}{\tau_c}\sum_{l=1}^L\log_2\det(p\mathbf{W}_l(\mathbf{Q}_{el}^d+\sigma^2\mathbf{I}_N)^{-1}+\mathbf{I}_{N}),
   \end{aligned}
    \label{sumRate_EC}  
\end{equation}
    % Following the same steps as in equations (\ref{eig_opt_Q_EC}) and (\ref{rec_sig_cov_mat}) and based on \nameref{apdb} and defining $\mathbf{T}_1=(\mathbf{H}^{\text{H}}{\mathbf{Z}_e^d}^{-1}\mathbf{H}+\frac{1}{p}\mathbf{I}_k)^{-1}$ and $\mathbf{T}_2=\mathbf{H}^{\text{H}}\mathbf{Z}_e^{-1}\mathbf{H}$, $\mathcal{I}(\mathbf{C}_e\hat{\mathbf{y}}_e;\mathbf{s},|\mathbf{C}_e,\mathbf{H})$ can be calculated as:
    % \begin{equation}
    % \begin{aligned}
    %     \mathcal{I}(\mathbf{C}_e\hat{\mathbf{y}}_e;\mathbf{s}|\mathbf{C}_e,\mathbf{H})&=\log_2\det(\mathbf{T}_2+\mathbf{I}_K)\\&=\log_2\det(p\mathbf{H}\mathbf{H}^{\text{H}}\mathbf{Z}_e^{-1}+\mathbf{I}_{NL})\\&\myineqla\sum_{l=1}^L\log_2\det(p\mathbf{H}_l\mathbf{H}_l^{\text{H}}\mathbf{Z}_{el}^{-1}+\mathbf{I}_{NL}).
    %  \end{aligned}
    %  \label{eig_opt_Q_PCA_EC}
    % \end{equation}
    where $\mathbf{W}_l$ is defined as
    $\mathbf{W}_l=\text{diag}(\lVert \mathbf{H}_{l[1,:]}\rVert^2,\hdots,\lVert \mathbf{H}_{l[N,:]}\rVert^2)
        $. In (\ref{sumRate_EC}), the upper bounds $\myineqla$ and $\myineqlb$ are proved similar to $\myineqlb$ in (\ref{sumRate_VC}). 
    % Note that only the diagonal elements of $\mathbf{Z}_{el}, \forall l\in\{1,\hdots,N\}$ are optimized as the compression of the received vector is done element-wise. Therefore, instead of the upper-bound function in (\ref{sumRate_EC}), we try to maximize an approximate function as follows:
    % \begin{equation} 
    % \sum_{l=1}^L\log_2\det(pdiag(\mathbf{H}_l\mathbf{H}_l^{\text{H}}){(\mathbf{Z}_{el}^{d}})^{-1}+\mathbf{I}_{N})=\sum_{l=1}^L\log_2\det(p\mathbf{W}_l(\mathbf{Q}_{el}^d+\sigma^2\mathbf{I})^{-1}+\mathbf{I}_{N}),
    % \label{sumRate_EC}
    %\end{equation} 
    % Note that in the simulation section, due to the lack of knowledge of the off-diagonal value of $\mathbf{Z}_{el},\forall l\in\{1,\hdots,L\}$, the sum-SE defined in (\ref{sumRate_EC}) is not calculable. Therefore, the lower-bound function in (\ref{eq31}) (proof in \nameref{apdd}) is plotted.
    
    %\robbert{I think you mean $\mathbf{Z}^d_{el}$ and <, no? In (30) it's upper-bound\vida{No Robbert, In (30) we had $Z_e$ which was block diagonal and we concluded that bound with $Z_{el}$. Here we have $Z_{e}^d$ which is only the diagonal part of $Z_{e}$ and keep in mind that $(diag(Z_e))^{-1}\neq(diag(Z_e^{-1})$ } Ok yes I get it now }
    % \begin{equation}
    %     R_{EC}>\log_2\det(p\mathbf{H}\mathbf{H}^{\text{H}}{({\mathbf{Z}^d_{e}})}^{-1}+\mathbf{I}_{NL})
    %     \label{eq31}
    % \end{equation}
    Similar to Section \ref{sec3_A}, the sub-problem at AP $l$ to find the diagonal element of $\mathbf{Q}_{el}^d$ and, subsequently, the number of bits to compress each of the elements of the received signal vector $\mathbf{y}_l$ is formulated as follows:
    \begin{equation}
     \begin{aligned}
\arg \max_{{{\mathbf{Q}_{el}^d}}^{-1}\succeq\mathbf{0}} \quad  &\log_2    \det(p\mathbf{W}_l(\mathbf{Q}_{el}^d+\sigma^2\mathbf{I}_N)^{-1}+\mathbf{I}_N)\\=&\log_2\det(p\mathbf{W}_l+\mathbf{Q}_{el}^d+\sigma^2\mathbf{I}_N)-\\&\log_2\det(\mathbf{Q}_{el}^d+\sigma^2\mathbf{I}_N)\\=& \log_2 \det(\mathbf{P}_l{{\mathbf{Q}_{el}^d}}^{-1}+\mathbf{I}_N)-\\&\log_2 \det(\sigma^2{{\mathbf{Q}_{el}^d}}^{-1}+\mathbf{I}_N)\\
\textrm{s.t.} \quad & C_{sc}=\log_2 \det(\mathbf{P}_l{{\mathbf{Q}_{el}^d}}^{-1}+\mathbf{I}_N),
\end{aligned} 
\label{sumRate_max_prob_EC}
\end{equation}
With $\mathbf{P}_l$ defined in (\ref{eq25}). Note that $\mathbf{P}_l=p\mathbf{W}_l+\sigma^2\mathbf{I}_N$.
Following the same steps as in \nameref{apdb}, the $i^{th}$ diagonal element of matrix ${({\mathbf{Q}_{el}^d}^{-1})}^o$, shown as $\lambda_{elqi}^o$,  is given as follows:
\begin{equation}
    \lambda^o_{elqi}=\max(0,\frac{1}{\mu_{el}^o}(\frac{1}{\sigma^2}-\frac{1}{{\mathbf{P}_l}_{[i,i]}})-\frac{1}{\sigma^2}), \forall i,
    \label{eig_opt_Q_EC}
\end{equation}
where $\mathbf{P}_{l[i,i]}$ denotes the $i^{th}$ diagonal element of $\mathbf{P}_{l}$. Finally, ${({\mathbf{Q}_{el}^d}^{-1})}^o=\text{diag}(\lambda^o_{elq1}, \hdots, \lambda^o_{elqN})$.
The parameter $\mu_{el}^o$ in (\ref{eig_opt_Q_EC}) is also calculated similarly to $\mu_{vl}^o$ in Section \ref{sec3_A}.
% Note that as sum-SE defined in (\ref{sumRate_EC}) is not calculable due to the lack of knowledge of the off-diagonal value of $\mathbf{Z}_{el},\forall l\in\{1,\hdots,L\}$, the lower-bound function (proof in \nameref{apdd}) $\log_2\det(p\mathbf{H}\mathbf{H}^{\text{H}}{({\mathbf{Z}^d_{e}})}^{-1}+\mathbf{I}_{NL})$ is instead plotted for performance evaluations. 
%The lower-bound functions in (\ref{sumRate_EC}) is expected to get tighter as antennas get more distributed in the area.
\subsection{Option 3: Element-wise compression of the PCA-pre-processed received signal vector}\label{sec3_C}
    % \subsection{Motivation}
    % In previous section, we considered buffering of the received vector in a vector-wise and an element-wise manner. Vector-wise compression takes advantage of the correlation between the elements of the received vector and consequently uses the buffering bits in a more efficient way which is supposed to give us a better performance in terms of spectral efficiency. However, vector-wise compression are of high complexity. On the other hand, element-wise compression are simpler but results in performance degradation. To take advantage of simple compression without compromising users' sum-SE, we can pre-process the received vector in each AP using some orthogonalization method such as PCA. In the following, the pre-processing operation followed by buffering of the preproced received vector is elaborated on. 
    % \subsection{Pre processing and buffering the received vector}
\textcolor{black}{Despite being efficient in bit usage, vector compression can be costly in practice. For example, using well-known quantization algorithms such as the generalized Lloyd algorithm to quantize a given set of $N$-dimensional vectors with $n_q$ bits have a complexity order of $\mathcal{O}(\text{number of vectors}\times N\times 2^{n_q}\times I)$ if the algorithm stops after $I$ iterations. 
On the other hand, element-wise quantization (sharing the total number of bits among the dimensions and then quantizing each dimension individually) is of lower complexity, i.e., $\mathcal{O}(\text{number of vectors}\times 2^{n_q\delta_i}\times I)$ for quantizing dimension $i$ and $\sum_{j=1}^N\delta_j=1$, at the cost of inefficient bit usage. To bridge this gap, it is proposed to de-correlate the vectors' dimensions before element-wise quantization to use the total bits efficiently.} With the above introduction in mind,
suppose each AP uses PCA to map its local received signal vector into another subspace. At AP $l$,
the covariance matrix of the received signal vector, using the SVD of $\mathbf{H}_l$ in (\ref{svdchan}), can be formulated as follows: 
\begin{equation}
\begin{aligned}
    \mathbf{R}_{yl}&=\mathbb{E}\{(\mathbf{y}_l-\mathbb{E}\{\mathbf{y}_l\})(\mathbf{y}_l-\mathbb{E}\{\mathbf{y}_l\})^{\text{H}}|\mathbf{H}_l\}\\&\myeq\mathbb{E}\{\mathbf{y}_l\mathbf{y}_l^{\text{H}}|\mathbf{H}_l\}=\mathbf{U}_l\underbrace{(p\mathbf{\Sigma}_l\mathbf{\Sigma}_l^{\text{H}}+\sigma^2\mathbf{I}_N)}_{\tilde{\mathbf{\Sigma}}_l}\mathbf{U}_l^{\text{H}},
    \label{rec_sig_cov_mat}
    \end{aligned}
\end{equation}
where $\myeq$ is due to the fact that $\mathbb{E}\{\mathbf{y}_l|\mathbf{H}_l\}=\mathbf{0}$ (as $\mathbb{E}\{\mathbf{s}\}=0$ and $\mathbb{E}\{\mathbf{n}_l\}=0$ ).
The mapped received signal vector is as follows:
\begin{equation}
    \begin{aligned} \tilde{\mathbf{y}}_l&=\mathbf{A}_l^{\text{H}}\mathbf{y}_l=\underbrace{\mathbf{A}_l^{\text{H}}\mathbf{H}_l}_{\tilde{\mathbf{H}}_l}\mathbf{s}+\underbrace{\mathbf{A}_l^{\text{H}}\mathbf{n}_l}_{\tilde{\mathbf{n}}_{l}},
    \label{PCA_on_rec_sig}
    \end{aligned}
\end{equation}
where $\mathbf{A}_l=\mathbf{U}_{l[:,1:x]}$, in which $x$ can be in the range $[1:min(N,K)]$. We select $x=min(N,K)$. Note that $\tilde{\mathbf{H}}_l$ and $\tilde{\mathbf{n}}_l$ are the local effective channel and receiver noise at AP $l$, respectively. After the mapping, the APs must store the pre-processed received signal vector. The vector $\{\tilde{\mathbf{y}}_l|\tilde{\mathbf{H}}_l\}$ is a circularly symmetric Gaussian random vector %(hence, each pair of its elements are jointly circularly symmetric Gaussian random variables) 
with zero mean and uncorrelated elements (with diagonal covariance matrix $\mathbb{E}\{\tilde{\mathbf{y}}_l\tilde{\mathbf{y}}_l^{\text{H}}|\tilde{\mathbf{H}}_l\}=\tilde{\mathbf{\Sigma}}_l$), hence the elements are also mutually independent as any two jointly Gaussian and uncorrelated random variables are independent as well. 
%Based on the independence of the pre-processed received signal vector elements and rate-distortion argument in (\ref{Rat_dis_fun}) and (\ref{mutualinfo}), 
The compressed pre-processed received signal vector at AP $l$ is as follows:
\begin{equation}
\begin{aligned}
\hat{\tilde{\mathbf{y}}}_{el}&=\tilde{\mathbf{y}}_l+\tilde{\mathbf{q}}_{el}=\tilde{\mathbf{H}}_l\mathbf{s}+\tilde{\mathbf{n}}_l+\tilde{\mathbf{q}}_{el},
    \label{rec_sig_PCA_pp_EC}
    \end{aligned}
\end{equation}
where $\tilde{\mathbf{z}}_{el}=\tilde{\mathbf{n}}_l+\tilde{\mathbf{q}}_{el}$
% with covariance matrix $\tilde{\mathbf{Z}}_{el}=\mathbb{E}\{\tilde{\mathbf{z}}_{el}\tilde{\mathbf{z}}_{el}^\text{H}\}$
and $\tilde{\mathbf{q}}_{el}\sim\mathcal{C}\mathcal{N}(0,\tilde{\mathbf{Q}}_{el})$. Note that $\tilde{\mathbf{Q}}_{el}$ is assumed to be diagonal as the elements of the $\{\tilde{\mathbf{y}}_l|\tilde{\mathbf{H}}_l\}$ are independent. Hence, they can be optimally compressed individually, \textcolor{black}{conditioned on the optimal allocation of a total number of bits.} 
% Similar to Sections \ref{sec3_A} and \ref{sec3_B}, a sum-SE optimization problem is solved to choose the number of compression bits for each element of the pre-processed received signal vector.
%and the corresponding $\tilde{\mathbf{Q}}_{el}$ should satisfy the limited memory capacity constraint at AP $l$. 
% Following the discussion in the previous sections, $\{\hat{\tilde{\mathbf{y}}}_l|\mathbf{G}_l\}$ is a multivariate Gaussian random vector with covariance matrix $\mathbf{R}_{\hat{\tilde{y}}_l}=\tilde{\mathbf{\Sigma}}_l+\tilde{\mathbf{Q}}_{vl}$. 
Matrix $\tilde{\mathbf{Q}}_{el}$ can be related to $C_{sc}$ as follows:
% \begin{equation}
% \begin{aligned}
% C_{sc}=&I(\tilde{\mathbf{y}}_l;\hat{\tilde{\mathbf{y}}}_l)\\
%     =&
%     \mathcal{H}_l(\hat{\tilde{\mathbf{y}}}_l)-\mathcal{H}_l(\hat{\tilde{\mathbf{y}}}_l|\tilde{\mathbf{y}}_l)\\
%     =&\log_2 2\pi e \det({\tilde{\mathbf{Q}}_{vl}+\mathbf{\Sigma}'_{l}})-\log_2 2\pi e \det({\tilde{\mathbf{Q}}_{vl}})\\
%     =& \log_2\det(\mathbf{I}_{N}+\tilde{\mathbf{Q}}_{vl}\mathbf{\Sigma}'_{l})\\
%     =& \log_2 \det(\mathbf{I}_{N}+\tilde{\mathbf{Q}}_{vl}^{-1}(\mathbf{\Sigma}_{l}\mathbf{\Sigma}_{l}^{\text{H}}+\sigma^2\mathbf{I}_{N})).
%     \label{rec_sig_apl_EC}
% \end{aligned}
\begin{equation}
\begin{aligned}
C_{sc}=&I(\tilde{\mathbf{y}}_{l};\hat{\tilde{\mathbf{y}}}_{el}|\tilde{\mathbf{H}}_l)\\
    =&\mathcal{H}_l(\hat{\tilde{\mathbf{y}}}_{el}|\tilde{\mathbf{H}}_l)-\mathcal{H}_l(\hat{\tilde{\mathbf{y}}}_{el}|\tilde{\mathbf{y}}_{l},\tilde{\mathbf{H}}_l)\\
    =& \log_2 \det(\tilde{\mathbf{Q}}_{el}^{-1}(p\tilde{\mathbf{H}}_l\tilde{\mathbf{H}}_l^{\text{H}}+\sigma^2\mathbf{I}_{x})+\mathbf{I}_{x})\\
    \textcolor{black}{=}&\textcolor{black}{\sum_{i=1}^{x}\log_2 ({\tilde{\lambda}_{elqi}}({p{\lambda}_{il}^2+\sigma^2})+1)},
    \label{Tot_bit_rel_to_mutlinfo_PCA_EC}
\end{aligned}
\end{equation}

\textcolor{black}{where $\tilde{\lambda}_{elqi}$ is the $i^{th}$ diagonal element of $\tilde{\mathbf{Q}}_{el}^{-1}$.} After pre-processing and compressing the local received signal vector, the network-wide compressed received signal vector is as follows:
\begin{equation}
\begin{aligned}
\hat{\tilde{\mathbf{y}}}_e=\tilde{\mathbf{y}}+\tilde{\mathbf{q}}_e=\tilde{\mathbf{H}}\mathbf{s}+\underbrace{\tilde{\mathbf{n}}+\tilde{\mathbf{q}}_e}_{\tilde{\mathbf{z}}_e},
    \label{Tot_rec_sig_EC}
    \end{aligned}
\end{equation}
where $\hat{\tilde{\mathbf{y}}}_e=\begin{bmatrix}
\hat{\tilde{\mathbf{y}}}_{e1}^{\text{T}},&\hdots,& \hat{\tilde{\mathbf{y}}}_{eL}^{\text{T}}
\end{bmatrix}^{\text{T}}$, ${\tilde{\mathbf{y}}}=\begin{bmatrix}
{\tilde{\mathbf{y}}}_{1}^{\text{T}},&\hdots,& {\tilde{\mathbf{y}}}_{L}^{\text{T}}
\end{bmatrix}^{\text{T}}$, $\tilde{\mathbf{H}}=\begin{bmatrix}
    \tilde{\mathbf{H}}_1^{\text{T}},&\hdots,& \tilde{\mathbf{H}}_L^{\text{T}}
\end{bmatrix}^{\text{T}}$. Furthermore, $\tilde{\mathbf{q}}_e=\begin{bmatrix}\tilde{\mathbf{q}}_{e1}^{\text{T}},&\hdots,& \tilde{\mathbf{q}}_{eL}^{\text{T}}\end{bmatrix}^{\text{T}}$ and $\tilde{\mathbf{z}}_e=\begin{bmatrix}\tilde{\mathbf{z}}_{e1}^{\text{T}},&\hdots,&\tilde{\mathbf{z}}_{eL}^{\text{T}}\end{bmatrix}^{\text{T}}$ with covariance matrix
$\tilde{\mathbf{Z}}_e=\mathbb{E}\{\tilde{\mathbf{z}}_e\tilde{\mathbf{z}}_e^\text{H}\}=\text{blkdiag}(\tilde{\mathbf{Z}}_{e1},\hdots,\tilde{\mathbf{Z}}_{eL})=\text{blkdiag}(\tilde{\mathbf{Q}}_{e1}+\sigma^2\mathbf{I}_{x},\hdots,\tilde{\mathbf{Q}}_{eL}+\sigma^2\mathbf{I}_{x})$.
Finally, \textcolor{black}{after RLS processing of the compressed PCA pre-processed signal vectors at the APs sequentially,} the estimate of users' signal vector \textcolor{black}{in the last AP} is as follows:
\begin{equation}
 \hat{\tilde{\mathbf{s}}}= \tilde{\mathbf{C}}_e\hat{\tilde{\mathbf{y}}}_e,
 \label{UE_sig_est_PCA_EC}
\end{equation}
where the combining matrix given $\tilde{\mathbf{H}}$ is formulated as follows:
\begin{equation}
\tilde{\mathbf{C}}_e=(\tilde{\mathbf{H}}^{\text{H}}\tilde{\mathbf{Z}}_e^{-1}\tilde{\mathbf{H}}+\frac{1}{p}\mathbf{I}_{xL})^{-1}\tilde{\mathbf{H}}^{\text{H}}\tilde{\mathbf{Z}}_e^{-1}.%% changed
\label{combVec_PCA_EC}
\end{equation}
% Based on the derivations in section \ref{sec4_C}, section C of \nameref{apdb} and following the same steps as the sum-SE derivation of two previous scenarios, we define $\mathbf{T}_1=(\mathbf{G}^{\text{H}}\tilde{\mathbf{Z}}_{v}^{-1}\mathbf{G}+\frac{1}{p}\mathbf{I}_K)^{-1}$ and $\mathbf{T}_2=\mathbf{G}^{\text{H}}\tilde{\mathbf{Z}}_{v}^{-1}\mathbf{G}$.
% The network wide effective channel is $\mathbf{G}=\begin{bmatrix}
%     \mathbf{G}_1^{\text{T}}&&\hdots&&\mathbf{G}_L^{\text{T}}
% \end{bmatrix}^{\text{T}}$
%  and network wide "compression+receiver noise" covarinace matrix is formulated as follows:
%  \begin{equation}
%      \mathbf{Z}=blkdiag(\mathbf{Q}_{v1}^{PCA}+\sigma^2\mathbf{I}_{x_1},\hdots,\tilde{\mathbf{Q}}_{vl}+\sigma^2\mathbf{I}_{x_L})
%  \end{equation}
% The network-wide MMSE combining vector given $\mathbf{G}$ on $\dbtilde{\mathbf{y}}_l$ is formulated as below:
% \begin{equation}
%     \mathbf{C}_{vl}^{PCA}=p(p\mathbf{G}\mathbf{G}^{\text{H}}+\mathbf{Z}_{vl}^{PCA})^{-1}\mathbf{G}
% \end{equation}
By defining the signal estimation error as $\tilde{\mathbf{e}}=\mathbf{s}-\hat{\tilde{\mathbf{s}}}$, and with the same reasoning as subsection \ref{sec3_A} the users' sum-SE is given as:
\begin{equation}
\begin{aligned}
   R_e^{PCA}&=\frac{\tau_u}{\tau_c}\mathcal{I}(\mathbf{s};\hat{\tilde{\mathbf{s}}}) 
     \\&=\frac{\tau_u}{\tau_c}\bigl(\mathcal{H}(\mathbf{s})-\mathcal{H}(\mathbf{s}|\hat{\tilde{\mathbf{s}}})\bigr)
     \\&
    \myeq\frac{\tau_u}{\tau_c}\log_2 \det(p\tilde{\mathbf{H}}\tilde{\mathbf{H}}^{\text{H}}\tilde{\mathbf{Z}}_e^{-1}+\mathbf{I}_{xL}) &\\ &\myineqlb \frac{\tau_u}{\tau_c}\sum_{l=1}^L \log_2 \det(p\tilde{\mathbf{H}}_l\tilde{\mathbf{H}}_l^{\text{H}}\tilde{\mathbf{Z}}_{el}^{-1}+\mathbf{I}_{x}),
    \end{aligned}
    \label{sumRate_PCA_EC}  
\end{equation}
where $\myeq$ and $\myineqlb$ are proved similarly to $\myeq$ and $\myineqlb$ in (\ref{sumRate_VC}), respectively.
The upper bound function defined in (\ref{sumRate_PCA_EC}) is a summation of $L$ functions, each dependent on a unique matrix $\tilde{\mathbf{Z}}_{el}$ and effective channel $\tilde{\mathbf{H}}_l$. The maximization sub-problem at AP $l$ is formulated as:
% To find the optimal value for the diagonal elements of the matrix $\tilde{\mathbf{Q}}_{vl}$, we try to maximize the sum-SE function in equation (\ref{eq49}) subject to the limited capacity buffer constraints in each AP based on equation (\ref{eig_opt_Qmat_VC}).
% As the sum-SE is a summation of multiple functions corresponding to different APs and the buffer constraints in each AP is independent from the other APs, the sum-SE maximization problem can be decomposed to $L$ independent maximization problems corresponding to $L$ APs. Therefore, to compute the diagonal elements of $\mathbf{Q}_l$ at AP $L$, AP $L$ solve the optimization problem as follows:
 \begin{equation}
     \begin{aligned}
\arg \max_{\tilde{\mathbf{Q}}_{el}^{-1}\succeq\mathbf{0} } \quad &\log_2 \det(p\tilde{\mathbf{H}}_l\tilde{\mathbf{H}}_l^{\text{H}}{\tilde{\mathbf{Z}}}_{el}^{-1}+\mathbf{I}_{x})\\ \quad  =&\log_2 \det({\tilde{\mathbf{Q}}}_{el}^{-1}(p\tilde{\mathbf{H}}_l\tilde{\mathbf{H}}_l^{\text{H}}+\sigma^2\mathbf{I}_{x})+\mathbf{I}_{x})-\\&\log_2 \det(\sigma^2\tilde{\mathbf{Q}}_{el}^{-1}+\mathbf{I}_{x}) \quad \\
    \textcolor{black}{=}&\textcolor{black}{\sum_{i=1}^{x}\log_2 ({\tilde{\lambda}_{elqi}}(p{\lambda}_{il}^2+\sigma^2)+1)-\sum_{i=1}^{x}\log_2 ({\tilde{\lambda}_{elqi}}{\sigma^2}+1)}\\
\textrm{s.t.} \quad & C_{sc}= \sum_{i=1}^{x}\log_2 ({\tilde{\lambda}_{elqi}}(p{\lambda}_{il}^2+\sigma^2)+1).
\end{aligned} 
\label{sumRate_max_prob_PCA_EC}
    \end{equation}
    Based on the discussion at the beginning of this section, we know that $\tilde{\mathbf{Q}}_{el}, \forall l$ is a diagonal matrix. The optimization problem follows the similar steps as \nameref{apdb}.
    Matrix ${(\tilde{\mathbf{Q}}_{el}^{-1})}^{o}$ with the $i^{th}$ diagonal element defined in (\ref{eig_opt_Q_PCA_EC}) maximizes the objective function in (\ref{sumRate_max_prob_PCA_EC}).
    \begin{equation}
\begin{aligned}
      \tilde{\lambda}^o_{elqi}=\max(0,\frac{1}{\tilde{\mu}_{el}^o}(\frac{1}{\sigma^2}-\frac{1}{p\lambda_{li}^2+\sigma^2})-\frac{1}{\sigma^2}), \forall i.
        \label{eig_opt_Q_PCA_EC}
        \end{aligned}
    \end{equation}
    Finally, matrix ${(\tilde{\mathbf{Q}}_{el}^{-1})}^o=\text{diag}(\tilde{\lambda}^o_{elq1},\hdots,\tilde{\lambda}^o_{elqx})$. The parameter $\tilde{\mu}_{el}^o$ in (\ref{eig_opt_Q_PCA_EC}) is also calculated similarly to ${\mu}_{vl}^o$ in Section \ref{sec3_A}.

\section{Memory capacity allocation models and their impact on the rate on the fronthaul links}\label{sec4}
The simulation section considers two specific fronthaul topologies, a
daisy chain topology, and a multi-branch tree topology, as shown in Fig. \ref{fig_FT}. The following discussion is based on the daisy chain topology and can be easily extended to the tree topology. 
\subsection{Memory capacity allocation models}
Regarding the memory capacity at the APs, we consider two reference scenarios. 
In the following, we assume a fixed number of antennas are distributed among the APs.
\begin{itemize}
    \item {\bf Fixed per AP (FAP) memory capacity}: Regardless of the number of antennas per AP, APs have the same processing chips and consequently the same processing power and fixed memory capacity $C_{AP}$. Therefore, the total memory capacity depends on the number of APs, i.e., $C_{T}=LC_{AP}$. %This memory capacity scenario is called fixed per AP (FAP).
    \item {\bf Fixed Total (FT) memory capacity: } There is a fixed total memory capacity $C_{T}$ that is divided between APs. %Therefore, the memory capacity per AP depends on the number of APs among which the antennas are distributed. 
    This total memory capacity can be allocated to APs in two ways:
    \begin{itemize}
        \item {\bf Equal Allocation (EA): } In this memory model, the total memory capacity is split equally among APs, so $C_{AP}=\frac{C_T}{L}$. 
        \item {\bf Linear Allocation (LA): }The memory capacity is distributed according to the 
        number of signal vectors stored in the APs. Therefore, the first APs in the sequence receive less memory capacity, and as we move along the sequence, the APs receive a larger share of the total memory capacity. Based on the discussion in Section \ref{sec2} 
        , the total number of the received signal vectors, i.e., $N_v$, in the whole network, is as follows:
        \begin{equation}
            N_{rsv}=N_{sc}+2N_{sc}+\hdots+(L-1)N_{sc}=\frac{L(L-1)N_{sc}}{2}.%,
            \label{tot_vect_to_buff}
        \end{equation}
    \end{itemize}
\end{itemize}
%\robbert{Vida, it's cache memory not cash :)}
%We assume that received signal vectors are stored in on-chip cache memory\cite{jesusthesis}. Cache memory is desirable for its fast accessibility and low energy consumption. 

In the case of FAP and FT-EA, the number of bits allocated to each received signal vector in AP $l$ is as follows:
        \begin{equation}
        C_{sc}=\frac{C_{AP}}{(l-1)N_{sc}}.
        \label{bit_to_each_sigvec}
\end{equation}
Furthermore, the number of bits allocated to each received signal vector in FT-LA is as follows:
        \begin{equation}
            C_{sc}=\frac{C_T}{N_{rsv}}=\frac{2C_T}{L(L-1)N_{sc}}.
            \label{bit_to_each_sigvec_FTLA}
        \end{equation}
In reality, local CSI should also be stored in the local memory, which makes efficient memory usage more critical. However, as the amount of data related to CSI is similar in each AP, we ignore the low precision storage of the local CSI. 
\subsection{The rate of the fronthaul links }
Compression in a CFmMIMO network can also happen in the fronthaul link connecting two APs, which occurs when the fronthaul capacity is the limiting factor. In what follows, we compute the data rate of the links connecting two subsequent APs using the memory model introduced earlier in this section. Each AP needs to send $K$ scalars to the next AP in RLS, as explained in Algorithm \ref{algSRLS}. To compute the estimated signal for a reference user, we need to multiply a combining vector with $\rho$ bits per element with a received signal vector with $\gamma_i$ bits for element $i$. %Hence, the fronthaul rate scales with the number of users. In an $N$-antenna APs, and  for a specific user, suppose that the number of bits allocated to each element of the combining vector is $\rho$. 
%Consider signal estimation of the user on a specific subcarrier. The bits allocated to store $i^{th}$ element of the corresponding received signal vector is $\gamma_i$ and the total bits to store the received signal vector $C_{sc}$, then we have:
 %   \begin{equation}
  %          C_{sc}=\sum_{i=1}^N\gamma_i,
  %      \end{equation}
        To measure the rate of the fronthaul links, it is worth knowing that: 
        \begin{itemize}
            \item The number of bits to represent the multiplication product of an $\rho-$bit multiplicand and an
            $\gamma_i-$bit multiplier (and ignoring the sign bit) is at most $\rho+\gamma_i$ bits \cite{comorg}.
            \item Furthermore, the number of bits to represent the summation of an $(\rho+\gamma_i)-$bit summand and an $(\rho+\gamma_j)-$bit summand can not exceed $\max(\rho+\gamma_i, \rho+\gamma_j)+1$ \cite{comorg}\footnote{We consider only the fixed point arithmetic in this section, due to its simplicity and practicality.}.
        \end{itemize}  
     Hence, \textcolor{black}{the number of bits to represent} each user's signal estimate, which is the result of $N$ multiplications and then $N$ summations (inner product of the combining vector with the compressed received signal vector), is at most as follows\footnote{\textcolor{black}{The upper bound in (\ref{eq47}) is for the case of adding the numbers sequentially from the smallest number to the largest.}}:
     \textcolor{black}{
     \begin{equation}
    \begin{aligned}
        \alpha&=\max_{i}(\rho+\gamma_i)+1,
        \end{aligned}
        \label{eq47}
            \end{equation}}            
    %  \textcolor{black}{
    % \begin{equation*}
    % \begin{aligned}
    %     \alpha\myineqla \lfloor\max_{i}(\rho+\gamma_i)+\log_2 N\rfloor+1&\myineqlb \sum_{i} (\rho+\gamma_i)\\&=N\rho+C_{sc},
    %     \end{aligned}
    %     \label{eq47}
    %         \end{equation*}}
 Scaling with the number of subcarriers and users, the rate of the fronthaul link connecting AP $l$ to AP ${l+1}$ is as follows:
 \begin{equation}
    \begin{aligned}
        C_{fl}=\frac{KN_{sc}\alpha}{T_s}&={\frac{KN_{sc}(\max_{i}(\rho+\gamma_i)+1)}{T_s}}\\&\myineqla \frac{KN_{sc}(\rho+C_{sc})}{T_s}, 
        \label{fronthaul_rate}
        \end{aligned}
    \end{equation}
    % \begin{equation}
    % \begin{aligned}
    %     C_{fl}=\frac{KN_{sc}\alpha}{T_s}&\myineqla\textcolor{black}{\frac{KN_{sc}(\lfloor\max_{i}(\rho+\gamma_i)+\log_2 N\rfloor+1)}{T_s}}\\&\myineqlb \frac{KN_{sc}(N\rho+C_{sc})}{T_s},
    %     \label{fronthaul_rate}
    %     \end{aligned}
    % \end{equation}
    where $T_s$ is OFDM symbol duration in second.
    %and $\myineqla$ is achieved if AP $l$ allocate all the $C_{sc}$ bits to the signal of one antenna. 
    Note that $C_{sc}$ in FAP and FT-EA depends on the AP index, as seen in (\ref{bit_to_each_sigvec}). Therefore, in these scenarios, the upper bound on the rate of the fronthaul link in (\ref{fronthaul_rate}) depends on the AP index, which is higher for the APs at the beginning of the sequence. However, \textcolor{black}{the upper bound} on the rate of the fronthaul link is the same between any two APs and independent of the AP index when FT-LA is used.

    It is worth mentioning that exchanging the CSI-related matrix, e.g., $\mathbf{\Gamma}_l$ in Algorithm \ref{algSRLS} from AP $l$ to the next AP, also contributes to the rate of the fronthaul link. However, this exchange only happens once in one coherence block and is assumed to be scenario and AP index independent. Therefore, it is neglected in the (\ref{fronthaul_rate}).
    % In this paper, two fronthaul topologies is used, as demonstrated in Fig.\ref{fig_FT}.
    Note that in case a multi-branch tree topology is used for the fronthaul, then the $C_{sc}$ depends on the level that a particular AP such as AP $l$ resides, e.g., in (\ref{bit_to_each_sigvec}) and also on the depth of the tree (total number of levels), e.g., in (\ref{bit_to_each_sigvec_FTLA}).

    In conclusion, the rate of each fronthaul link scales with $C_{sc}$. Therefore, optimizing memory size to minimize the memory cost in the APs positively impacts the rate of the fronthaul links.
\section{simulation results}\label{sec5}
This section presents simulation results, which give insight into how the limited memory capacity in each AP affects the average per-user SE.
The average per-user SE i.e., $R^u_{sub}$ is calculated as follows:
\begin{equation}
    R^u_{sub}=\mathbb{E}_{\mathbf{H}}\{R_{sub}\}/K,
    \label{eqSumSE}
\end{equation}
where $sub\in\{v,e,ePCA\}$ indicates one of the three local compression options and $R_{sub}\in\{R_v, R_e, R_{ePCA}\}$ is the sum-SE of the users corresponding to one of the three options which are defined in section\ref{section3}. The expectation in \ref{eqSumSE} is on the different realization of channel matrix $\mathbf{H}$.
The simulation area is square with a perimeter of $D=500 m$ \cite{Shaik2021}. \textcolor{black}{The total number of antennas is ${M=128}$ (unless otherwise stated) which are distributed in ${L=\{2,4,8,16,32,64,128\}}$ APs}. The APs are distributed around the area in a daisy chain or a
multi-branch tree with at most two branches per node.  
The path loss model of an urban microcell with 2GHz carrier frequency is considered \cite{Shaik2021},\cite{3gpp_PL}.
\begin{equation}
    \beta_{kl}=-30.5-36.7\log_{10}(\frac{d_{kl}}{1m}),
    \label{large_sacle_fading_coef}
\end{equation}
where $d_{kl}$ is the distance (in meters) between user $k$ and AP $l$.
The communication bandwidth is $B=100\text{MHz}$ and an FFT size of $N_{sc}=4096$ \cite{3gpp_NR}. Noise variance is $\sigma^2=-85\text{dBm}$. In addition, the transmitted power of the users is $p=10\text{mWatt}$.
Table \ref{tab2} summarizes a list of abbreviations introduced before and used in this section.

\begin{table}[!t]
\caption{Abbreviations}
\centering
\begin{tabular}{||c c||} 
 \hline
 Parameter Description & Abbreviation \\ [0.5ex] 
 \hline\hline
 Vector-wise Compression & VC  \\ 
 \hline
 Element-wise compression & EC  \\
 \hline
 Fixed memory capacity per AP & FAP \\
 \hline
 Fixed total memory capacity &FT   \\
 \hline
 Equal allocation& EA\\[1ex]
 \hline
 Linear allocation& LA \\  
 \hline
\end{tabular}
 \label{tab2}
\end{table}
% From this total memory capacity and considering our setup with $NL=128$ and the maximum number of users $K=50$, $5MB$ is reserved for CSI-related information. As mentioned previously, for larger setups, the reserved memory for CSI should be increased.
% \begin{center}
% \begin{tabular}{||c c c c||} 
%  \hline
% Total cache memory &CSI& Received vectors& Per subcarrier  \\ [0.2ex] 
%  \hline\hline
% 16MB & 5MB & 11MB& 3.67KB \\ 
%  \hline
%  32MB & 5MB & 27MB&9KB \\
%  \hline
%  64MB & 5MB & 59MB&19.67KB \\
%   \hline
%   128MB & 5MB & 123MB&41KB\\
%   \hline
% \end{tabular}
% \end{center}
% It is worth mentioning that total memory capacity $B$ is divided between APs so each APs share of memory is relative to its number of antennas, i.e.  $C_{f}=\frac{BN}{M}$ with $N$ being the number of antennas per AP.
%In Fig. \ref{fig2}, the average SE per-user  while using option 1 for received data compression is plotted. The number of users is $K=5$. 
% \begin{figure}[!h]
%     \centering
%     \input{figures/fig2.tex}
%     \caption{Impact of l%L
%     imited memory capacity %affects 
%     on average SE per-user. %The compression option is vector-wise compression of raw received vector 
%     VC which is 
%     elaborated in section\ref{sec3_A} is used for different memory capacity allocation methods. (Left) $K=4$. (Middle) $K=16$. (Right) $K=64$.}
%     \label{figlimited_vs_inf}
% \end{figure}
\begin{figure}[!h]
    \centering
    \pgfplotsset{width=8.4cm,compat=1.18}
\pgfplotsset{every x tick label/.append style={font=\small, yshift=0.5ex},every y tick label/.append style={font=\small, xshift=0.5ex},
every axis legend/.append style={
at={(5,1)},
anchor=north west,font=\large
}}

%\begin{tikzpicture}
\begin{tikzpicture}[scale=0.7]
\begin{scope}
\begin{axis}[
domain=0:4,
grid=major,
ylabel= Average per-user SE (bit/sec/Hz),
%xlabel= Number of the APs,
xmin=1,
xmax=7,
xtick style={color=clr3},
 xtick={1,2,3,4,5,6,7},
 xticklabels={L=2,L=4,L=8,L=16,L=32,L=64,L=128},
 xticklabel style={yshift=-2.5pt},
ymin=2.5,
ymax=8,
ytick={3,4,5,6,7,8},
mark size=4.0 pt,
legend columns=2,
legend style={nodes={scale=0.9, transform shape},at={(0.5,-1.4)},anchor=north},
%legend entries={
%$Option 1$,$Option 2$,$Option 3$,$Option4$,$ OPtion 5$,
%{[text width=25pt,text depth=]Neg. Sign:},
%},
% same effect:
% legend style={
% nodes={text width=25pt,text depth=},}
]

        \addplot [red, mark=asterisk]
        table[x expr=\coordindex+1,y=B,col sep=comma] {Data_for_figures/data_K4_cfinf.csv};
        \addlegendentry{Inf memory size}

        \addplot [orange, mark=o]
        table[x expr=\coordindex+1,y=B,col sep=comma] {Data_for_figures/data_K4_cf256KB.csv};
        \addlegendentry{FAP with $C_{AP}=256KB$}
        
         \addplot [orange, mark=triangle]
        table[x expr=\coordindex+1,y=B,col sep=comma] {Data_for_figures/data_K4_cf64KB.csv};
        \addlegendentry{FAP with $C_{AP}=64KB$}

        \addplot [blue, mark=o]
        table[x expr=\coordindex+1,y=B,col sep=comma] {Data_for_figures/data_K4_cf32MB.csv};
        \addlegendentry{FT-EA with $C_{T}=32MB$}

        \addplot [blue, mark=triangle]
        table[x expr=\coordindex+1,y=B,col sep=comma] {Data_for_figures/data_K4_cf8MB.csv};
        \addlegendentry{FT-EA with $C_{T}=8MB$}
        
        % \addplot [brown, mark=diamond]
        % table[x=D,y=B,col sep=comma] {Data/Data_forthrun/data_K5_cf128.csv};
        % \addlegendentry{$B=128MB$}

\end{axis}
\end{scope}
\begin{scope}[xshift=0cm,yshift=-6.8cm]
\begin{axis}[
domain=0:4,
grid=major,
% ylabel= users sum-rate,
 xlabel= Number of the APs,
 ylabel= Average per-user SE (bit/sec/Hz),
xmin=1,
xmax=7,
xtick style={color=clr3},
xtick={1,2,3,4,5,6,7},
 xticklabels={L=2,L=4,L=8,L=16,L=32,L=64,L=128},
 xticklabel style={yshift=-2.5pt},
ymin=2.5,
ymax=8,
ytick={3,4,5,6,7,8},
mark size=4.0 pt
]

         \addplot [red, mark=asterisk]
        table[x expr=\coordindex+1,y=B,col sep=comma] {Data_for_figures/data_K64_cfinf.csv};
       % \addlegendentry{Infinte Buffer size}

        \addplot [orange, mark=o]
        table[x expr=\coordindex+1,y=B,col sep=comma] {Data_for_figures/data_K64_cf256KB.csv};
        %\addlegendentry{$B=256KB$}
        
         \addplot [orange, mark=triangle]
        table[x expr=\coordindex+1,y=B,col sep=comma] {Data_for_figures/data_K64_cf64KB.csv};
        \addplot [blue, mark=o]
        table[x expr=\coordindex+1,y=B,col sep=comma] {Data_for_figures/data_K64_cf32MB.csv};
        %\addlegendentry{$B=128KB$}
        \addplot [blue, mark=triangle]
        table[x expr=\coordindex+1,y=B,col sep=comma] {Data_for_figures/data_K64_cf8MB.csv};
        %\addlegendentry{$B=128KB$}

\end{axis}
\end{scope}
\end{tikzpicture}
    \caption{Impact of limited memory capacity %affects 
    on average per-user SE in a daisy chain fronthaul with two different numbers of users: (Top) $K=4$, (Bottom) $K=64$. %The compression option is vector-wise compression of raw received vector 
    VC is used for compression.}
    \label{figlimited_vs_inf}
\end{figure}
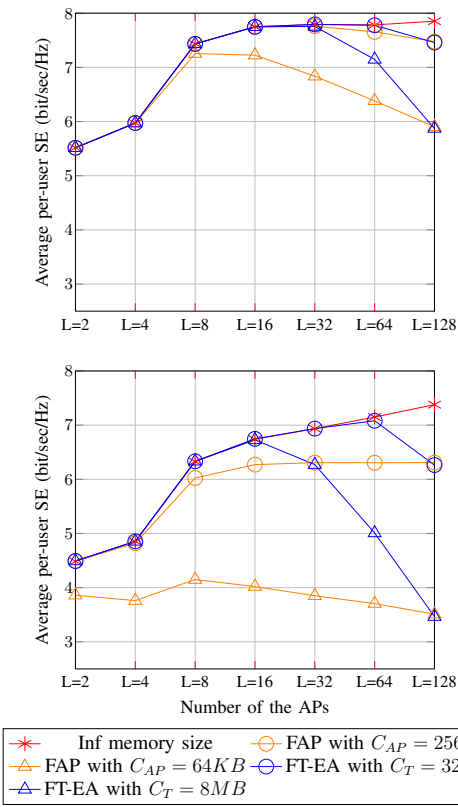
% \begin{figure}[!h]
%     \centering
%     \input{figures/fig3_2.tex}
%     \caption{Limited memory capacity affects on average SE per-user. (Left) $K=8$. (Middle) $K=16$. (Right) $K=64$. }
%     \label{figlimited_vs_inf}
% \end{figure}
% \begin{figure}[h]
% \centering
% \begin{subfigure}{.5\textwidth}
% \input{figures/fig1.tex}
% \caption{$K=5$}
% \end{subfigure}%
% \begin{subfigure}{.5\textwidth}
% \input{figures/fig1_2.tex}
% \caption{$K=15$}
% \end{subfigure}
% \caption{Limited memory capacity effect on optimal number of APs maximizing sum SE of users. Total number of antennas are 150 and they are distributed in $L=\{2,3,5,6,10,15,25,30,50,75,150\}$ APs.}
% \label{fig2}

% \end{figure}
% \begin{figure}[t!]
%   %\centering
%         %\makebox[\textwidth][c]{\includegraphics[width=1.2\textwidth]{SE_CDF_sim_L100_N4_K25_100_TC100_td90_v4.png}}
%         \input{figures/fig1.tex}
% 		\caption{Average squared error vs APs, $L=25$, $N=4$, $K=5$, $W=16$.}
% 		\label{fig1}
% \end{figure}

% Two ways of dividing the Total memory between APs have been considered:
% \begin{itemize}
%     \item The allocated memory to the APs are same.
%     \item The allocated memory capacity to any vector that needs to be memoryed is the same. Therefore, as the APs with larger indexes have more data to memory, the capacity memory capacity is increasing linearly with the AP index in the sequence. 
% %    \item The APs are connected based on a multi-branch tree and the APs are allocated same memory capacity.
% \end{itemize}
\subsection{\textcolor{black}{Optimal number of APs in the presence of limited memory}}
In Fig. \ref{figlimited_vs_inf}, the average per-user SE\footnote{Note that for simulation results, as 1) we don't consider any specific values for $\tau_u$ and $\tau_c$ and 2) the factor $\frac{\tau_u}{\tau_c}$ appears sum-SE of all the compression options, we neglect it in plotting the figures. Also, the average per-user SE is the average users' sum-SE divided by the number of users.} versus the number of APs is plotted. Two scenarios of memory capacity allocation are considered. Furthermore, the sub-figures are associated with different numbers of users. It is worth mentioning that FAP and FT-EA with the selected $C_{AP}$ and $C_T$ represent the same scenario
in the case of the full distribution of antennas. We also considered infinite memory size (no compression) as a benchmark.
%\sofie{We also compare with a scenario without compression. } 
It is observed that:
\begin{itemize}
\item Having memory constraints always limits the performance of distributed antenna systems with sequential processing. While the average per-user SE improves when distributing the fixed number of antennas over multiple APs with infinite memory, this is no longer true when there are memory constraints.
\item When limiting the memory capacity so that all APs have the same memory constraint, the performance is dominated by the memory requirements of the last AP. In all sub-figures in Fig. \ref{figlimited_vs_inf}, the FAP model with $C_{AP}=64KB$ performs the same as the FT-LA with $C_T=8MB$ when $L=128$. However, when the number of AP is halved, i.e., $L=64$, the increase in the average per-user SE is about $20\%$ in FT-EA and $6\%$ in FAP, e.g., in the case of $K=4$. This is because, in FT-EA, when the number of APs is halved, the allocated memory to each AP gets doubled as the same total memory is now distributed among 64 instead of 128 APs. While in FAP, the memory per AP remains the same in the case of $L=64$ or $L=128$.
\item When comparing the different results in Fig. \ref{figlimited_vs_inf}, it is clear that the impact of compression is much more severe when the number of users $K$ is higher, as with increasing users, the entropy of the local received signal vector in each AP increases, which demands more bits to keep a certain distortion level. This is especially visible for the scenario of FAP, and the performance degradation is already visible for a low number of APs in the sequence. Even without severe memory requirements for sequential processing, limited AP memory degrades multiuser performance. Thus, when the number of users increases and the memory capacity is limited,  
it pays off to have collocated antennas.
\item Figs. \ref{fig5_cfvar_FAP} testify to the aforementioned claim. It is observed that for a fixed number of APs $L$, the performance degradation of limited memory capacity cases compared to the infinite memory case gets more severe when the number of users increases. 
\begin{figure}[!h]
    \centering
    \pgfplotsset{width=8.4cm,compat=1.18}
\pgfplotsset{every x tick label/.append style={font=\small, yshift=0.5ex},every y tick label/.append style={font=\small, xshift=0.5ex},
every axis legend/.append style={
at={(5,1)},
%anchor=north west,
font=\large
}}

%\begin{tikzpicture}
\begin{tikzpicture}[scale=0.7]
\begin{scope}[xshift=0cm,yshift=0]
\begin{axis}[
domain=0:4,
grid=major,
% ylabel= users sum-rate,
ylabel= Average per-user SE (bit/sec/Hz),
 xlabel= FAP memory size ($C_{AP}$)(KB),
xmin=1,
xmax=4,
xtick style={color=clr3},
xtick={1,2,3,4},
 xticklabels={64,128,192,256},
 xticklabel style={yshift=-2.5pt},
ymin=3,
ymax=8,
ytick={4,5,6,7,8},
mark size=4.0 pt,
legend columns=2,
legend style={nodes={scale=0.9, transform shape},at={(0.5,-1.4)},anchor=north},
]       
       \addplot [blue, mark=o]
        table[x expr=\coordindex+1,y=A,col sep=comma] {Data_for_figures/data_K4_L32_FAP_cfvar_uncorrCh_pCSI.csv};
        \addlegendentry{Finite memory size, $K=4$}
        %\addlegendentry{Infinite memory size}
        \addplot [blue, mark=triangle]
        table[x expr=\coordindex+1,y=D,col sep=comma] {Data_for_figures/data_K4_L32_FAP_cfvar_uncorrCh_pCSI.csv};
        \addlegendentry{Inf memory size, $K=4$}
      
        \addplot [orange, mark=o]
        table[x expr=\coordindex+1,y=A,col sep=comma] {Data_for_figures/data_K64_L32_FAP_cfvar_uncorrCh_pCSI.csv};
        \addlegendentry{Finite memory size, $K=64$}
        %\addlegendentry{FAP with $C_{AP}=256KB$}
        
         \addplot [orange, mark=triangle]
        table[x expr=\coordindex+1,y=D,col sep=comma] {Data_for_figures/data_K64_L32_FAP_cfvar_uncorrCh_pCSI.csv};
        \addlegendentry{Inf memory size, $K=64$}

\end{axis}
\end{scope}
\begin{scope}[xshift=0cm,yshift=-6.8cm]
\begin{axis}[
domain=0:4,
grid=major,
ylabel= Average per-user SE (bit/sec/Hz),
xlabel=FT memory size ($C_T$) (MB),
xmin=1,
xmax=4,
xtick style={color=clr3},
 xtick={1,2,3,4},
 xticklabels={{8,16,24,32}},
 xticklabel style={yshift=-3pt},
ymin=3,
ymax=8,
ytick={4,5,6,7,8},
yticklabel style={xshift=-3pt},
mark size=4.0 pt,
legend columns=2,
legend style={nodes={scale=0.9, transform shape},at={(0.5,-1.5)},anchor=north},
%legend entries={
%$Option 1$,$Option 2$,$Option 3$,$Option4$,$ OPtion 5$,
%{[text width=25pt,text depth=]Neg. Sign:},
%},
% same effect:
% legend style={
% nodes={text width=25pt,text depth=},}
]

       \addplot [blue, mark=o]
        table[x expr=\coordindex+1,y=A,col sep=comma] {Data_for_figures/data_K4_L32_FT_cfvar_uncorrCh_pCSI.csv};
       % \addlegendentry{Finite memory size, $K=4$}
        \addplot [blue, mark=triangle]
        table[x expr=\coordindex+1,y=D,col sep=comma] {Data_for_figures/data_K4_L32_FT_cfvar_uncorrCh_pCSI.csv};
      %\addlegendentry{Inf memory size, $K=4$}
        \addplot [orange, mark=o]
        table[x expr=\coordindex+1,y=A,col sep=comma] {Data_for_figures/data_K64_L32_FT_cfvar_uncorrCh_pCSI.csv};
      % \addlegendentry{Finite memory size, $K=64$}
        
         \addplot [orange, mark=triangle]
        table[x expr=\coordindex+1,y=D,col sep=comma] {Data_for_figures/data_K64_L32_FT_cfvar_uncorrCh_pCSI.csv};
        %\addlegendentry{Inf memory size, $K=64$}

\end{axis}
\end{scope}
% \begin{scope}[xshift=6.2cm,yshift=0]
% \begin{axis}[
% %title={$L=64$},
% domain=0:4,
% grid=major,
% %ylabel= Average per user SE (bit/sec/Hz),
% xlabel=FAP memory size ($C_{AP}$)(KB),
% xmin=1,
% xmax=4,
% xtick style={color=clr3},
%  xtick={1,2,3,4},
%  xticklabels={64,128,192,256},
%  xticklabel style={yshift=-2.5pt},
% ymin=3,
% ymax=8,
% ytick={4,5,6,7,8},
% mark size=4.0 pt,
% %yticklabels={5,6,7,8},
% %legend columns=2,
% %legend style={nodes={scale=0.9, transform shape},at={(0,-1)},anchor=north},
% %legend entries={
% %$Option 1$,$Option 2$,$Option 3$,$Option4$,$ OPtion 5$,
% %{[text width=25pt,text depth=]Neg. Sign:},
% %},
% % same effect:
% % legend style={
% % nodes={text width=25pt,text depth=},}
% ]

%        \addplot [blue, mark=o]
%         table[x expr=\coordindex+1,y=A,col sep=comma] {Data_for_figures/data_K4_L64_FAP_cfvar_uncorrCh_pCSI.csv};
%        % \addlegendentry{Finite memory size, $K=4$}
%         \addplot [orange, mark=o]
%         table[x expr=\coordindex+1,y=D,col sep=comma] {Data_for_figures/data_K4_L64_FAP_cfvar_uncorrCh_pCSI.csv};
%       %\addlegendentry{infinite memory size, $K=4$}
%         \addplot [blue, mark=triangle]
%         table[x expr=\coordindex+1,y=A,col sep=comma] {Data_for_figures/data_K64_L64_FAP_cfvar_uncorrCh_pCSI.csv};
%       % \addlegendentry{Finite memory size, $K=64$}
        
%          \addplot [orange, mark=triangle]
%         table[x expr=\coordindex+1,y=D,col sep=comma] {Data_for_figures/data_K64_L64_FAP_cfvar_uncorrCh_pCSI.csv};
%         %\addlegendentry{Infinite memory size, $K=64$}

% \end{axis}
% \end{scope}
\end{tikzpicture}
    \caption{Average per-user SE with (Top) varying the value of $C_{AP}$ in FAP memory model and (Bottom) varying the value of $C_{T}$ in FT memory model. The number of AP is fixed to $L=32$. The compression option for all plots is VC.}
    \label{fig5_cfvar_FAP}
% \end{figure}
% \begin{figure}[h!]
%\rule{16cm}{0.002mm}\\
% \vspace{7pt}
%     \centering
%     \input{figures/figures_fr/cfvar_FT}
%     \caption{Average per-user SE variation with the value of $C_{T}$ in FT memory model. The number of AP is fixed to (Left) $L=32$ and (Right) $L=64$. The compression option for all plots is VC.}
%     \label{fig6_cfvar_FT}
\end{figure}
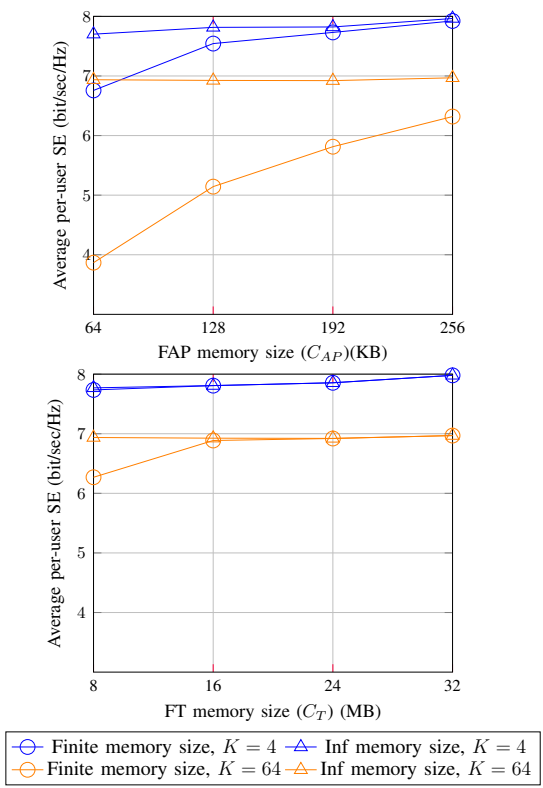
%(your last bullet was also quite good actually)
    % \item Conversely, %with limited memory capacity, 
    % the effect of limited memory capacity has a more %detrimental 
    % prominent %significant impact
    % when the number of users is large. With increasing users, the entropy of the local received vector in each AP increases, which demands more bits to keep a certain distortion level. Thus, when the number of users increases and the memory capacity is limited, %there is a tendency 
    % it pays off to have collocated antennas.
 \item Furthermore, in Fig. \ref{fig6_M256}, we plotted the average per-user SE for the case of $M=256$ and compared it with the case of $M=128$ when the FT-EA memory model with $C_T=32MB$ is used. By increasing the total number of antennas from $M=128$ to $M=256$, we observe that the performance degradation compared to infinite memory gets worse. For the case of $K=64$, the optimal number of APs decreases. This is because for a fixed number of APs, with an increasing total number of antennas $M$, we are increasing the number of antennas per AP ($N$), and hence, a higher dimensional received signal vector needs to be stored in each AP.  
 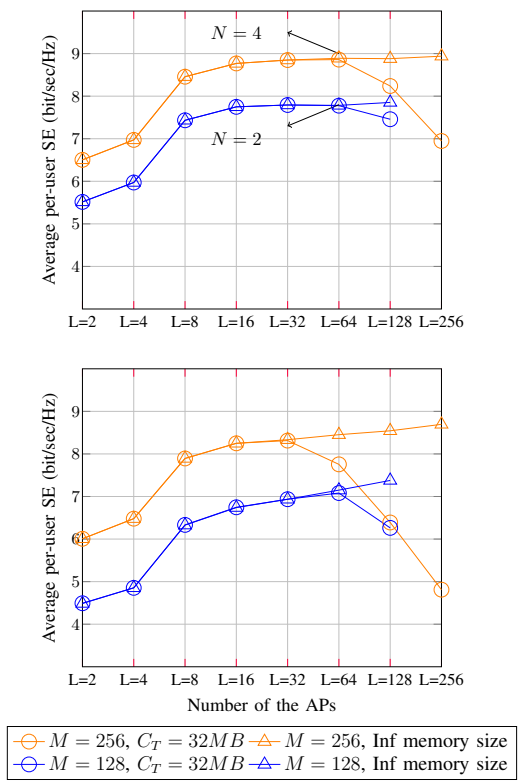
\begin{figure}[!ht]
\centering
\begin{center}
    \pgfplotsset{width=8.4cm,compat=1.18}
\pgfplotsset{every x tick label/.append style={font=\small, yshift=0.5ex},every y tick label/.append style={font=\small, xshift=0.5ex},
every axis legend/.append style={
at={(5,1)},
anchor=north west,font=\large
}}

%\begin{tikzpicture}
\begin{tikzpicture}[scale=0.7]
\begin{scope}
\begin{axis}[
domain=0:4,
grid=major,
ylabel= Average per-user SE (bit/sec/Hz),
% ylabel= users sum-rate,
% xlabel= Number of the APs,
xmin=1,
xmax=8,
xtick style={color=clr3},
xtick={1,2,3,4,5,6,7,8},
 xticklabels={L=2,L=4,L=8,L=16,L=32,L=64,L=128,L=256},
 xticklabel style={yshift=-2.5pt},
ymin=3,
ymax=10,
ytick={4,5,6,7,8,9},
mark size=4.0 pt,
legend columns=2,
legend style={nodes={scale=0.9, transform shape},at={(0.5,-1.4)},anchor=north},
]       
       \addplot [orange, mark=o]
        table[x expr=\coordindex+1,y=A,col sep=comma] {Data_for_figures/data_K4_cf32MB_M256.csv};\label{K4_M256_vc}
       % \addlegendentry{$M=256$, $C_T=32MB$}
       \addlegendentry{$M=256$, $C_T=32MB$}

        \addplot [orange, mark=triangle]
        table[x expr=\coordindex+1,y=D,col sep=comma] {Data_for_figures/data_K4_cf32MB_M256.csv};\label{K4_M256_INF}
        %\addlegendentry{$M=256$, Infinite memory size}
        \addlegendentry{$M=256$, Inf memory size}

        \addplot [blue, mark=o]
        table[x expr=\coordindex+1,y=B,col sep=comma] {Data_for_figures/data_K4_cf32MB.csv};\label{K4_M128_INF}
        \addlegendentry{$M=128$, $C_T=32MB$}
      %  \addlegendentry{$M=256$, Infinite memory size}
      \addplot [blue, mark=triangle]
        table[x expr=\coordindex+1,y=B,col sep=comma] {Data_for_figures/data_K4_cfinf.csv};\label{K4_M128_vc}
       % \addlegendentry{$M=256$, $C_T=32MB$}
        \addlegendentry{$M=128$, Inf memory size}
     
      \draw [->] (6,9) -- (5,9.5) ;
      \node at (4,9.5) (nodetxt) {$N=4$};

       \draw [->] (6,7.8) -- (5,7.3) ;
      \node at (4,7) (nodetxt) {$N=2$};

\end{axis}
\end{scope}

\begin{scope}[xshift=0cm,yshift=-6.8cm]
\begin{axis}[
domain=0:4,
grid=major,
ylabel= Average per-user SE (bit/sec/Hz),
xlabel= Number of the APs,
xmin=1,
xmax=8,
xtick style={color=clr3},
 xtick={1,2,3,4,5,6,7,8},
 xticklabels={L=2,L=4,L=8,L=16,L=32,L=64,L=128,L=256},
 xticklabel style={yshift=-2.5pt},
ymin=3,
ymax=10,
ytick={4,5,6,7,8,9},
mark size=4.0 pt,
%yticklabels={5,6,7,8},
%legend entries={
%$Option 1$,$Option 2$,$Option 3$,$Option4$,$ OPtion 5$,
%{[text width=25pt,text depth=]Neg. Sign:},
%},
% same effect:
% legend style={
% nodes={text width=25pt,text depth=},}
]

        \addplot [orange, mark=o]
        table[x expr=\coordindex+1,y=A,col sep=comma] {Data_for_figures/data_K64_cf32MB_M256.csv};%\label{K64_M256_vc}
        %\addlegendentry{Infinite memory size}
        %\addlegendentry{$M=256$, $C_T=32MB$}
      
        \addplot [orange, mark=triangle]
        table[x expr=\coordindex+1,y=D,col sep=comma] {Data_for_figures/data_K64_cf32MB_M256.csv};%\label{K64_M256_INF}
        
        %\addlegendentry{$M=256$, Infinite memory size}
        %\addlegendentry{FAP with $C_{AP}=256KB$}
        
       % \addlegendentry{Infinte Buffer size}
       %\addlegendentry{$M=128$, $C_T=32MB$}

        \addplot [blue, mark=o]
        table[x expr=\coordindex+1,y=B,col sep=comma] {Data_for_figures/data_K64_cf32MB.csv};%\label{K64_M128_INF}
        %\addlegendentry{$B=128KB$}
        %\addlegendentry{$M=128$, Infinite memory size}
       
        \addplot [blue, mark=triangle]
        table[x expr=\coordindex+1,y=B,col sep=comma] {Data_for_figures/data_K64_cfinf.csv};%\label{K64_M128_vc}

\end{axis}
\end{scope}
%{nodes={scale=0.9, transform shape},at={(3,0)},anchor=north}
% \node[draw,fill=white,inner sep=0pt,above left=0.5em] at (25.5,0) 
% {\small
%     \begin{tabular}{ccl}
%     $M=128$ & $M=256$ \\
%     \ref{K4_M128_vc} & \ref{K4_M256_vc} & $C_T=32MB$\\
%     \ref{K4_M128_INF} & \ref{K4_M256_INF} & $ C_T=inf$\\
%     \end{tabular}};

\end{tikzpicture}
    \caption{Average per-user SE when having a different total number of antennas, i.e., $M=128$ and $M=256$ and a different number of users: (Top) $K=4$ and (Bottom) $K=64$. In all plots, the memory model is FT-EA with $CT=32MB$. The compression option for all plots is VC.}
    \label{fig6_M256}
% \end{figure}
% \begin{figure}[h!]
\end{center}
\end{figure}
\end{itemize}

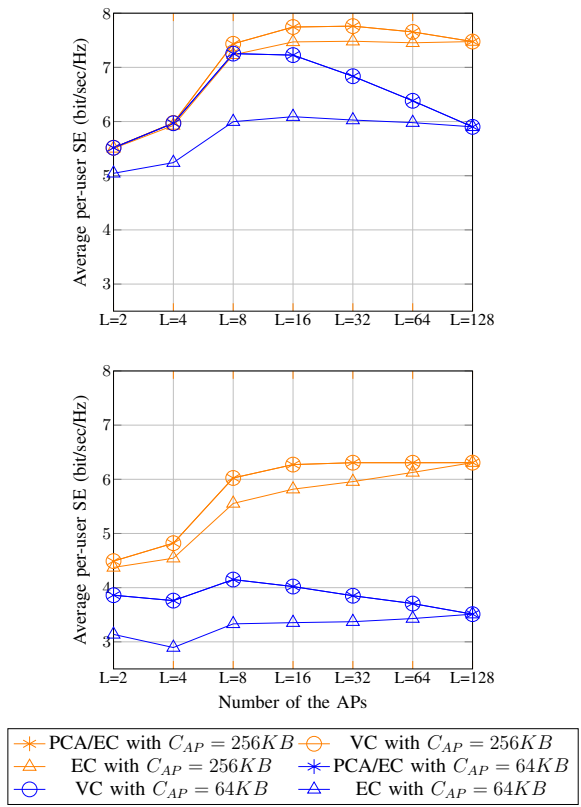
\begin{figure}[h]
    \centering
    \pgfplotsset{width=8.4cm,compat=1.18}
\pgfplotsset{every x tick label/.append style={font=\small, yshift=0.5ex},every y tick label/.append style={font=\small, xshift=0.5ex},
every axis legend/.append style={
at={(1.02,1)},
anchor=north west,font=\large
}}

%\begin{tikzpicture}
\begin{tikzpicture}[scale=0.7]
\centering
\begin{scope}
\begin{axis}[
domain=0:4,
grid=major,
ylabel= Average per-user SE (bit/sec/Hz),
%xlabel= Number of the APs,
xmin=1,
xmax=7,
xtick style={color=orange},
 xtick={1,2,3,4,5,6,7},
 xticklabels={L=2,L=4,L=8,L=16,L=32,L=64,L=128},
ymin=2.5,
ymax=8,
ytick={3,4,5,6,7,8},
mark size=4.0pt,
legend columns=2,
legend style={nodes={scale=0.9, transform shape},at={(0.5,-1.4)},anchor=north},
]

       \addplot [orange, mark=asterisk]
        table[x expr=\coordindex+1,y=A,col sep=comma] {Data_for_figures/data_K4_cf256KB.csv};
        \addlegendentry{PCA/EC with $C_{AP}=256KB$}
       \addplot [orange, mark=o]
        table[x expr=\coordindex+1,y=B,col sep=comma] {Data_for_figures/data_K4_cf256KB.csv};
        \addlegendentry{VC with $C_{AP}=256KB$ };

        \addplot [orange, mark=triangle]
        table[x expr=\coordindex+1,y=C,col sep=comma] {Data_for_figures/data_K4_cf256KB.csv};
        \addlegendentry{EC with $C_{AP}=256KB$ };
        
        %  \addplot [orange, mark=o]
        % table[x expr=\coordindex+1,y=B,col sep=comma] {Data_for_figures/data_K4_cf128KB.csv};
        % \addlegendentry{$B=128KB$}

        % \addplot [orange, mark=o]
        % table[x expr=\coordindex+1,y=C,col sep=comma] {Data_for_figures/data_K4_cf128KB.csv};
        % \addlegendentry{$B=256KB$}
        \addplot [blue, mark=asterisk]
        table[x expr=\coordindex+1,y=A,col sep=comma] {Data_for_figures/data_K4_cf64KB.csv};
        \addlegendentry{PCA/EC with $C_{AP}=64KB$}
        \addplot [blue, mark=o]
        table[x expr=\coordindex+1,y=B,col sep=comma] {Data_for_figures/data_K4_cf64KB.csv};
        \addlegendentry{VC with $C_{AP}=64KB$ }
        \addplot [blue, mark=triangle]
        table[x expr=\coordindex+1,y=C,col sep=comma] {Data_for_figures/data_K4_cf64KB.csv};
        \addlegendentry{EC with $C_{AP}=64KB$ }
        
        % \addplot [brown, mark=o]
        % table[x=D,y=C,col sep=comma] {Data/Data_forthrun/data_K5_cf128.csv};
        % \addlegendentry{$B=128MB$}

\end{axis}
\end{scope}

\begin{scope}[xshift=0cm,yshift=-6.8cm]
\begin{axis}[
domain=0:4,
grid=major,
% ylabel= users sum-rate,
 xlabel= Number of the APs,
 ylabel= Average per-user SE (bit/sec/Hz),
xmin=1,
xmax=7,
xtick style={color=orange},
xtick={1,2,3,4,5,6,7},
 xticklabels={L=2,L=4,L=8,L=16,L=32,L=64,L=128},
ymin=2.5,
ymax=8,
ytick={3,4,5,6,7,8},
mark size=4.0pt,
]

       \addplot [orange, mark=asterisk]
        table[x expr=\coordindex+1,y=A,col sep=comma] {Data_for_figures/data_K64_cf256KB.csv};
        
          \addplot [orange, mark=o]
        table[x expr=\coordindex+1,y=B,col sep=comma] {Data_for_figures/data_K64_cf256KB.csv};
        %\addlegendentry{Infinite Buffer size}

        \addplot [orange, mark=triangle]
        table[x expr=\coordindex+1,y=C,col sep=comma] {Data_for_figures/data_K64_cf256KB.csv};
        %\addlegendentry{$B=256KB$}
        
        %  \addplot [orange, mark=o]
        % table[x expr=\coordindex+1,y=B,col sep=comma] {Data_for_figures/data_K64_cf128KB.csv};
        % \addlegendentry{$B=128KB$}

        % \addplot [orange, mark=o]
        % table[x expr=\coordindex+1,y=C,col sep=comma] {Data_for_figures/data_K64_cf128KB.csv};
        % \addlegendentry{$B=256KB$}
        \addplot [blue, mark=asterisk]
        table[x expr=\coordindex+1,y=A,col sep=comma] {Data_for_figures/data_K64_cf64KB.csv};
        
        \addplot [blue, mark=o]
        table[x expr=\coordindex+1,y=B,col sep=comma] {Data_for_figures/data_K64_cf64KB.csv};
       % \addlegendentry{$B=64KB$}
        \addplot [blue, mark=triangle]
        table[x expr=\coordindex+1,y=C,col sep=comma] {Data_for_figures/data_K64_cf64KB.csv};
        %\addlegendentry{$B=64KB$}

\end{axis}
\end{scope}
\end{tikzpicture}
    \caption{Average per-user SE in a daisy chain fronthaul with three compression options and two different numbers of users: (Top) $K=4$, (Bottom) $K=64$. FAP memory model is used.   
    %\sofie{the different number of K are not different compression options? try to write more directly what you see. I would also describe here that all have a fixed per AP compression and you compare VC with EC with and without PCA.}. 
    }
    \label{fig7_3comp_option}
\end{figure}

\subsection{\textcolor{black}{PCA-pre-processing before element-wise compression}}
In Fig. \ref{fig7_3comp_option}, the average per-user SE 
under three compression options %of compression,
 and a FAP memory capacity scenario with an increasing
 number of users over sub-figures is studied.
 It is observed that:
 \begin{itemize}
 \item When the number of users is $K=4$ and $C_{AP}=64KB$, the performance improvement of VC compared to EC is around $1.2$ bit/sec/Hz - at the optimal number of AP %, i.e.,
 $L=8$. 
 % However, the gap between EC and VC decreases as the number of users 
 % increases, as more distinct user signals result in a decrease in the  
 % correlation coefficient between the elements of $\{\mathbf{y}_l|\mathbf{H}_l\}$. 
 % Together with the fact that $\{\mathbf{y}_l|\mathbf{H}_l\}$ has a multivariate circularly symmetric Gaussian distribution, this means that its elements become asymptotically independent, enabling a % near-optimal
 % good compression with EC.
 \item In the two sub-figures, Element-wise compression of the PCA pre-processed vector results in the same performance as vector-wise compression of the received signal vector. This is because pre-processing the circularly symmetric Gaussian received signal vector with PCA de-correlates 
    the elements of the vector, and for a circularly symmetric Gaussian vector, the de-correlation of the elements means their independence from each other. Our analysis shows that only in some particular scenarios, such as strict memory constraints and a low number of users, vector-wise compression improves over element-wise compression. Moreover, by pre-processing the data, there is no motivation for vector-wise compression.  %Hence the element of the pre-processed signal vector can be optimally compressed element-wisely. % corresponding to different numbers of users \sofie{check grammar of this last sentence}

\end{itemize} 
% \begin{figure}[h]
%     \centering
%     \input{figures/fig5.tex}
%     \caption{Average per-user SE Comparison of the FT-EA and FT-LA in a daisy chain fronthaul with an increasing number of users:  
%     (Left) $K=4$, (Middle) $K=16$, (Right) $K=64$.  
%     % \sofie{Vida, for me, if you have a ":", the cases after the : should be examples of what you see before the :. So, I would expect the caption to read as: 
%     % Average per-user SE as function of the number of APs in a scenario with daisy chain fronthaul topology and increasing number of users:   
%     % (Left) $K=4$, (Middle) $K=16$, (Right) $K=64$. The memory modet is FT-EA and FT-LA }
%    }
%     \label{fig_lin}
% \end{figure}
\subsection{\textcolor{black}{Impact of imperfect CSI}}
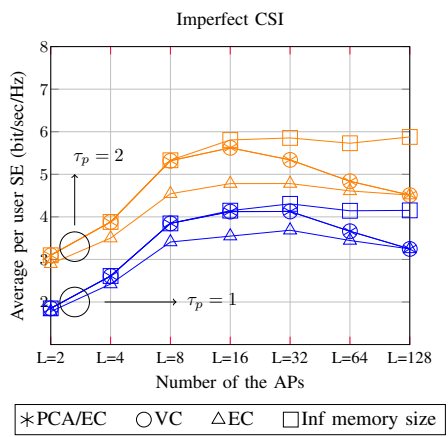
\begin{figure}[h!]
    \centering
    \pgfplotsset{width=8.4cm,compat=1.18}
\pgfplotsset{every x tick label/.append style={font=\small, yshift=0.5ex},every y tick label/.append style={font=\small, xshift=0.5ex},
every axis legend/.append style={
at={(5,1)},
anchor=north west,font=\large
}}

%\begin{tikzpicture}
\begin{tikzpicture}[scale=0.7]

\begin{scope}
\begin{axis}[
title=Imperfect CSI,
domain=0:4,
grid=major,
ylabel= Average per user SE (bit/sec/Hz),
xlabel= Number of the APs,
xmin=1,
xmax=7,
xtick style={color=clr3},
 xtick={1,2,3,4,5,6,7},
 xticklabels={L=2,L=4,L=8,L=16,L=32,L=64,L=128},
 xticklabel style={yshift=-2.5pt},
ymin=1,
ymax=8,
ytick={2,3,4,5,6,7,8},
mark size=4.0 pt,
legend columns=-1,
legend style={nodes={scale=0.9, transform shape},at={(0.5,-0.2)},anchor=north, /tikz/every even column/.append style={column sep=0.5cm}},
%yticklabels={5,6,7,8},
% legend columns=3,
% legend style={nodes={scale=0.9, transform shape},at={(-0.5,-1.8)},anchor=north},
%legend entries={
%$Option 1$,$Option 2$,$Option 3$,$Option4$,$ OPtion 5$,
%{[text width=25pt,text depth=]Neg. Sign:},
%},
% same effect:
% legend style={
% nodes={text width=25pt,text depth=},}
]
\addlegendimage{mark=asterisk,black,only marks}
 \addlegendimage{mark=o,black,only marks}
 \addlegendimage{mark=triangle,black,only marks}
 \addlegendimage{mark=square,black,only marks}
       \addplot [orange, mark=asterisk]
        table[x expr=\coordindex+1,y=A,col sep=comma] {Data_for_figures/data_K4_cf64KB_tauc_0.5K_uncorrCh_impCSI.csv};
        \addlegendentry{PCA/EC}

        \addplot [orange, mark=o]
        table[x expr=\coordindex+1,y=B,col sep=comma] {Data_for_figures/data_K4_cf64KB_tauc_0.5K_uncorrCh_impCSI.csv};
        \addlegendentry{VC}
        %\addlegendimage{mark=o,only marks}

        \addplot [orange, mark=triangle]
        table[x expr=\coordindex+1,y=C,col sep=comma] {Data_for_figures/data_K4_cf64KB_tauc_0.5K_uncorrCh_impCSI.csv};
        \addlegendentry{EC}
       % \addlegendentry{$C_{AP}=64KB$, Vector-wise compression}
        
          \addplot [orange, mark=square]
        table[x expr=\coordindex+1,y=D,col sep=comma] {Data_for_figures/data_K4_cf64KB_tauc_0.5K_uncorrCh_impCSI.csv};
        \addlegendentry{ Inf memory size}
        %\addlegendentry{Infinite memory size}
        \addplot [blue, mark=asterisk]
        table[x expr=\coordindex+1,y=A,col sep=comma] {Data_for_figures/data_K4_cf64KB_tauc_0.25K_uncorrCh_impCSI.csv};
       \addplot [blue, mark=o]
        table[x expr=\coordindex+1,y=B,col sep=comma] {Data_for_figures/data_K4_cf64KB_tauc_0.25K_uncorrCh_impCSI.csv};
        %\addlegendentry{Infinite memory size}

        \addplot [blue, mark=triangle]
        table[x expr=\coordindex+1,y=C,col sep=comma] {Data_for_figures/data_K4_cf64KB_tauc_0.25K_uncorrCh_impCSI.csv};
        %\addlegendentry{FAP with $C_{AP}=256KB$}
        
          \addplot [blue, mark=square]
        table[x expr=\coordindex+1,y=D,col sep=comma] {Data_for_figures/data_K4_cf64KB_tauc_0.25K_uncorrCh_impCSI.csv};
        %\addlegendentry{FAP with $C_{AP}=64KB$}
  \draw [->] (1.4,3.8) -- (1.4,5) ;
  \draw (1.4,3.3) circle (8pt);
      \node at (1.8,5.4) (nodetxt) {$\tau_p=2$};

      \draw [->] (1.9,2) -- (3.1,2) ;
  \draw (1.4,2) circle (8pt);
      \node at (3.7,2) (nodetxt) {$\tau_p=1$};
        % \addplot [brown, mark=diamond]
        % table[x=D,y=B,col sep=comma] {Data/Data_forthrun/data_K5_cf128.csv};
        % \addlegendentry{$B=128MB$}

\end{axis}
\end{scope}
%\begin{scope}[xshift=0,yshift=-7.2cm]
%\begin{axis}[
%title=Perfect CSI,
%domain=0:4,
%grid=major,
% ylabel= users sum-rate,
% xlabel= Number of the APs,
% xmin=1,
% xmax=7,
% xtick style={color=clr3},
% xtick={1,2,3,4,5,6,7},
%  xticklabels={L=2,L=4,L=8,L=16,L=32,L=64,L=128},
% ymin=1,
% ymax=7,
% ytick={0,1,2,3,4,5,6,7,8},
% mark size=4.0 pt,
% ]       
% \addplot [orange, mark=asterisk]
%         table[x expr=\coordindex+1,y=A,col sep=comma] {Data_for_figures/data_K4_cf64KB_tauc_0.25K_uncorrCh_impCSI.csv};
%        \addplot [orange, mark=o]
%         table[x expr=\coordindex+1,y=B,col sep=comma] {Data_for_figures/data_K4_cf64KB_tauc_0.25K_uncorrCh_impCSI.csv};
%         %\addlegendentry{Infinite memory size}

%         \addplot [orange, mark=triangle]
%         table[x expr=\coordindex+1,y=C,col sep=comma] {Data_for_figures/data_K4_cf64KB_tauc_0.25K_uncorrCh_impCSI.csv};
%         %\addlegendentry{FAP with $C_{AP}=256KB$}
        
%           \addplot [red, mark=asterisk]
%         table[x expr=\coordindex+1,y=D,col sep=comma] {Data_for_figures/data_K4_cf64KB_tauc_0.25K_uncorrCh_impCSI.csv};
%         %\addlegendentry{FAP with $C_{AP}=64KB$}

% \end{axis}
% \end{scope}
\end{tikzpicture}
    \caption{Average per-user SE vs number of APs when $K=4$, $C_{AP}=64KB$. The channel model is uncorrelated Rayleigh fading, and imperfect CSI is assumed with pilot signal length to $\tau_p=2$ or $\tau_p=1$.}
    \label{fig_impcsi}
\end{figure}
 \textcolor{black}{Throughout the whole paper, we assumed perfect CSI for the mathematical tractibility. Specifically, assuming imperfect CSI in the presence of correlated Rayleigh fading, the transformations in problem (\ref{sumRate_max_prob_VC}) do not hold. However, we considered the uncorrelated Rayleigh fading scenario to show that the paper's conclusion also holds for imperfect CSI cases (due to pilot contamination, for example). We assumed that the pilot signal length is $\tau_p<K$. Then, the channel is estimated in the presence of pilot contamination, and the matrix $\mathbf{H}$ is replaced by its estimate. The impact of channel estimation error is reflected by adding the channel estimation error noise to $\sigma^2$. The simulation results in Fig. \ref{fig_impcsi} validate the earlier discussion on the limited memory effects on optimal fronthaul architecture.}
% \begin{figure}[hb]
%     \centering
%     \input{figures/figures_R2/imp_CSI_K4_cf64KB}
%     \caption{Average per-user SE vs number of APs when $K=4$, $C_{AP}=64KB$. The channel model is uncorrelated Rayleigh fading, and imperfect CSI is assumed with pilot signal length to $\tau_p=2$ or $\tau_p=1$.}
%     \label{fig_impcsi}
% \end{figure}

\subsection{\textcolor{black}{Equal vs linear allocation of a total memory budget}}
Intuitively, the memory capacity of the last AP becomes a bottleneck in the network with sequential topology. Therefore, it may seem like a solution that the memory capacity of the AP increases linearly (FT-LA), similar to the number of stored received signal vectors. Consequently, the APs at the end of the sequence get a larger share of the total memory. It is observed from Fig. \ref{fig_lin} that FT-LA memory capacity allocation improves the average per-user SE only slightly around the optimal AP length where the memory constraint is not yet limiting. For a larger length of the sequential fronthaul, FT-LA even results in a lower performance compared to the FT-EA.  
With FT-LA, even though the received signal vectors of the APs at the end of the sequence can be compressed 
less, the compression noise power of the stored received signal vectors of the APs at the beginning of the sequence increases compared to the case of FT-EA. FT-LA will hence suffer from more noisy
user signal estimation at first APs, upon which the subsequent APs should build their own estimation. Because of this error propagation in the sequence, giving more memory capacity to the last APs does not pay off. Thus, we conclude EA is highly preferred for longer sequences.
\begin{figure}[h]
    \centering
    \pgfplotsset{width=8.4cm,compat=1.18}
\pgfplotsset{every x tick label/.append style={font=\small, yshift=0.5ex},every y tick label/.append style={font=\small, xshift=0.5ex},
every axis legend/.append style={
at={(1.02,1)},
anchor=north west,font=\large
}}

%\begin{tikzpicture}
\begin{tikzpicture}[scale=0.7]
\begin{scope}[xshift=0cm,yshift=0cm]
\begin{axis}[
domain=0:4,
grid=major,
ylabel= Average per-user SE (bit/sec/Hz),
%xlabel= Number of the APs,
xmin=1,
xmax=7,
xtick style={color=clr3},
 xtick={1,2,3,4,5,6,7},
 xticklabels={L=2,L=4,L=8,L=16,L=32,L=64,L=128},
 xticklabel style={yshift=-2.5pt},
ymin=1.5,
ymax=8,
ytick={2,3,4,5,6,7,8},
mark size=4.0 pt,
legend columns=2,
legend style={nodes={scale=0.9, transform shape},at={(0.5,-1.4)},anchor=north},
%legend entries={
%$Option 1$,$Option 2$,$Option 3$,$Option4$,$ OPtion 5$,
%{[text width=25pt,text depth=]Neg. Sign:},
%},
% same effect:
% legend style={
% nodes={text width=25pt,text depth=},}
]
% \addplot [clr3, mark=x]
%        % add a plot from table; you select the columns by using the actual name in
%         %the .csv file (on top)
%         table[x=Aps,y=K15_W16_init_0.01,col sep=comma]  {Data/Data_forthrun/SRLSData/Data_forthrun.csv};
%         \addlegendentry{SRLS with $\delta=10^{2}$}
        
%         \addplot [clr3, mark=square]
%         table[x=Aps,y=K15_W16_init_1000,col sep=comma] {Data/Data_forthrun/SRLSData/Data_forthrun.csv};
%         \addlegendentry{SRLS with $\delta=10^{-3}$}
        % \addplot [clr3, mark=o]
        % table[x=D,y=A,col sep=comma] {Data/Data_forthrun/SR_K25cf7.csv};
        % \addlegendentry{SRLS with $\delta=10^{-9}$}
        
        % \addplot [brown, mark=diamond]
        % table[x=D,y=C,col sep=comma] {Data/Data_forthrun/SR_K25cf7.csv};
        % \addlegendentry{SRLS with $\delta=10^{-9}$}

        \addplot [blue, mark=o]
        table[x expr=\coordindex+1,y=B,col sep=comma] {Data_for_figures/datal_K4_cf8MB.csv};
        \addlegendentry{FT-LA $C_{T}=8MB$}
        
         \addplot [blue, mark=triangle]
        table[x expr=\coordindex+1,y=B,col sep=comma] {Data_for_figures/data_K4_cf8MB.csv};
        \addlegendentry{FT-EA $C_{T}=8MB$}
        \addplot [orange, mark=o]
        table[x expr=\coordindex+1,y=B,col sep=comma] {Data_for_figures/datal_K4_cf32MB.csv};
        \addlegendentry{FT-LA $C_{T}=32MB$}
        \addplot [orange, mark=triangle]
        table[x expr=\coordindex+1,y=B,col sep=comma] {Data_for_figures/data_K4_cf32MB.csv};
        \addlegendentry{FT-EA $C_{T}=32MB$}

        % \addplot [blue, mark=asterisk]
        % table[x expr=\coordindex+1,y=B,col sep=comma] {Data_for_figures_TT/dataTT_K4_cf8MB.csv};
        % \addlegendentry{FT-NAEA $C_{T}=8MB$ in multi-branch tree fronthaul }
        % \addplot [orange, mark=asterisk]
        % table[x expr=\coordindex+1,y=B,col sep=comma] {Data_for_figures_TT/dataTT_K4_cf32MB.csv};
        % \addlegendentry{FT-NAEA $C_{T}=32MB$ in multi-branch tree fronthaul }
        % \addplot [brown, mark=diamond]
        % table[x=D,y=B,col sep=comma] {Data/Data_forthrun/data_K5_cf128.csv};
        % \addlegendentry{$B=128MB$}

\end{axis}
\end{scope}

\begin{scope}[xshift=0,yshift=-6.8cm]
\begin{axis}[
domain=0:4,
grid=major,
% ylabel= users sum-rate,
 xlabel= Number of the APs,
 ylabel= Average per-user SE (bit/sec/Hz),
xmin=1,
xmax=7,
xtick style={color=clr3},
xtick={1,2,3,4,5,6,7},
 xticklabels={L=2,L=4,L=8,L=16,L=32,L=64,L=128},
 xticklabel style={yshift=-2.5pt},
ymin=1.5,
ymax=8,
ytick={2,3,4,5,6,7,8},
mark size=4.0 pt,
% legend columns=2,
% legend style={nodes={scale=0.8, transform shape},at={(0.5,-0.3)},anchor=north},
%legend entries={
%$Option 1$,$Option 2$,$Option 3$,$Option4$,$ OPtion 5$,
%{[text width=25pt,text depth=]Neg. Sign:},
%},
% same effect:
% legend style={
% nodes={text width=25pt,text depth=},}
]
% \addplot [clr3, mark=x]
%        % add a plot from table; you select the columns by using the actual name in
%         %the .csv file (on top)
%         table[x=Aps,y=K15_W16_init_0.01,col sep=comma]  {Data/Data_forthrun/SRLSData/Data_forthrun.csv};
%         \addlegendentry{SRLS with $\delta=10^{2}$}
        
%         \addplot [clr3, mark=square]
%         table[x=Aps,y=K15_W16_init_1000,col sep=comma] {Data/Data_forthrun/SRLSData/Data_forthrun.csv};
%         \addlegendentry{SRLS with $\delta=10^{-3}$}
        % \addplot [clr3, mark=o]
        % table[x=D,y=A,col sep=comma] {Data/Data_forthrun/SR_K25cf7.csv};
        % \addlegendentry{SRLS with $\delta=10^{-9}$}
        
        % \addplot [brown, mark=diamond]
        % table[x=D,y=C,col sep=comma] {Data/Data_forthrun/SR_K25cf7.csv};
        % \addlegendentry{SRLS with $\delta=10^{-9}$}

       % \addlegendentry{Infinite Buffer size}
        
       \addplot [blue, mark=o]
        table[x expr=\coordindex+1,y=B,col sep=comma] {Data_for_figures/datal_K64_cf8MB.csv};
        %\addlegendentry{FAP with $BC_{f}=256KB$}
        
         \addplot [blue, mark=triangle]
        table[x expr=\coordindex+1,y=B,col sep=comma] {Data_for_figures/data_K64_cf8MB.csv};

        \addplot [orange, mark=o]
        table[x expr=\coordindex+1,y=B,col sep=comma] {Data_for_figures/datal_K64_cf32MB.csv};

        \addplot [orange, mark=triangle]
        table[x expr=\coordindex+1,y=B,col sep=comma] {Data_for_figures/data_K64_cf32MB.csv};

        % \addplot [clr1, mark=asterisk]
        % table[x expr=\coordindex+1,y=B,col sep=comma] {Data_for_figures_TT/dataTT_K64_cf8MB.csv};

        % \addplot [clr2, mark=diamond]
        % table[x expr=\coordindex+1,y=B,col sep=comma] {Data_for_figures_TT/dataTT_K64_cf32MB.csv};

\end{axis}
\end{scope}

\end{tikzpicture}
    \caption{Average per-user SE comparison  between FT-EA and FT-LA memory model in a daisy chain fronthaul with two different numbers of users:  
    (Top) $K=4$, (Bottom) $K=64$. VC is used for compression.
    % \sofie{Vida, for me, if you have a ":", the cases after the : should be examples of what you see before the :. So, I would expect the caption to read as: 
    % Average per-user SE as function of the number of APs in a scenario with daisy chain fronthaul topology and increasing number of users:   
    % (Left) $K=4$, (Middle) $K=16$, (Right) $K=64$. The memory modet is FT-EA and FT-LA }
   }
    \label{fig_lin}
\end{figure}
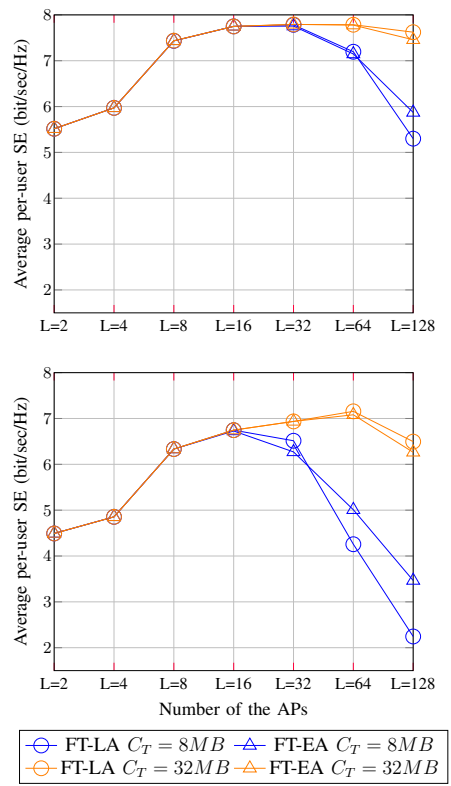
% \begin{figure}
%     \centering
%     \input{figures/fig7.tex}
%     \caption{FT-EA memory allocation among APs connected in a daisy chain or multi-branch tree topology (depicted in Fig.\ref{fig_FT}) with three different number of users. The compression option is VC.}
%     \label{fig7}
% \end{figure}

\begin{figure}
    \centering
    \pgfplotsset{width=8.4cm,compat=1.18}
\pgfplotsset{every x tick label/.append style={font=\small, yshift=0.5ex},every y tick label/.append style={font=\small, xshift=0.5ex},
every axis legend/.append style={
at={(1.02,1)},
anchor=north west,font=\large
}}

%\begin{tikzpicture}
\begin{tikzpicture}[scale=0.7]
\begin{scope}[xshift=0cm,yshift=0cm]
\begin{axis}[
domain=0:4,
grid=major,
ylabel= Average per-user SE (bit/sec/Hz),
%xlabel= Number of the APs,
xmin=1,
xmax=7,
xtick style={color=clr3},
 xtick={1,2,3,4,5,6,7},
 xticklabels={L=2,L=4,L=8,L=16,L=32,L=64,L=128},
 xticklabel style={yshift=-2.5pt},
ymin=3,
ymax=8,
ytick={4,5,6,7,8},
mark size=4.0 pt,
legend columns=2,
legend style={nodes={scale=0.9, transform shape},at={(0.5,-1.4)},anchor=north},
%legend entries={
%$Option 1$,$Option 2$,$Option 3$,$Option4$,$ OPtion 5$,
%{[text width=25pt,text depth=]Neg. Sign:},
%},
% same effect:
% legend style={
% nodes={text width=25pt,text depth=},}
]
% \addplot [clr3, mark=x]
%        % add a plot from table; you select the columns by using the actual name in
%         %the .csv file (on top)
%         table[x=Aps,y=K15_W16_init_0.01,col sep=comma]  {Data/Data_forthrun/SRLSData/Data_forthrun.csv};
%         \addlegendentry{SRLS with $\delta=10^{2}$}
        
%         \addplot [clr3, mark=square]
%         table[x=Aps,y=K15_W16_init_1000,col sep=comma] {Data/Data_forthrun/SRLSData/Data_forthrun.csv};
%         \addlegendentry{SRLS with $\delta=10^{-3}$}
        % \addplot [clr3, mark=o]
        % table[x=D,y=A,col sep=comma] {Data/Data_forthrun/SR_K25cf7.csv};
        % \addlegendentry{SRLS with $\delta=10^{-9}$}
        
        % \addplot [brown, mark=diamond]
        % table[x=D,y=C,col sep=comma] {Data/Data_forthrun/SR_K25cf7.csv};
        % \addlegendentry{SRLS with $\delta=10^{-9}$}

        \addplot [blue, mark=triangle]
        table[x expr=\coordindex+1,y=B,col sep=comma] {Data_for_figures/dataTT_K4_cf512KB.csv};
        \addlegendentry{$C_{T}=512KB$, multi-branch tree}

        \addplot [blue, mark=o]
        table[x expr=\coordindex+1,y=B,col sep=comma] {Data_for_figures/dataTT_K4_cf8MB.csv};
        \addlegendentry{$C_{T}=8MB$, multi-branch tree}
        
         \addplot [orange, mark=o]
        table[x expr=\coordindex+1,y=B,col sep=comma] {Data_for_figures/data_K4_cf8MB.csv};
        \addlegendentry{$C_{T}=8MB$, Daisy chain}
        
        % \addplot [orange, mark=triangle]
        % table[x expr=\coordindex+1,y=B,col sep=comma] {Data_for_figures/data_K4_cf32MB.csv};
        % \addlegendentry{FT-EA $C_{T}=32MB$}

        % \addplot [blue, mark=asterisk]
        % table[x expr=\coordindex+1,y=B,col sep=comma] {Data_for_figures_TT/dataTT_K4_cf8MB.csv};
        % \addlegendentry{FT-NAEA $C_{T}=8MB$ in multi-branch tree fronthaul }
        % \addplot [orange, mark=asterisk]
        % table[x expr=\coordindex+1,y=B,col sep=comma] {Data_for_figures_TT/dataTT_K4_cf32MB.csv};
        % \addlegendentry{FT-NAEA $C_{T}=32MB$ in multi-branch tree fronthaul }
        % \addplot [brown, mark=diamond]
        % table[x=D,y=B,col sep=comma] {Data/Data_forthrun/data_K5_cf128.csv};
        % \addlegendentry{$B=128MB$}

\end{axis}
\end{scope}

\begin{scope}[xshift=0,yshift=-6.8cm]
\begin{axis}[
domain=0:4,
grid=major,
% ylabel= users sum-rate,
 xlabel= Number of the APs,
 ylabel= Average per-user SE (bit/sec/Hz),
xmin=1,
xmax=7,
xtick style={color=clr3},
xtick={1,2,3,4,5,6,7},
 xticklabels={L=2,L=4,L=8,L=16,L=32,L=64,L=128},
 xticklabel style={yshift=-2.5pt},
ymin=3,
ymax=8,
ytick={4,5,6,7,8},
mark size=4.0 pt,
% legend columns=2,
% legend style={nodes={scale=0.8, transform shape},at={(0.5,-0.3)},anchor=north},
%legend entries={
%$Option 1$,$Option 2$,$Option 3$,$Option4$,$ OPtion 5$,
%{[text width=25pt,text depth=]Neg. Sign:},
%},
% same effect:
% legend style={
% nodes={text width=25pt,text depth=},}
]
% \addplot [clr3, mark=x]
%        % add a plot from table; you select the columns by using the actual name in
%         %the .csv file (on top)
%         table[x=Aps,y=K15_W16_init_0.01,col sep=comma]  {Data/Data_forthrun/SRLSData/Data_forthrun.csv};
%         \addlegendentry{SRLS with $\delta=10^{2}$}
        
%         \addplot [clr3, mark=square]
%         table[x=Aps,y=K15_W16_init_1000,col sep=comma] {Data/Data_forthrun/SRLSData/Data_forthrun.csv};
%         \addlegendentry{SRLS with $\delta=10^{-3}$}
        % \addplot [clr3, mark=o]
        % table[x=D,y=A,col sep=comma] {Data/Data_forthrun/SR_K25cf7.csv};
        % \addlegendentry{SRLS with $\delta=10^{-9}$}
        
        % \addplot [brown, mark=diamond]
        % table[x=D,y=C,col sep=comma] {Data/Data_forthrun/SR_K25cf7.csv};
        % \addlegendentry{SRLS with $\delta=10^{-9}$}

       % \addlegendentry{Infinite Buffer size}
        
      \addplot [blue, mark=triangle]
        table[x expr=\coordindex+1,y=B,col sep=comma] {Data_for_figures/dataTT_K64_cf512KB.csv};
        %\addlegendentry{FT-LA $C_{T}=8MB$}

        \addplot [blue, mark=o]
        table[x expr=\coordindex+1,y=B,col sep=comma] {Data_for_figures/dataTT_K64_cf8MB.csv};
        
         \addplot [orange, mark=o]
        table[x expr=\coordindex+1,y=B,col sep=comma] {Data_for_figures/data_K64_cf8MB.csv};
        %\addlegendentry{FT-EA $C_{T}=8MB$}

        % \addplot [orange, mark=triangle]
        % table[x expr=\coordindex+1,y=B,col sep=comma] {Data_for_figures/data_K64_cf32MB.csv};

        % \addplot [clr1, mark=asterisk]
        % table[x expr=\coordindex+1,y=B,col sep=comma] {Data_for_figures_TT/dataTT_K64_cf8MB.csv};

        % \addplot [clr2, mark=diamond]
        % table[x expr=\coordindex+1,y=B,col sep=comma] {Data_for_figures_TT/dataTT_K64_cf32MB.csv};

\end{axis}
\end{scope}

\end{tikzpicture}
    \caption{Average per-user SE comparison between a daisy chain and multi-branch tree fronthaul topologies with two different numbers of users: (Top) $K=4$, (Bottom) $K=64$. FT-EA memory model is used, and the compression option is VC.}
    \label{fig10_tree}
\end{figure}
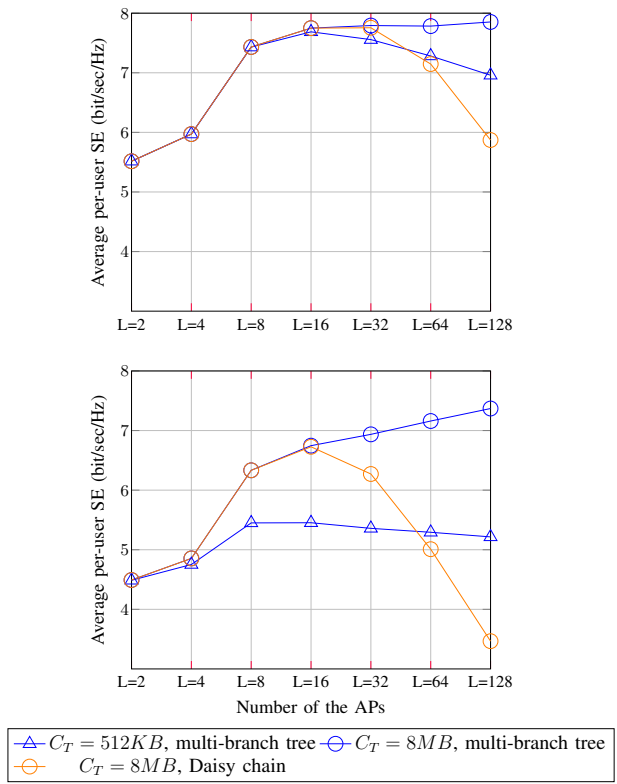
\subsection{\textcolor{black}{An alternative fronthaul topology to daisy chain topology: multi-branch tree topology}}
Finally, in Fig. \ref{fig10_tree}, a multi-branch tree topology %depicted in 
is compared with the daisy chain fronthaul topology. As in a multi-branch tree, the depth of the tree is less than in the case of the daisy chain topology (the depth of a daisy chain with $L=128$ is $128$ while the depth of a multi-branch tree with $L=128$ and APs arranged as shown in Fig. \ref{fig_FT} is $8$ which is $16$ times smaller.), we first consider a smaller total memory size, i.e., $C_T=512KB$ (almost $16$ times smaller than $C_T=8MB$).
As shown in Fig. \ref{fig10_tree}, for $K=4$, using a multi-branch tree topology with $C_T=512KB$ have comparable peak performance to the daisy chain topology with $C_T=8MB$, e.g., for $K=4$, daisy chain with $C_T=8MB$ has its peak performance of $7.8$ bit/sec/Hz at $L=32$, which is almost the same as peak performance of multi-branch tree topology with $C_T=512KB$. Therefore, multi-branch topology can save around $0.94\%$ memory cost compared to daisy chain topology. For a higher number of users, e.g., $K=64$, multi-branch tree topology can improve the average per-user SE performance for around $9\%$ (at $L=128$) compared to the peak performance of the daisy chain topology (at $L=16$) when the same memory capacity, i.e., $C_T=8MB$, is used in both topologies.
However, assuming the same total memory, i.e., $C_T=8MB$, in the multi-branch tree topology, the average incoming fronthaul rate to the APs is larger than the daisy chain topology.
\section{Conclusions}\label{sec6}
% \begin{itemize}
%     \item When the number of users is decreased, the condition number of $\mathbf{H}_l*\mathbf{H}_l'$ will increase which means that not all the eigen vector are equally important so we can try only to compress the first few strong eigen direction. However, seems that equation \ref{eq1.15} is adaptive to the eigen vectors importance, so maybe no need to ignore some eigen vector in matrix $\mathbf{A}_l$.
%     \item Consider compressing the raw received vector element wise and compare it with this method.
%     \item was looking at the simulation and how the length of the tree will affect memory capacity problem. Assuming fixed memory capacity (independent of the number of antennas per AP or it is dependant to AP?)
% \end{itemize}
In this paper, we considered the 
performance scaling with the  
number of APs in a CFmMIMO network under a realistic assumption of limited working memory in each AP. We considered two models for the memory capacity of the APs, the fixed per AP memory capacity model and the fixed total memory capacity model, and investigated the impact of using different memory models at the APs on the rate of the fronthaul link. Furthermore, the simulation result shows that the limited memory capacity at the APs limits the optimal number of APs in the sequence. %like 
This is in sharp contrast to the infinite memory case 
, which favors the maximal distribution of the antennas.

We further analyzed multiple compression options and showed that vector-wise compression, which is joint compression of all the antenna signals per AP, results in larger average per-user spectral efficiency gains. However, element-wise compression is used in practice for its simplicity. More importantly, by simply pre-processing the received signal vectors at each AP using PCA, element-wise compression achieves the same performance as rather expensive vector-wise compression in all considered scenarios. 

Furthermore, we analyzed different memory allocation models.
As the delay and number of samples to be stored at the APs increase linearly, a good option to allocate a total memory capacity may seem to be the linearly increasing memory allocation to the APs. However, the simulation results show that this memory allocation method doesn't bring much of a benefit as we allocate less memory capacity to the APs at the beginning of the sequence. Consequently, we start with a noisy compression of the received signal vector and a low-quality local estimate of the users' signal vector in the first AP, upon which the rest of the APs build their refined version of the users' signal vector estimate.

Finally, we show that when the number of users is relatively low, i.e., $K=4$, using a multi-branch tree topology can save the memory cost more than $90\%$, compared to daisy chain topology, with a similar average per-user SE peak performance. When the number of users gets larger, i.e., $K=64$, multi-branch tree topology can improve the average per-user SE for around $9\%$ compared to the daisy chain fronthaul with the same total memory, i.e., $C_T=8MB$, \textcolor{black}{however, at the cost of increase average incoming fronthaul rate to the APs.}

\nocite{HardwareBjörnson}
\nocite{jointKang}
\nocite{masoumiperformance}
\nocite{Shaik2021}

\section*{Acknowledgments}
This work is supported by European Union’s
Horizon 2020 research and innovation program under grant
agreements: 101013425 (REINDEER) and 101017171 (MARSAL)
and from Research Foundation – Flanders (FWO) under project number G0C0623N
.
The resources and services used in this work were provided by the VSC (Flemish Supercomputer Center), funded by the Research Foundation - Flanders (FWO) and the Flemish Government.

The authors thank Panagiotis (Panos) Patrinos for improving and simplifying the paper's appendix.

\begin{appendices}
\section*{Appendix A}\label{apda}
Before proving the equality $\myeq$ and inequality $\myineqlb$ in (\ref{sumRate_VC}, Theorem $1$ is provided.
\paragraph*{Theorem 1}: For matrices $\mathbf{A}$ and $\mathbf{B}$ with size $n\times m$ and $m\times n$ respectively, the following equality holds:
\begin{equation}
\det(\mathbf{I}_n+\mathbf{A}\mathbf{B})=\det(\mathbf{I}_m+\mathbf{B}\mathbf{A}).
    \label{eqsylvester}
\end{equation}
\textit{Proof of Theorem 1}: 
Theorem $1$ is Sylvester's determinant theorem and can be proved easily \cite{poz_book}.
\subsection{\textcolor{black}{proof of $\myeq$ in (\ref{sumRate_VC})}}
We know that the vector $\{\hat{\mathbf{y}}_v|\mathbf{H}\}$ is a circularly symmetric Gaussian random vector because:
\begin{equation}
    \{\hat{\mathbf{y}}_v|\mathbf{H}\}=\{\mathbf{y}| \mathbf{H}\}+\mathbf{q}_v
    \label{eq49}
\end{equation}
where $\{\mathbf{y}|\mathbf{H}\}$ and $\mathbf{q}_v$ are two independent circularly symmetric Gaussian random vectors and based on the definition of the circularly symmetric Gaussian random vectors in \cite[Appendix A]{marzetta_larsson_yang_ngo_2016}, their summation is also circularly symmetric Gaussian random vector. Furthermore, any linear transformation of a circularly symmetric Gaussian vector results in a vector that is a circularly symmetric Gaussian vector \cite[Appendix A]{marzetta_larsson_yang_ngo_2016}, which concludes that $\hat{\mathbf{s}}_v=\hat{\mathbf{C}}_v\hat{\mathbf{y}}_v$, defined in (\ref{UE_sig_est_using_quantz}) is a circularly symmetric Gaussian vector. Similarly, 
the 
signal estimation error vector $\mathbf{e}$ can also be proved to be a circularly symmetric Gaussian vector, independent of $\hat{\mathbf{s}}_v$. Therefore,

\begin{equation}
    \begin{aligned}
        \mathcal{I}(\hat{\mathbf{s}}_v;\mathbf{s})&=\mathcal{H}(\mathbf{s})-\mathcal{H}(\mathbf{s}|\hat{\mathbf{s}}_v)\\&=\mathcal{H}(\mathbf{s})-\mathcal{H}(\underbrace{\mathbf{s}-\hat{\mathbf{s}}_v}_{\mathbf{e}})\\&=\log_2\det(\pi e p\mathbf{I}_K)-\log_2\det(\pi e \mathbf{\Gamma}_L)\\&=\log_2\det(p\mathbf{\Gamma}_L^{-1})\\&=\textcolor{black}{\log_2\det(p\mathbf{H}^{\text{H}}\mathbf{Z}_v^{-1}\mathbf{H}+\mathbf{I}_K)}.
    \end{aligned}
    \label{eqR_v_proof}
\end{equation}
where we substitute the $\mathbf{\Gamma}_L$ in (\ref{eqR_v_proof}) with its equivalent matrix as in (\ref{sig_est_error}) and used Theorem $1$ to conclude the proof of $\myeq$ in (\ref{sumRate_VC}).

% Finally, using the differential entropy of a circularly symmetric Gaussian vector, $\myeqb$ is proved.
\subsection{\textcolor{black}{Proof of $\myineqlb$ in (\ref{sumRate_VC})}}
To prove inequality $\myineqlb$ in (\ref{sumRate_VC}), Theorem 1 is used.\newline
% \paragraph*{Theorem 1}: For matrices $\mathbf{A}$ and $\mathbf{B}$ with size $n\times m$ and $m\times n$ respectively, the following equality holds:
% \begin{equation}
% \det(\mathbf{I}_n+\mathbf{A}\mathbf{B})=\det(\mathbf{I}_m+\mathbf{B}\mathbf{A}).
%     \label{eqsylvester}
% \end{equation}
% \textit{Proof of Theorem 1}: 
% Theorem $1$ is Sylvester's determinant theorem and can be proved easily \cite{poz_book}.
% \begin{equation}
% \begin{aligned}
%     \det(\mathbf{I}_n+\mathbf{A}\mathbf{B})&=\det(\mathbf{A}(\mathbf{A}^{-1}+\mathbf{B}))\\&=\det(\mathbf{A}^{-1}+\mathbf{B})\det(\mathbf{A})\\&=\det(\mathbf{I}_n+\mathbf{B}\mathbf{A})
%     \end{aligned}
%     \label{eq51}
% \end{equation}
Using Theorem 1:
\begin{equation}
\begin{aligned}
&\log_2\det(p\mathbf{H}\mathbf{H}^{\text{H}}\mathbf{Z}_v^{-1}+\mathbf{I}_{NL})\\&=\log_2\det(p\mathbf{H}\mathbf{H}^{\text{H}}\mathbf{Z}_v^{-1/2}\mathbf{Z}_v^{-H/2}+\mathbf{I}_{NL})\\&=\log_2\det(p\mathbf{Z}_v^{-\text{H}/2}\mathbf{H}\mathbf{H}^{\text{H}}\mathbf{Z}_v^{-1/2}+\mathbf{I}_{NL}).
\end{aligned}
\label{eq54}
\end{equation}
where $\mathbf{Z}_v^{-1/2}=\text{blkdiag}(\mathbf{Z}_{v1}^{-1/2},\hdots,\mathbf{Z}_{vL}^{-1/2})$ is the square root of $\mathbf{Z}_v^{-1}$. Matrix $\mathbf{Z}_v^{-\text{H}/2}\mathbf{H}\mathbf{H}^{\text{H}}\mathbf{Z}_v^{-1/2}$ is positive semi-definite and for positive semi-definite matrices, invoking the Hadamard's inequality \cite{mcook} and subsequently Theorem 1 leads to :
\begin{equation}
\begin{aligned}
&\log_2\det(p\mathbf{Z}_v^{-\text{H}/2}\mathbf{H}\mathbf{H}^{\text{H}}\mathbf{Z}_v^{-1/2}+\mathbf{I}_{NL})\\&\leq \log_2\prod_{l=1}^{L}\det(p\mathbf{Z}_{vl}^{-\text{H}/2}\mathbf{H}_l\mathbf{H}_l^{\text{H}}\mathbf{Z}_{vl}^{-1/2}+\mathbf{I}_{N})\\&=\sum_{l=1}^{L}\log_2\det(p\mathbf{H}_l\mathbf{H}_l^{\text{H}}\mathbf{Z}_{vl}^{-1}+\mathbf{I}_{N}).
\end{aligned}
\label{eqhadamard}
\end{equation}
%\robbert{Where equality in the bound is achieved by having many users. To conclude, we have proven that the problem can be perfectly decomposed into many smaller, local subproblems per AP. }
\section*{Appendix B}\label{apdb}
To prove that the solution derived in (\ref{eig_opt_Qmat_VC}) is globally optimal,  
% \robbert{we first define the Lagrangian of aforementioned Problem \ref{sumRate_max_prob_VC} as:
% \begin{equation}
% \begin{aligned}
%     &\mathcal{L}(\mathbf{Q}_{vl}^{-1},\mu_{vl})=\\&(1-\mu_{vl})\log_2 \det({\mathbf{Q}_{vl}^{-1}(p\mathbf{H}_l\mathbf{H}_l^{\text{H}}+\sigma^2\mathbf{I}_N)+\mathbf{I}_N})-\\&\log_2 \det(\sigma^2\mathbf{Q}_{vl}^{-1}+\mathbf{I}_N)+\mu_{vl} C_{sc},
%     \label{eqlagrange}
%     \end{aligned}
% \end{equation}
% where $\mu_{vl}^o$ depicts the dual variable for the equality constraint.
% }
\textcolor{black}{Theorem 2 is used, which states under which conditions the proposed solution is optimal. In the following, we first prove why satisfying the conditions in Theorem 2 makes a solution optimal. Then, we prove that the solution provided in (\ref{eig_opt_Qmat_VC}) satisfies the condition in Theorem 2 and hence, is optimal. }
\paragraph*{Theorem 2}
Consider the problem defined in (\ref{sumRate_max_prob_VC}). $(\mathbf{Q}_{vl}^{-1})^o$ and $\mu_{vl}^o$ are optimal and strong duality holds, if and only if
\begin{equation}
    \begin{split}
    &(\text{Primal feasibility})\\
     &C_{sc}=\log_2 \det\bigl({{(\mathbf{Q}_{vl}^{-1})}^o(p\mathbf{H}_l\mathbf{H}_l^{\text{H}}+\sigma^2\mathbf{I}_N)+\mathbf{I}_N}\bigr), \hspace{4pt}{(\mathbf{Q}_{vl}^{-1})^o\succeq\mathbf{0}}\hspace{10pt} \\
     &(\text{Lagrange optimality})\\
    &{(\mathbf{Q}_{vl}^{-1})}^o=\arg \max_{\mathbf{Q}_{vl}^{-1}\succeq\mathbf{0}}\mathcal{L}(\mathbf{Q}_{vl}^{-1},\mu_{vl}^o), \hspace{80pt}  \\
    \end{split}
    \label{eq56}
\end{equation}
where the Lagrange function is defined as follows:
\begin{equation}
\begin{aligned}
    &\mathcal{L}(\mathbf{Q}_{vl}^{-1},\mu_{vl})=\\&(1-\mu_{vl})\log_2 \det({\mathbf{Q}_{vl}^{-1}(p\mathbf{H}_l\mathbf{H}_l^{\text{H}}+\sigma^2\mathbf{I}_N)+\mathbf{I}_N})-\\&\log_2 \det(\sigma^2\mathbf{Q}_{vl}^{-1}+\mathbf{I}_N)+\mu_{vl} C_{sc}.
    \label{eqlagrange}
    \end{aligned}
\end{equation}
\textit{Proof of Theorem 2}: Theorem 2 can be proved based on Proposition 6.2.5 in \cite{bertsekas2003}. 
Before starting the proof, we 
define the objective function as:
\begin{equation}
\begin{aligned}
    F(\mathbf{Q}_{vl}^{-1})=&\log_2 \det({\mathbf{Q}_{vl}^{-1}(p\mathbf{H}_l\mathbf{H}_l^{\text{H}}+\sigma^2\mathbf{I}_N)+\mathbf{I}_N})-\\&\log_2 \det(\sigma^2\mathbf{Q}_{vl}^{-1}+\mathbf{I}_N)
    \label{objfun}
    \end{aligned}
\end{equation}
% \textcolor{black}{and equality constrain function as:
% \begin{equation}
%     \begin{split}
%      &h(\mathbf{Q}_{vl}^{-1})=C_{sc}-\log_2 \det\bigl({{(\mathbf{Q}_{vl}^{-1})}^o(p\mathbf{H}_l\mathbf{H}_l^{\text{H}}+\sigma^2\mathbf{I}_N)+\mathbf{I}_N}\bigr), \hspace{4pt} \\
%     \end{split}
%     \label{eq56}
% \end{equation}
% }

In the following, we prove that Theorem $2$ provides necessary and sufficient conditions for the optimality of our solution. 
\begin{itemize}
    \item If $(\mathbf{Q}_{vl}^{-1})^o$ and $\mu_{vl}^o$ are optimal primal and dual variables for which strong duality holds, they must satisfy the following conditions:
    \begin{itemize}
        \item For any $(\mathbf{Q}_{vl}^{-1})^o$ to be optimal, it must be in the feasible set obviously, so the primal feasibility of Theorem $2$ is achieved.
        %\robbert{For any $(\mathbf{Q}_{vl}^{-1})^o$ to be optimal, it must be in the feasible set, so Primal Feasibility is a necessary condition for optimality. }
        \item For the
        Lagrange optimality condition, as $(\mathbf{Q}_{vl}^{-1})^o$ and $\mu_{vl}^o$ are primal and dual optimal and based on the
          Lagrange function definition in (\ref{eqlagrange}),
       % \begin{equation}
       %  \begin{aligned}
       %    F^o&=F((\mathbf{Q}_{vl}^{-1})^o)\\&\myeq\max_{\mathbf{Q}_{vl}^{-1}\succeq\mathbf{0}} \mathcal{L}((\mathbf{Q}_{vl}^{-1}),\mu_{vl}^o)\\&\myineqb\mathcal{L}((\mathbf{Q}_{vl}^{-1})^o,\mu_{vl}^o)\\&\myeqc F^o,
       %     \end{aligned}
       %     \label{lagopt_prf}
       %  \end{equation}
       \textcolor{black}{
        \begin{equation}
        \begin{aligned}
          F^o&=F((\mathbf{Q}_{vl}^{-1})^o)\\&\myeq\mathcal{L}((\mathbf{Q}_{vl}^{-1})^o,\mu_{vl}^o)\\&=\max_{\mathbf{Q}_{vl}^{-1}\succeq\mathbf{0}, h(\mathbf{Q}_{vl}^{-1})=0} \mathcal{L}(\mathbf{Q}_{vl}^{-1},\mu_{vl}^o)\\&\leq\max_{\mathbf{Q}_{vl}^{-1}\succeq\mathbf{0}} \mathcal{L}((\mathbf{Q}_{vl}^{-1}),\mu_{vl}^o)\myeqb F^o\\
          % &\myineqb\max_{\mathbf{Q}_{vl}^{-1}\succeq\mathbf{0}, h(\mathbf{Q}_{vl}^{-1})=0} \mathcal{L}((\mathbf{Q}_{vl}^{-1}),\mu_{vl}^o)\\&=\max_{\mathbf{Q}_{vl}^{-1}\succeq\mathbf{0}, h(\mathbf{Q}_{vl}^{-1})=0} F(\mathbf{Q}_{vl}^{-1})=F^o,
           \end{aligned}
           \label{lagopt_prf}
        \end{equation}}
        where $\myeq$ holds due to the primal and dual optimality of $(\mathbf{Q}_{vl}^{-1})^o$ and $\mu_{vl}^o$ respectively and $\myeqb$ holds due to strong duality. Based on (\ref{lagopt_prf}), equality holds throughout (\ref{lagopt_prf}) and we have:
        \begin{equation}
        \begin{aligned}
         (\mathbf{Q}_{vl}^{-1})^o=\arg \max_{\mathbf{Q}_{vl}^{-1}\succeq\mathbf{0}} \mathcal{L}((\mathbf{Q}_{vl}^{-1}),\mu_{vl}^o).
           \end{aligned}
           \label{lagopt_eq}
        \end{equation}
        % \begin{equation}
        % \begin{aligned}
        %   (\mathbf{Q}_{vl}^{-1})^o=\arg \max_{\mathbf{Q}_{vl}^{-1}\succeq\mathbf{0}}F(\mathbf{Q}_{vl}^{-1})=\arg \max_{\mathbf{Q}_{vl}^{-1}\succeq\mathbf{0}}\mathcal{L}((\mathbf{Q}_{vl}^{-1}),\mu_{vl}^o)
        %    \end{aligned}
        %    \label{lagrangopt}
        % \end{equation}
         which proves the Lagrange optimality condition in Theorem $2$.
        \end{itemize}
    \item Conversely, 
    If the primal feasibility and Lagrange optimality conditions in Theorem $2$ hold, then we can prove that $(\mathbf{Q}_{vl}^{-1})^o$ and $\mu_{vl}^o$ are optimal primal and dual variables for which strong duality holds. This can be easily proved with the help of the fact that weak duality always holds.
    The proof is omitted due to its simplicity, and interested readers are referred to 6.2.5 in \cite{bertsekas2003}.
    % \begin{itemize}
    %     \item We know that weak duality always holds, which means that: 
    %     \begin{equation}
    %     F((\mathbf{Q}_{vl}^{-1})^o)\leq \mathcal{L}((\mathbf{Q}_{vl}^{-1}),\mu_{vl}^o).
    %     \label{dualitygap}
    %     \end{equation}
    %     Furthermore, based on the assumption that the condition in theorem $1$ holds, we have the following inequality:
    %     \begin{equation}
    %     \begin{aligned}
    %       \max_{\mathbf{Q}_{vl}^{-1}} F(\mathbf{Q}_{vl}^{-1})&= F((\mathbf{Q}_{vl}^{-1})^o)\\&=\max_{\mathbf{Q}_{vl}^{-1}}\mathcal{L}((\mathbf{Q}_{vl}^{-1}),\mu_{vl}^o)\\&\geq \mathcal{L}((\mathbf{Q}_{vl}^{-1}),\mu_{vl}^o)
    %        \end{aligned}
    %        \label{primal_dual_ineq}
    %     \end{equation}
    %     Based on eqs. (\ref{dualitygap}) and (\ref{primal_dual_ineq}), it is proven that:
    %     \begin{equation}
    %     \begin{aligned}
    %        (\mathbf{Q}_{vl}^{-1})^o=\arg \max_{\mathbf{Q}_{vl}^{-1}} \mathcal{L}((\mathbf{Q}_{vl}^{-1}),\mu_{vl}^o)
    %        \end{aligned}
    %        \label{cond2}
    %     \end{equation}
    % \end{itemize}
    % \item Conversely, now we prove that if optimality conditions in theorem $1$ hold, then we can prove that $(\mathbf{Q}_{vl}^{-1})^o$ and $\mu_{vl}^o$ are optimal primal and dual variables.
\end{itemize}

% The derivative of the Lagrange function with respect to $\mathbf{Q}_{vl}^{-1} $ is given as:
% \begin{equation}
%    \frac{\partial \mathcal{L}(\mathbf{Q}_{vl}^{-1},\mu_{vl})}{\partial \mathbf{Q}_{vl}^{-1}}= (1-\mu_{vl})(p\mathbf{H}_l\mathbf{H}_l^{\text{H}}+\sigma^2\mathbf{I}_N)({\mathbf{Q}_{vl}^{-1}(p\mathbf{H}_l\mathbf{H}_l^{\text{H}}+\sigma^2\mathbf{I}_N)+\mathbf{I}_N})^{-1}-\sigma^2(\sigma^2\mathbf{Q}_{vl}^{-1}+\mathbf{I}_N)^{-1}
%    \label{eq56}
% \end{equation}

% The matrix given in (\ref{opt_Qmat_VC}) is the matrix for which the derivative of the Lagrange function in (\ref{eq56}) is zero.
% Furthermore, inserting it into primal feasibility condition in (\ref{eq54}) (i.e., the equality constraint in (\ref{sumRate_max_prob_VC})) gives $\mu_{vl}^o$. Furthermore,
Based on Theorem $2$ and its proof, to prove that the matrix in (\ref{opt_Qmat_VC}) is globally optimal (over the range of all positive semi-definite matrices), it should satisfy the optimality conditions in (\ref{eq56}). To prove that the matrix in (\ref{opt_Qmat_VC}) optimizes the Lagrange function, Lemma $1$ is used.
% The KKT conditions of the problem defined in (\ref{sumRate_max_prob_VC}), is defined as below:
% \begin{equation}
%     \begin{split}
%     \frac{\partial \mathcal{L}(\mathbf{Q}_{vl}^{-1},\mu_{vl}^o)}{\partial \mathbf{Q}_{vl}^{-1}} & = 0, \\ \mu_{vl}^o(\log_2 \det({\mathbf{Q}_{vl}^{-1}(p\mathbf{H}_l\mathbf{H}_l^{\text{H}}+\sigma^2\mathbf{I}_N)+\mathbf{I}_N}-C_{sc})& =  0,\\
%     \mu_{vl}^o& \geq 0.\end{split}
%     \label{eq56}
% \end{equation}

% The ${\mathbf{Q}_{vl}^{-1}}^{*}$ with eigenvalues defined in (\ref{eig_opt_Qmat_VC}) satisfy the KKT conditions, which are necessary conditions for optimality. The sufficient condition for optimality of ${\mathbf{Q}_{vl}^{-1}}^{*}$ is given as below:
%  \begin{equation}
%      \begin{aligned}
%   {\mathbf{Q}_{vl}^{-1}}^{*} \in \arg \max_{\mathbf{Q}_{vl} } \quad  \mathcal{L}(\mathbf{Q}_{vl}^{-1},\mu_{vl}^o) 
%   \label{eq57}
% \end{aligned} 
%     \end{equation}
%     To prove that the ${\mathbf{Q}_{vl}^{-1}}^{*}$ satisfies the sufficiency condition, Lemma 1 is used.
    
\paragraph*{Lemma $1$}: \cite[Appendix B]{dis_comp} If matrix $\mathbf{A}$ and $\mathbf{B}$ are positive (semi-)definite matrix and $\mathbf{\Gamma}_{\mathbf{A}}$ and $\mathbf{\Gamma}_{\mathbf{B}}$ have the ordered eigenvalues of $\mathbf{A}$ and $\mathbf{B}$ on their diagonal respectively, then:
\begin{equation} \det(\mathbf{A}\mathbf{B}+\mathbf{I}_n)\leq\det(\mathbf{\Gamma}_{\mathbf{A}}\mathbf{\Gamma}_{\mathbf{B}}+\mathbf{I}_n),
    \label{eq61}
\end{equation}
with equality if the eigenvectors of $\mathbf{A}$ is conjugate transpose of eigen vectors of $\mathbf{B}$.

Using Lemma $1$, the Lagrange function in (\ref{eqlagrange}) can be upper bounded as follows:
\begin{equation}
\begin{aligned}
    &\mathcal{L}(\mathbf{Q}_{vl}^{-1},\mu_{vl}^o)\leq\\& (1-\mu_{vl}^o)\log_2 \det({\mathbf{\Sigma}_{vlq}^{-1}(p\mathbf{\Sigma}_l\mathbf{\Sigma}_l^{\text{H}}+\sigma^2\mathbf{I}_N)+\mathbf{I}_N})-\\&\log_2 \det(\sigma^2\mathbf{\Sigma}_{vlq}^{-1}+\mathbf{I}_N)+\mu_{vl}^o C_{sc},
    \label{eq62}
    \end{aligned}
\end{equation}
where $\mathbf{\Sigma}_{vlq}^{-1}$ is the matrix whose $i^{th}$ diagonal element is the $i^{th}$ eigenvalues of $\mathbf{Q}_{vl}^{-1}$, $i^{th}$ diagonal elements of $\mathbf{\Sigma}_l\mathbf{\Sigma}_l^{\text{H}}$ is given as $\lambda_{il}^2$. Therefore (\ref{eq62}) can be re-written as follows \cite{mcook}:
\begin{equation}
\begin{aligned}
    \mathcal{L}(\mathbf{Q}_{vl}^{-1},\mu_{vl}^o)\leq& (1-\mu_{vl}^o)\sum_{i=1}^N \log_2 ({\lambda_{vlqi}(p{\lambda}_{li}^2+\sigma^2)+1})-\\&\log_2 (\lambda_{vlqi}\sigma^2+1)+\mu_{vl}^o C_{sc}\\=&\sum_{i=1}^{N} f_i(\lambda_{vlqi})+\mu_{vl}^oC_{sc},
    \label{eq63}
    \end{aligned}
\end{equation}
where each of the functions $f_i, \forall i\in{1,\hdots,N}$ depends only on one variable $\lambda_{vlqi}, \forall i\in {1,\hdots,N}$,
\begin{equation}
\begin{aligned}
    f_i(\lambda_{vlqi})=&(1-\mu_{vl}^o)\log_2 ({\lambda_{vlqi}(p{\lambda}_{li}^2+\sigma^2)+1})-\\&\log_2 (\lambda_{vlqi}\sigma^2+1).
    \end{aligned}
\end{equation}
The derivative of $f_i(\lambda_{vlqi})$ with respect to $\lambda_{vlqi}$ is zero at $\lambda_{vlqi}^o$, defined as follows:
\begin{equation}
    \begin{aligned}
       \lambda^o_{vlqi}=\frac{1}{\mu_{vl}^o}(\frac{1}{\sigma^2}-\frac{1}{p\lambda_{li}^2+\sigma^2})-\frac{1}{\sigma^2}, \forall i.
        \label{eq65}
        \end{aligned}
    \end{equation}
Following the discussion on \cite[Appendix. B]{dis_comp}, $\lambda_{vlqi}^o$ in (\ref{eq65}) globally maximizes $f_i, \forall i\in{1,\hdots,N}$. However, restricting $\lambda^o_{vlqi}\geq0$, the maximizer of $f_i, \forall i\in{1,\hdots,N}$ is as follows:
 \begin{equation}
    \begin{aligned}
       \lambda^o_{vlqi}=\max(0, \frac{1}{\mu_{vl}^o}(\frac{1}{\sigma^2}-\frac{1}{p\lambda_{li}^2+\sigma^2})-\frac{1}{\sigma^2}), \forall i,
        \label{eq66}
        \end{aligned}
    \end{equation}
    where $\mu_{vl}^o$ is derived by inserting (\ref{eq66}) into equality constraint of (\ref{sumRate_max_prob_VC}).
 Inserting the optimal values given in (\ref{eq66}) into the upper bound in (\ref{eq63}), the Lagrange function will be upper bounded as follows:
\begin{equation}
\begin{aligned}
    \mathcal{L}(\mathbf{Q}_{vl}^{-1},\mu_{vl}^o)\leq& (1-\mu_{vl}^o)\sum_{i=1}^N \log_2 ({\lambda_{vlqi}^o(p{\lambda}_{li}^2+\sigma^2)+1})-\\&\log_2 (\lambda_{vlqi}^o\sigma^2+1)+\mu_{vl}^o C_{sc}\\=&\mathcal{L}({(\mathbf{Q}_{vl}^{-1})}^{o},\mu_{vl}^o).
    \label{eq67}
    \end{aligned}
\end{equation}
Hence, we have $(\mathbf{Q}_{vl}^{-1})^o=\arg \max_{\mathbf{Q}_{vl}^{-1}\succeq\mathbf{0}} \mathcal{L}((\mathbf{Q}_{vl}^{-1}),\mu_{vl}^o)$.
% \begin{equation}
% \begin{aligned}
%     \mathcal{L}(\mathbf{Q}_{vl}^{-1},\mu_{vl}^o)\leq& (1-\mu_{vl}^o)\sum_{i=1}^N \log_2 ({\lambda_{vlqi}^o(p{\lambda}_{li}^2+\sigma^2)+1})-\\&\log_2 (\lambda_{vlqi}^o\sigma^2+1)+\mu_{vl}^o C_{sc}\\=&\mathcal{L}({(\mathbf{Q}_{vl}^{-1})}^{o},\mu_{vl}^o)
%     \label{eq62}
%     \end{aligned}
% \end{equation}
Consequently, both optimality conditions in (\ref{eq56}) are satisfied for ${(\mathbf{Q}_{vl}^{-1})}^o,\mu_{vl}^o$, which means that ${(\mathbf{Q}_{vl}^{-1})}^o,\mu_{vl}^o$ are the global maximizer of the maximization problem defined in (\ref{sumRate_max_prob_VC}).

\end{appendices}

\bibliographystyle{ieeetr}
\bibliography{refs}
%TC: and ignore
\end{document}